\documentstyle[12pt,epsf]{article}
\epsfverbosetrue
\def\figlabel#1{\xdef#1{\thefigure}}

\def\figalign#1#2#3#4#5#6{
\begin{figure}
\centerline{
\hbox to 2.5truein{\vtop{\hsize=2.5truein\epsfxsize=6cm
\centerline{\epsfbox{#1} }
\caption[]{#3}
\figlabel{#2} }}
\qquad\hbox to 2.5truein{\vtop{\hsize=2.5truein\epsfxsize=6cm
\centerline{\epsfbox{#4} }
\caption[]{#6}
\figlabel{#5} }} }
\end{figure} }

\newcommand{\RR}{{\mbox{{\bf R}}}}

\newcommand{\CS}{{\scriptstyle {\rm CS}}}

\newcommand{\beq}{\begin{equation}}
\newcommand{\eeq}{\end{equation}}
\newcommand{\bear}{\begin{eqnarray}}
\newcommand{\eear}{\end{eqnarray}}
\newcommand{\W}{{\cal W}}

\def\tr{{\rm Tr}}

\def\ex{{\hbox{\rm e}}}

\newcommand{\k}{{\cal K}}
\newcommand{\ie}{{\it i.e.}}

\linethickness{.05 pt}

\newsavebox{\diba}
\newsavebox{\dibb}
\newsavebox{\dibc}
\newsavebox{\dibd}
\newsavebox{\dibe}
\newsavebox{\dibf}
\newsavebox{\dibg}
\newsavebox{\dibh}
\newsavebox{\dibi}
\newsavebox{\dibj}
\newsavebox{\dibk}
\newsavebox{\dibl}
\newsavebox{\dibm}
\newsavebox{\dibn}
\newsavebox{\dibo}
\newsavebox{\dibp}
\newsavebox{\dibq}
\newsavebox{\dibr}
\newsavebox{\dibs}
\newsavebox{\dibt}
\newsavebox{\dibu}
\newsavebox{\dibv}
\newsavebox{\dibw}
\newsavebox{\dibx}
\newsavebox{\diby}
\newsavebox{\dibz}

\savebox{\diba}(0,2){
\put(2,2) {\circle{4}}
\put(2,0){\line(0,1){4}} }

\savebox{\dibb}(0,2){
\put(2,2) {\circle{4}}
\put(1.5,0.1){\line(0,1){3.9}}
\put(2.5,0.1){\line(0,1){3.9}} }

\savebox{\dibc}(0,2){
\put(2,2) {\circle{4}}
\put(2,0){\line(0,1){4}} 
\put(0,2){\line(1,0){4}} }

\savebox{\dibd}(0,2){
\put(2,2) {\circle{4}}
\put(.2,1){\line(1,0){3.45}}
\put(0.1,1.7){\line(2,1){3.5}}
\put(.6,3.4){\line(2,-1){3.5}}  }

\savebox{\dibe}(0,2){
\put(2,2) {\circle{4}}
\put(2,0){\line(0,1){4}}
\put(1,0.2){\line(0,1){3.5}}
\put(3,0.2){\line(0,1){3.5}}  }

\savebox{\dibf}(0,2){
\put(2,2) {\circle{4}}
\put(0 ,2){\line(1,0){4}}
\put(1,0.2){\line(0,1){3.5}}
\put(3,0.2){\line(0,1){3.5}}  }

\savebox{\dibg}(0,2){
\put(2,2) {\circle{4}}
\put(2,0){\line(0,1){4}}
\put(0.2,1.2){\line(2,1){3.6}}
\put(.2,2.9){\line(2,-1){3.6}} }

\savebox{\dibh}(0,0){
\put(2,2) {\circle{4}}
\put(0,2){\line(1,0){4}}
\put(2,0){\line(0,1){2.0}} }

\savebox{\dibi}(0,2){
\put(2,2) {\circle{4}}
\put(0,2){\line(1,0){4}}
\put(2,0){\line(0,1){2.0}}
\put(.6,3.4){\line(1,0){2.8}} }

\savebox{\dibj}(0,0){
\put(2,2) {\circle{4}}
\put(0 ,2){\line(1,0){4}}
\put(1,.25){\line(0,1){1.8}}
\put(3,.25){\line(0,1){1.8}}  }

\savebox{\dibk}(0,2){
\put(2,2) {\circle{4}}
\put(.5,.65){\line(0,1){2.7}}
\put(1.5,.2){\line(0,1){3.8}}
\put(2.5,.2){\line(0,1){3.8}}
\put(3.5,.65){\line(0,1){2.7}} }

\savebox{\dibl}(0,2){
\put(2,2) {\circle{4}}
\put(.6,.6){\line(1,0){2.75}}
\put(.25,1){\line(1,0){3.5}}
\put(0.1,1.7){\line(2,1){3.5}}
\put(.55,3.35){\line(2,-1){3.5}} }

\savebox{\dibm}(0,2){
\put(2,2) {\circle{4}}
\put(0,2){\line(1,0){4}}
\put(0.1,1.5){\line(1,0){3.9}}
\put(.25,.9){\line(2,1){3.7}}
\put(.65,3.4){\line(1,0){2.75}} }

\savebox{\dibn}(0,2){
\put(2,2) {\circle{4}}
\put(0.1,2.2){\line(3,1){3.5}}
\put(.55,3.35){\line(3,-1){3.5}}
\put(0.6,.6){\line(3,1){3.5}}
\put(.15,1.75){\line(3,-1){3.5}} }

\savebox{\dibo}(0,2){
\put(2,2) {\circle{4}}
\put(2,0){\line(0,1){4}}
\put(1,0.2){\line(0,1){3.5}}
\put(3,0.2){\line(0,1){3.5}}
\put(0 ,2){\line(1,0){4}}  }

\savebox{\dibp}(0,2){
\put(2,2) {\circle{4}}
\put(.45,.7){\line(0,1){2.6}}
\put(0,2){\line(1,0){4}}
\put(.65,.6){\line(1,0){2.75}}
\put(2,0){\line(0,1){4}} }

\savebox{\dibq}(0,2){
\put(2,2) {\circle{4}}
\put(.6,.65){\line(1,0){2.8}}
\put(2,0){\line(0,1){4}}
\put(0.1,1.7){\line(2,1){3.5}}
\put(.55,3.35){\line(2,-1){3.5}} }

\savebox{\dibr}(0,2){
\put(2,2) {\circle{4}}
\put(1.5,.1){\line(0,1){3.9}}
\put(2.5,.1){\line(0,1){3.9}}
\put(.1,1.5){\line(1,0){3.9}}
\put(.1,2.5){\line(1,0){3.9}} }

\savebox{\dibs}(0,2){
\put(2,2) {\circle{4}}
\put(1.5,.1){\line(0,1){3.9}}
\put(2.5,.1){\line(0,1){3.9}}
\put(0.25,1.1){\line(2,1){3.5}}
\put(.25,2.9){\line(2,-1){3.5}} }

\savebox{\dibt}(0,2){
\put(2,2) {\circle{4}}
\put(0.25,1.1){\line(2,1){3.5}}
\put(.25,2.9){\line(2,-1){3.5}}
\put(0,2){\line(1,0){4}}
\put(2,0){\line(0,1){4}} }

\savebox{\dibu}(0,0){
\put(2,2) {\circle{4}}
\put(.2,1.5){\line(1,0){3.8}}
\put(2,0){\line(0,1){1.5}}
\put(.2,2.2){\line(1,0){3.8}}
\put(.4,3){\line(1,0){3.2}}}

\savebox{\dibv}(0,0){
\put(2,2) {\circle{4}}
\put(.2,1.5){\line(1,0){3.8}}
\put(2,0){\line(0,1){1.5}}
\put(.2,2.2){\line(1,0){3.8}}
\put(2,2.2){\line(0,1){1.8}}}

\savebox{\dibw}(0,0){
\put(2,2) {\circle{4}}
\put(0,2){\line(1,0){4}}
\put(2,0){\line(0,1){2.0}}
\put(1,.3){\line(0,1){1.7}}
\put(3,.25){\line(0,1){1.7}} }

\savebox{\dibx}(0,0){
\put(2,2) {\circle{4}}
\put(.3,1.2){\line(1,0){3.6}}
\put(1.2,1.2){\line(0,1){1.6}}
\put(.3,2.8){\line(1,0){3.6}}
\put(2.8,1.2){\line(0,1){1.6}}}

\savebox{\diby}(0,0){
\put(2,2) {\circle{4}}
\put(0 ,2){\line(1,0){4}}
\put(1,.25){\line(0,1){1.8}}
\put(3,.25){\line(0,1){1.8}}
\put(.3,3){\line(1,0){3.4}} }

\savebox{\dibz}(0,2){
\put(2,2) {\circle{4}}
\put(0,2){\line(1,0){4}}
\put(.6,3.4){\line(1,0){2.7}}
\put(0.25,1.1){\line(2,1){3.5}}
\put(.25,2.9){\line(2,-1){3.5}} }

\newsavebox{\faca}
\newsavebox{\facy}
\newsavebox{\facb}
\newsavebox{\facc}
\newsavebox{\facd}
\newsavebox{\face}
\newsavebox{\facf}
\newsavebox{\facg}
\newsavebox{\fach}
\newsavebox{\faci}
\newsavebox{\facj}
\newsavebox{\fack}
\newsavebox{\facl}
\newsavebox{\facm}
\newsavebox{\facn}
\newsavebox{\faco}
\newsavebox{\facp}
\newsavebox{\facq}
\newsavebox{\facr}
\newsavebox{\facs}
\newsavebox{\fact}
\newsavebox{\facu}
\newsavebox{\facv}
\newsavebox{\facw}
\newsavebox{\facx}

\savebox{\faca}(0,2){
\put(2,2) {\circle{4}}
\put(2,0){\line(0,1){4}} }

\savebox{\facb}(0,2){
\put(2,2) {\circle{4}}
\put(1.5,0.1){\line(0,1){3.9}}
\put(2.5,0.1){\line(0,1){3.9}} }

\savebox{\facc}(0,2){
\put(2,2) {\circle{4}}
\put(2,0){\line(0,1){4}} 
\put(0,2){\line(1,0){4}} }

\savebox{\facd}(0,2){
\put(2,2) {\circle{4}}
\put(.2,1){\line(1,0){3.45}}
\put(0.1,1.7){\line(2,1){3.5}}
\put(.6,3.4){\line(2,-1){3.5}}  }

\savebox{\face}(0,2){
\put(2,2) {\circle{4}}
\put(2,0){\line(0,1){4}}
\put(1,0.2){\line(0,1){3.5}}
\put(3,0.2){\line(0,1){3.5}}  }

\savebox{\facf}(0,2){
\put(2,2) {\circle{4}}
\put(0 ,2){\line(1,0){4}}
\put(1,0.2){\line(0,1){3.5}}
\put(3,0.2){\line(0,1){3.5}}  }

\savebox{\facg}(0,2){
\put(2,2) {\circle{4}}
\put(2,0){\line(0,1){4}}
\put(0.2,1.2){\line(2,1){3.6}}
\put(.2,2.9){\line(2,-1){3.6}} }

\savebox{\fach}(0,0){
\put(2,2) {\circle{4}}
\put(0,2){\line(1,0){4}}
\put(2,0){\line(0,1){2.0}} }

\savebox{\faci}(0,2){
\put(2,2) {\circle{4}}
\put(0,2){\line(1,0){4}}
\put(2,0){\line(0,1){2.0}}
\put(.6,3.4){\line(1,0){2.8}} }

\savebox{\facj}(0,0){
\put(2,2) {\circle{4}}
\put(0 ,2){\line(1,0){4}}
\put(1,.25){\line(0,1){1.8}}
\put(3,.25){\line(0,1){1.8}}  }

\savebox{\fack}(0,2){
\put(2,2) {\circle{4}}
\put(.5,.65){\line(0,1){2.7}}
\put(1.5,.2){\line(0,1){3.8}}
\put(2.5,.2){\line(0,1){3.8}}
\put(3.5,.65){\line(0,1){2.7}} }

\savebox{\facl}(0,2){
\put(2,2) {\circle{4}}
\put(.6,.6){\line(1,0){2.75}}
\put(.25,1){\line(1,0){3.5}}
\put(0.1,1.7){\line(2,1){3.5}}
\put(.55,3.35){\line(2,-1){3.5}} }

\savebox{\facm}(0,2){
\put(2,2) {\circle{4}}
\put(0,2){\line(1,0){4}}
\put(0.1,1.5){\line(1,0){3.9}}
\put(.25,.9){\line(2,1){3.7}}
\put(.65,3.4){\line(1,0){2.75}} }

\savebox{\facn}(0,2){
\put(2,2) {\circle{4}}
\put(0.1,2.2){\line(3,1){3.5}}
\put(.55,3.35){\line(3,-1){3.5}}
\put(0.6,.6){\line(3,1){3.5}}
\put(.15,1.75){\line(3,-1){3.5}} }

\savebox{\faco}(0,2){
\put(2,2) {\circle{4}}
\put(2,0){\line(0,1){4}}
\put(1,0.2){\line(0,1){3.5}}
\put(3,0.2){\line(0,1){3.5}}
\put(0 ,2){\line(1,0){4}}  }

\savebox{\facp}(0,2){
\put(2,2) {\circle{4}}
\put(.45,.7){\line(0,1){2.6}}
\put(0,2){\line(1,0){4}}
\put(.65,.6){\line(1,0){2.75}}
\put(2,0){\line(0,1){4}} }

\savebox{\facq}(0,2){
\put(2,2) {\circle{4}}
\put(.6,.65){\line(1,0){2.8}}
\put(2,0){\line(0,1){4}}
\put(0.1,1.7){\line(2,1){3.5}}
\put(.55,3.35){\line(2,-1){3.5}} }

\savebox{\facr}(0,2){
\put(2,2) {\circle{4}}
\put(1.5,.1){\line(0,1){3.9}}
\put(2.5,.1){\line(0,1){3.9}}
\put(.1,1.5){\line(1,0){3.9}}
\put(.1,2.5){\line(1,0){3.9}} }

\savebox{\facs}(0,2){
\put(2,2) {\circle{4}}
\put(1.5,.1){\line(0,1){3.9}}
\put(2.5,.1){\line(0,1){3.9}}
\put(0.25,1.1){\line(2,1){3.5}}
\put(.25,2.9){\line(2,-1){3.5}} }

\savebox{\fact}(0,2){
\put(2,2) {\circle{4}}
\put(0.25,1.1){\line(2,1){3.5}}
\put(.25,2.9){\line(2,-1){3.5}}
\put(0,2){\line(1,0){4}}
\put(2,0){\line(0,1){4}} }

\savebox{\facu}(0,0){
\put(2,2) {\circle{4}}
\put(.2,1.5){\line(1,0){3.8}}
\put(2,0){\line(0,1){1.5}}
\put(.2,2.2){\line(1,0){3.8}}
\put(.4,3){\line(1,0){3.2}}}

\savebox{\facv}(0,0){
\put(2,2) {\circle{4}}
\put(.2,1.5){\line(1,0){3.8}}
\put(2,0){\line(0,1){1.5}}
\put(.2,2.2){\line(1,0){3.8}}
\put(2,2.2){\line(0,1){1.8}}}

\savebox{\facw}(0,0){
\put(2,2) {\circle{4}}
\put(0,2){\line(1,0){4}}
\put(2,0){\line(0,1){2.0}}
\put(1,.3){\line(0,1){1.7}}
\put(3,.25){\line(0,1){1.7}} }

\savebox{\facx}(0,0){
\put(2,2) {\circle{4}}
\put(.3,1.2){\line(1,0){3.6}}
\put(1.2,1.2){\line(0,1){1.6}}
\put(.3,2.8){\line(1,0){3.6}}
\put(2.8,1.2){\line(0,1){1.6}}}

\savebox{\facy}(0,0){
\put(2,2) {\circle{4}}
\put(0 ,2){\line(1,0){4}}
\put(1,.25){\line(0,1){1.8}}
\put(3,.25){\line(0,1){1.8}}
\put(.3,3){\line(1,0){3.4}} }

\newcommand{\lld}{\mbox{ {\Large {\it D}}} }
\newcommand{\lldi}{\mbox{ {\Large {\it D$^i$}}} }
\newcommand{\lldii}{\mbox{ {\Large {\it D$^{ii}$}}} }
\newcommand{\lldij}{\mbox{ {\Large {\it D$^{ij}$}}} }
\newcommand{\lldia}{\mbox{ {\Large {\it D$^{ij,a}$}}} }
\newcommand{\lldib}{\mbox{ {\Large {\it D$^{ij,b}$}}} }
\newcommand{\lls}{\mbox{ {\Large {\it S}}} }

\newcommand{\dda}{\mbox{ $_{\, \usebox{\diba} \hskip 12pt }$ } }

\newcommand{\ddb}{\mbox{ $_{\, \hskip -4pt\usebox{\dibb} \hskip 12pt }$ } }
\newcommand{\ddc}{\mbox{ $_{\, \hskip -4pt \usebox{\dibc} \hskip 12pt }$ } }
\newcommand{\ddd}{\mbox{ $_{\, \hskip -4pt\usebox{\dibd} \hskip 12pt }$ } }
\newcommand{\dde}{\mbox{ $_{\, \hskip -4pt\usebox{\dibe} \hskip 12pt }$ } }
\newcommand{\ddf}{\mbox{ $_{\, \hskip -4pt\usebox{\dibf} \hskip 12pt }$ } }
\newcommand{\ddg}{\mbox{ $_{\, \hskip -4pt\usebox{\dibg} \hskip 12pt }$ } }

\newcommand{\ddi}{\mbox{ $_{\, \hskip -4pt\usebox{\dibi} \hskip 12pt }$ } }
\newcommand{\ddj}{\mbox{ $_{\, \hskip -4pt\usebox{\dibj} \hskip 12pt }$ } }
\newcommand{\ddk}{\mbox{ $_{\, \hskip -4pt\usebox{\dibk} \hskip 12pt }$ } }
\newcommand{\ddl}{\mbox{ $_{\, \hskip -4pt\usebox{\dibl} \hskip 12pt }$ } }
\newcommand{\ddm}{\mbox{ $_{\, \hskip -4pt\usebox{\dibm} \hskip 12pt }$ } }
\newcommand{\ddn}{\mbox{ $_{\, \hskip -4pt\usebox{\dibn} \hskip 12pt }$ } }
\newcommand{\ddo}{\mbox{ $_{\, \hskip -4pt\usebox{\dibo} \hskip 12pt }$ } }
\newcommand{\ddp}{\mbox{ $_{\, \hskip -4pt\usebox{\dibp} \hskip 12pt }$ } }
\newcommand{\ddq}{\mbox{ $_{\, \hskip -4pt\usebox{\dibq} \hskip 12pt }$ } }
\newcommand{\ddr}{\mbox{ $_{\, \hskip -4pt\usebox{\dibr} \hskip 12pt }$ } }
\newcommand{\dds}{\mbox{ $_{\, \hskip -4pt\usebox{\dibs} \hskip 12pt }$ } }
\newcommand{\ddt}{\mbox{ $_{\, \hskip -4pt\usebox{\dibt} \hskip 12pt }$ } }
\newcommand{\ddu}{\mbox{ $_{\, \hskip -4pt\usebox{\dibu} \hskip 12pt }$ } }
\newcommand{\ddv}{\mbox{ $_{\, \hskip -4pt\usebox{\dibv} \hskip 12pt }$ } }
\newcommand{\ddw}{\mbox{ $_{\, \hskip -4pt\usebox{\dibw} \hskip 12pt }$ } }
\newcommand{\ddx}{\mbox{ $_{\, \hskip -4pt\usebox{\dibx} \hskip 12pt }$ } }
\newcommand{\ddy}{\mbox{ $_{\, \hskip -4pt\usebox{\diby} \hskip 12pt }$ } }
\newcommand{\ddz}{\mbox{ $_{\, \hskip -4pt\usebox{\dibz} \hskip 12pt }$ } }

\newcommand{\eea}{\mbox{ {\Large {\it E}}$_{\, \usebox{\diba} \hskip 12pt }$ 
} }
\newcommand{\eeb}{\mbox{ {\Large {\it E}}$_{\, \usebox{\dibb} \hskip 12pt }$ 
} }
\newcommand{\eec}{\mbox{ {\Large {\it E}}$_{\, \usebox{\dibc} \hskip 12pt }$ 
} }
\newcommand{\eed}{\mbox{ {\Large {\it E}}$_{\, \usebox{\dibd} \hskip 12pt }$ 
} }
\newcommand{\eee}{\mbox{ {\Large {\it E}}$_{\, \usebox{\dibe} \hskip 12pt }$ 
} }
\newcommand{\eef}{\mbox{ {\Large {\it E}}$_{\, \usebox{\dibf} \hskip 12pt }$ 
} }
\newcommand{\eeg}{\mbox{ {\Large {\it E}}$_{\, \usebox{\dibg} \hskip 12pt }$ 
} }

\newcommand{\eek}{\mbox{ {\Large {\it E}}$_{\, \usebox{\dibk} \hskip 12pt }$ 
} }
\newcommand{\eel}{\mbox{ {\Large {\it E}}$_{\, \usebox{\dibl} \hskip 12pt }$ 
} }
\newcommand{\eem}{\mbox{ {\Large {\it E}}$_{\, \usebox{\dibm} \hskip 12pt }$ 
} }
\newcommand{\een}{\mbox{ {\Large {\it E}}$_{\, \usebox{\dibn} \hskip 12pt }$ 
} }
\newcommand{\eeo}{\mbox{ {\Large {\it E}}$_{\, \usebox{\dibo} \hskip 12pt }$ 
} }
\newcommand{\eep}{\mbox{ {\Large {\it E}}$_{\, \usebox{\dibp} \hskip 12pt }$ 
} }
\newcommand{\eeqq}{\mbox{ {\Large {\it E}}$_{\, \usebox{\dibq} \hskip 12pt 
}$ 
} }
\newcommand{\eer}{\mbox{ {\Large {\it E}}$_{\, \usebox{\dibr} \hskip 12pt }$ 
} }
\newcommand{\ees}{\mbox{ {\Large {\it E}}$_{\, \usebox{\dibs} \hskip 12pt }$ 
} }
\newcommand{\eet}{\mbox{ {\Large {\it E}}$_{\, \usebox{\dibt} \hskip 12pt }$ 
} }

\newcommand{\eez}{\mbox{ {\Large {\it E}}$_{\, \usebox{\dibz} \hskip 12pt }$ 
} }

\newcommand{\ffaa}{\mbox{  \usebox{\faca} \hskip 15pt  } } 
\newcommand{\ffab}{\mbox{  \usebox{\facb} \hskip 15pt  } } 
\newcommand{\ffac}{\mbox{  \usebox{\facc} \hskip 15pt  } } 
 
\newcommand{\ffae}{\mbox{  \usebox{\face} \hskip 15pt  } }

\newcommand{\ffah}{\mbox{  \usebox{\fach} \hskip 15pt  } } 
\newcommand{\ffai}{\mbox{  \usebox{\faci} \hskip 15pt  } } 
\newcommand{\ffaj}{\mbox{  \usebox{\facj} \hskip 15pt  } }
\newcommand{\ffak}{\mbox{  \usebox{\fack} \hskip 15pt  } }

\newcommand{\ffau}{\mbox{  \usebox{\facu} \hskip 15pt  } } 
\newcommand{\ffav}{\mbox{  \usebox{\facv} \hskip 15pt  } } 
\newcommand{\ffaw}{\mbox{  \usebox{\facw} \hskip 15pt  } }  
\newcommand{\ffax}{\mbox{  \usebox{\facx} \hskip 15pt  } } 
\newcommand{\ffay}{\mbox{  \usebox{\facy} \hskip 15pt  } } 

\begin{document}

\begin{titlepage}
\begin{flushright} { ~}\vskip -1in CERN-TH/98-193\\ US-FT-11/98\\
hep-th/9807155\\ 
\end{flushright}
\vspace*{20pt}
\bigskip
\begin{center}
 {\Large Combinatorial Formulae for Vassiliev Invariants}
\vskip 0.2truecm
{\Large from Chern-Simons Gauge Theory}
\vskip 0.9truecm

{J. M. F. Labastida$^{a,b}$ and  Esther P\'erez$^{b}$}

\vspace{1pc}

{\em $^a$ Theory Division, CERN,\\
 CH-1211 Geneva 23, Switzerland.\\
 \bigskip
  $^b$ Departamento de F\'\i sica de Part\'\i culas,\\ Universidade de
Santiago de Compostela,\\ E-15706 Santiago de Compostela, Spain.\\}

\vspace{5.5pc}

{\large \bf Abstract}
\end{center} 
We analyse the perturbative series expansion of the vacuum  expectation
value of a Wilson loop in Chern-Simons gauge theory in the  temporal gauge.
From the analysis emerges the notion of the {\it kernel} of a  Vassiliev
invariant. The kernel of a Vassiliev invariant of order $n$ is not a knot
invariant, since it depends on the regular knot projection chosen, but it 
differs from a Vassiliev invariant by terms that vanish on knots with $n$
singular crossings. We conjecture that Vassiliev invariants can be
reconstructed from their kernels. We present the general form of the kernel
of a Vassiliev invariant and we  describe the reconstruction of the full
primitive Vassiliev invariants at orders two, three and four. At orders two
and three we recover known combinatorial expressions for these invariants.
At order four we present new combinatorial expressions for the two primitive
Vassiliev invariants present at this  order.

\vspace{5pc}

\begin{flushleft} { ~}\vskip -1in CERN-TH/98-193\\ June 1998\\
\end{flushleft}


\end{titlepage}

\def\theequation{\thesection.\arabic{equation}}

\section{Introduction}
\setcounter{equation}{0}

Topological quantum field theories have provided important connections 
between different types of topological invariants. These connections are
obtained by exploiting the multiple approaches inherent in quantum field
theory. Chern-Simons gauge theory constitutes a very successful case in
this  respect. Its analysis, from both the perturbative and the
non-perturbative points of view, has provided numerous important insights
in the theory of knot and  link invariants. Non-perturbative methods
\cite{csgt,nbos,torus,king,martin,kaul} have established the connection of
Chern-Simons gauge theory with polynomial invariants as the Jones
polynomial \cite{jones} and its generalizations
\cite{homfly,kauffman,aku}. Perturbative methods
\cite{gmm,natan,vande,alla,torusknots,alts,lcone,singular} have provided
representations of Vassiliev invariants.

Gauge theories can be analysed by performing different gauge fixings. Vacuum
expectation values of gauge-invariant operators are gauge-independent and
 they can therefore be computed in different gauges.  Covariant gauges are 
simple to treat and its analysis in the case of perturbative Chern-Simons
gauge  theory has shown to lead to covariant combinatorial formulae for
Vassiliev  invariants
\cite{gmm,natan,alla,torusknots,alts}. These formulae involve 
multidimensional space and path integrals which, in general, are rather
involved to carry out explicit computations of Vassiliev invariants.
Non-covariant gauges seem  to lead to simpler formulae. However, the
subtleties inherent in non-covariant gauges \cite{leibrew} plague their
analysis with difficulties. The two non-covariant gauges that have been
more widely studied are the light-cone gauge and the temporal gauge
\cite{cata,cmartin,leibbrandt}. Both belong to  the general category of
axial gauges. In the light-cone gauge the resulting expressions for the
Vassiliev invariants turn out to be the ones involving Kontsevich integrals 
\cite{kont}.  These integrals, although  simpler than the ones appearing in
covariant gauges,  are still too complicated to carry out explicit
computations of Vassiliev invariants. Simpler expressions in which no
integrals are involved, \ie combinatorial ones, are desirable. The aim of
this paper is to reach this  goal  by studying the theory in the temporal
gauge.

From the analysis in the light-cone gauge we have learned an important 
lesson:  a naive treatment of the perturbative series expansion in a
non-covariant  gauge leads to non-invariant quantities. One needs to
introduce correction terms  to render the perturbative series invariant. At
present it is not known what  the `physical' reasons are for being forced
to introduce a correction term,  but
 it certainly must be inherent in the subtleties involved in the use of
non-covariant gauges. It is very likely that the understanding of this 
problem  in  Chern-Simons gauge theory will shed some light on the general
solution to these problems. 

In this paper we concentrate our attention on the analysis of the 
perturbative series expansion of the vacuum expectation value of a Wilson
loop in the temporal gauge. Some aspects of this gauge have been studied in
\cite{vande,cata}. In our analysis we encounter all the problems, inherent 
in non-covariant gauges, which were present in the light-cone gauge. A key
ingredient in the analysis of the perturbative series expansion is the gauge
propagator. The computation of the gauge propagator in non-covariant gauges 
is plagued with ambiguities, which are solved by demanding some properties
for  the correlation functions of the theory. These properties are usually
based on physical grounds. In our case we must demand invariance of the
vacuum expectation values of Wilson loops. As we encounter these
ambiguities in our analysis we are forced to work with a rather general
propagator in which  some of the terms are not known explicitly.
Fortunately, the complete explicit form  of the propagator is not needed to
compute vacuum expectation values of Wilson loops. Consistency, however,
forces the introduction of a correction term  similar to the one needed in
the light-cone gauge. Since in our analysis some  hypotheses are introduced
to cope with the ambiguities, we must check that our final expressions are
indeed knot invariants. We will prove this to be the case  for the terms of
the perturbative series expansion  under consideration.

The propagator possesses two terms, one whose explicit form is known and 
that depends on the signatures at the crossings, and one whose complete
explicit  form is not known but  is independent of the signatures at the
crossings. Taking  into account only the first term, we construct what we
call the kernels of the Vassiliev invariants. These are quantities that are
not knot invariants but depend on the regular projection chosen. These
kernels have the property  that they differ from an invariant by terms that
vanish on singular knots with a  high enough number of singular crossings.
More precisely, if one considers an order-$m$ kernel, it differs from an
order-$m$ Vassiliev invariant by terms  that vanish after performing the
$m$ subtractions needed to get the invariant for a singular knot with $m$
singular crossings. 

\vspace{0.1cm}

The three main goals of this paper are:

\vspace{0.1cm}

$-$ to provide the general formulae for the kernels of the Vassiliev
invariants,

$-$ to conjecture that the information contained in the kernels is 
sufficient
to reconstruct all the Vassiliev invariants at a given order,

$-$ to sustain this conjecture by showing how the reconstruction procedure 
is
implemented at orders two, three and four.

\vspace{0.1cm}

\noindent
Our results agree with the known combinatorial expression for Vassiliev 
invariants at orders two and three. The combinatorial formulae obtained for
the two primitive Vassiliev invariants of order four are new. At present we
lack an all-order reconstruction theorem but, as it will become clear  from
our analysis, the reconstruction procedure can be generalized. The key 
ingredients of our analysis are the structure of Chern-Simons gauge theory
and the factorization theorem proved in  \cite{factor}. Thus our approach
is valid  only for Vassiliev invariants based on Lie algebras.  

The paper is organized as follows. In sect. 2 we formulate the perturbative
series expansion of the vacuum expectation value of a Wilson loop in the
temporal gauge. In sect. 3 we present the kernels of Vassiliev invariants 
and  we analyse their properties. In sect. 4 we carry out the
reconstruction  procedure at order two, three and four.  In sect. 5 we
prove that the quantities  obtained at order four are invariant under
Reidemeister moves. Finally, in sect. 6 we state our conclusions. An
appendix contains  tables where the output of our combinatorial expressions
for the primitive Vassiliev invariants of orders  two to four for prime
knots up to nine crossings is compiled.

\vfill
\newpage

\section{Chern-Simons perturbation theory in the temporal gauge}
\setcounter{equation}{0}

In this section we formulate Chern-Simons gauge theory in the temporal 
gauge.
Let us consider a semi-simple gauge group $G$ and a $G$-connection on a
three-space $M$. The action of the theory is the integral of the 
Chern-Simons
form: 
\beq S_\CS (A)={k\over 4\pi}\int_{M} \tr  \Big(A\wedge {\rm d} A +  {2\over 
3}
A\wedge A\wedge A\Big), \label{action} \eeq where Tr denotes the trace over 
the
fundamental representation of $G$, and $k$ is a real parameter. The 
exponential
$\exp(i S_\CS  )$ of  this action is invariant under the gauge 
transformation
\beq A_\mu \rightarrow h^{-1}A_\mu h + h^{-1} \partial_\mu h,
\label{gauge}
\eeq 
where $h$ is a map from $M$ to $G$, when the parameter $k$ is an integer. 
Of special interest in Chern-Simons gauge theory are the Wilson loops. These 
are
gauge-invariant operators labelled by a loop $C$ embedded in $M$ and a
representation $R$ of the gauge group $G$. They are defined by the holonomy 
along the loop $C$ of the gauge connection $A$: 
\beq \W_R(C,G)=\tr\left[{\hbox{\rm
P}}_R \exp g \oint A \right], \label{wilsonloop} \eeq 
where ${\hbox{\rm P}}_R$ denotes that the integral is path-ordered and that 
$A$
must be considered in the representation $R$ of $G$. As shown in  
\cite{csgt}, 
the vacuum expectation values of products of these operators lead to 
invariants
associated to links corresponding to sets of non-intersecting loops. 

Gauge-invariant theories need to undergo a gauge-fixing procedure to make 
their
associated functional integrals well defined. Different choices of gauge 
fixing
lead to different representations of the same quantities. For vacuum 
expectation
values of products of Wilson lines one obtains different expressions for 
knot
and link invariants. The aim of this paper is to study the perturbative 
series
expansion corresponding to these quantities when one chooses the temporal 
gauge.
In the temporal gauge the condition imposed on the gauge connection $A$ is  
\beq
n^\mu A_\mu = 0, \label{tempo} 
\eeq 
where $n$ is the vector $n^\mu=(1,0,0)$. This gauge is a particular case of 
a
more general class of non-covariant gauges called axial gauges in which just
(\ref{tempo}) is imposed, $n$ being a constant vector satisfying some
condition. The light-cone gauge studied in \cite{lcone} is another 
particular
case of this type of gauges. We showed in \cite{lcone} that in the 
light-cone
gauge the perturbative series expansion of the vacuum expectation value of a
Wilson line leads to the Kontsevich integral \cite{kont} representation for 
Vassiliev
invariants.

Condition (\ref{tempo}) is imposed in the functional integral, adding the
following gauge-fixing term to the action: 
\beq S_{\scriptstyle {\rm gf}} =
\int_{M}  {\rm d}^3 x \tr ( d n^{\mu} A_{\mu} + b n^{\mu} D_{\mu} c +\alpha 
d^2),
\label{fixing}
\eeq 
where $d$ is an auxiliary field, $c$ and $b$ are ghost fields, and $\alpha$
is an arbitrary  constant. In defining perturbative series expansions, it is
convenient to rescale the fields by $A\rightarrow g A$, where
$g=\sqrt{{4\pi\over k}}$, and to integrate out $d$. The quantum action 
becomes: 
\beq  S =  -{1 \over 2}  \int_{M} {\rm d}^3 x \Bigl[\epsilon^{\mu\nu\rho}  
\Big(A_{\mu}^a \partial_{\nu} A_{\rho}^a  - {g \over 3} f_{abc}  A_{\mu}^a 
A_{\nu}^b  A_{\rho}^c \Big)
 -{1\over \alpha}(n^\mu A_\mu^a)^2  +
b^a  n^{\mu}  D_{\mu}^{ab} c^b \Bigr].
\label{component}
\eeq 

We will study the theory in the gauge $\alpha\rightarrow 0$. In this case we 
can
impose the condition (\ref{tempo}) for the terms in the action and it turns 
out
that all terms but the quadratic ones vanish. Thus, the corresponding 
Feynman
rules do not have vertices and all the information is contained in the form 
of
the propagator. This observation might not hold for some types of
three-manifolds $M$ since there can be zero modes that cannot be gauged away
and some interaction terms could remain. For the case of $M=\RR^3$, the one 
on 
which we concentrate our attention, this does not occur. However, they may 
play
an important role in other situations \cite{vande}. We are  therefore left 
with
the quadratic part of (\ref{component}). The ghost contribution is trivial 
and
for the gauge fields one obtains the following propagator in momentum space:
\beq \Delta_{\mu\nu}(p) = {\alpha\over (np)^2} \big( p_\mu p_\nu - {i\over
\alpha}(np)\epsilon_{\mu\rho\nu} n^\rho \big), \label{fuprop} 
\eeq 
which, in the
limit $\alpha \rightarrow 0$, becomes: 
\beq \Delta_{\mu\nu}(p) \rightarrow
-{i}\epsilon_{\mu\rho\nu}{ n^\rho\over np}. \label{propi} 
\eeq 
This propagator presents a pole at $np=0$ and a prescription to regulate it 
is
needed. This type of problems is standard in non-covariant gauges and 
several
prescriptions have been proposed to avoid the  pole (see \cite{leibrew}
for a review on the subject). To construct the perturbative series expansion 
of
the vacuum expectation value of a Wilson loop, we need the Fourier transform 
of
(\ref{propi}) and therefore the problem related to the presence of the pole 
is
unavoidable. In the temporal gauge, the momentum-space integral that has to 
be
carried out has the form: 
\beq \Delta(x_0,x_1,x_2) = \int_M {{\rm d} ^3p\over (2\pi)^3}
{\ex^{i(p^0 x_0 +  p^1 x_1 + p^2 x_2)} \over p^0}. 
\label{espacio} 
\eeq This
integral is ill-defined due to the pole at $p^0=0$. To make sense of it a
prescription has to be given to circumvent the pole. But, before studying 
possible prescriptions, let us first analyse the dependence of
$\Delta(x_0,x_1,x_2)$ in (\ref{espacio}) on $x_0$. The pole in $p^0$ is 
avoided
if, instead of (\ref{espacio}), one analyses the derivative of
$\Delta(x_0,x_1,x_2)$ with respect to $x_0$. Considering 
$\Delta(x_0,x_1,x_2)$ as a
distribution one obtains: 
\beq {\partial \Delta \over \partial x_0} = i
\delta(x_0) \delta(x_1) \delta(x_2). 
\label{lader} 
\eeq Integrating this
expression with respect to $x_0$, one finds that any prescription would lead 
to a
result of the following form: 
\beq \Delta(x_0,x_1,x_2)= {i\over 2} {\hbox{\rm
sign}}(x_0) \delta(x_1)\delta(x_2) +f(x_1,x_2), 
\label{solu} 
\eeq where
$f(x_1,x_2)$ is a prescription-dependent function. The important consequence 
of
the result (\ref{solu}) is that the  dependence of $\Delta(x_0,x_1,x_2)$ on
$x_0$ has to be in the form  ${\hbox{\rm sign}}(x_0) 
\delta(x_1)\delta(x_2)$. 
This observation will be crucial in our analysis. We will actually work with
the rather general formula (\ref{solu}) for $\Delta(x_0,x_1,x_2)$. This form 
of
the propagator will allow us to introduce the notion of kernel of a 
Vassiliev
invariant and to design a procedure to compute combinatorial expressions for
these invariants.

Although we will not use an explicit prescription to compute the propagator
(\ref{espacio}) let us analyse one of them to check that indeed it has the 
form (\ref{solu}). We will choose a Mandelstam-like prescription 
\cite{leibrew}
to show that the propagator in the right-hand side of (\ref{propi}) has the 
form 
advocated in \cite{cata}. Let
us consider: 
\beq 
\Delta_\epsilon(x_0,x_1,x_2) = \int_M {{\rm d} ^3p\over (2\pi)^3}
{\ex^{i(p^0 x_0 +  p^1 x_1 + p^2 x_2)} \over p^0- i\epsilon {\hbox{\rm
sign}}(p^2)}. 
\label{espacioep}
\eeq
The integral in the left-hand side of (\ref{espacioep}) has poles at 
$p^0=i\epsilon 
{\hbox{\rm sign}}(p^2)$. To
carry out the $p^0$-integration,  we close for $x_0>0$ (for $x_0<0$) the 
contour
integration in the upper (lower) half plane: 
\bear \Delta_\epsilon &=&
i\Theta(x_0) \delta(x_1) \int_{p^2>0} {{\rm d} p^2\over 2\pi} \, \ex^{ip^2( 
x_2 +
i\epsilon x_0)} -i\Theta(-x_0) \delta(x_1) \int_{p^2<0} {{\rm d} p^2\over 
2\pi} \,
\ex^{ip^2( x_2 + i\epsilon x_0)} \nonumber \\ &=& -{\delta(x_1) \over 2\pi}
{1\over x_2+i\epsilon x_0}.  \label{greta} 
\eear
Using the relation:
\beq
{1\over x_2 + i \epsilon x_0} = {\hbox{\rm P}} \Bigl( {1\over x_2}\Bigr)
- i\pi{\hbox{\rm sign}}(x_0)\delta(x_2),
\label{laparte}
\eeq
one finally obtains:
\beq
\Delta_\epsilon = 
{i\over 2} {\hbox{\rm sign}}(x_0) \delta(x_1)\delta(x_2)
-{1\over 2\pi}{\hbox{\rm P}} \Bigl( {1\over x_2}\Bigr) \delta(x_1),
\label{sormeih}
\eeq
which has the general form (\ref{solu}). This propagator is the one used in 
the
analysis performed in \cite{cata}. Notice that the prescription that we have
used breaks the symmetry under rotations in the $x_1$, $x_2$ plane, which is
present in the temporal gauge. A more symmetric prescription in which this
symmetry is kept would be preferable. Although such a prescription could be
constructed easily, we will not do it here. As stated above, we will not 
need 
in
our analysis an explicit form of the distribution $f$ in (\ref{solu}).

Taking the expression (\ref{solu}) for $\Delta(x_0,x_1,x_2)$ and 
(\ref{propi})
we can easily obtain the form of all the components of the propagator: 
\bear
\langle A_{0}^a(x) A_\mu^b(x') \rangle &=& 0, \nonumber
\\  \langle A^a_m(x) A^b_n(x') \rangle &=&
 {i\over 2} \delta^{ab} \varepsilon_{mn}  {\hbox{\rm sign}}(x_0-x_0') 
\delta(x_1-x_1')\delta(x_2-x_2') \nonumber \\
& & +f(x_1-x_1',x_2-x_2'),
\label{prop}
\eear
where $m,n=1,2$ and $\varepsilon_{mn}$ is antisymmetric with 
$\varepsilon_{12} 
=
1$. This propagator contains the basic information of the theory and 
constitutes
the essential ingredient in the construction of the perturbative series
expansion of the vacuum expectation value of a Wilson loop.

\vfill
\newpage

\section{Kernels of Vassiliev invariants}
\setcounter{equation}{0}

 Wilson loops are the gauge-invariant operators (\ref{wilsonloop}) whose 
vacuum
expectation value leads to knot invariants. Our aim is to compute the 
normalized
vacuum expectation value: 
\beq\langle \W_{R}(C,G)\rangle 
 = {1\over Z_k}
\int [DA] \W_{R}(C,G) \; {\rm e}^{iS_{CS}(A)}, 
\label{vev}
\eeq 
where $Z_k$ is the partition function of the theory:
\begin{equation} 
Z_k=\int [DA]\, {\rm e}^{iS_{CS}(A)}.
\label{parfun}
\end{equation}
This quantity leads to knot invariants and possesses a perturbative series
expansion in the coupling constant $g$. This series can be constructed
diagrammatically  from the Feynman rules of the theory. One assigns an 
external
circle to the loop $C$ carrying a representation $R$, and internal lines to 
the
propagator (\ref{prop}). These internal lines are attached to the external
circle by the vertex dictated by the form of the Wilson loop 
(\ref{wilsonloop}):
\beq V_{i}^{j\, \mu a}(x) = g (T^a_{(R)})_i^j \int {\rm d} x^\mu. 
\label{frules}
\eeq 

The perturbative series is constructed by expanding the Wilson loop operator
(\ref{wilsonloop}) and contracting the gauge fields with the propagator 
(\ref{prop}). It has the following general form \cite{alla}:
\beq
\langle \W_R(C,G)  \rangle={\rm dim}\,R\sum_{i=0}^{\infty}\sum_{j=1}^{d_i}
\alpha_{ij}(C)\,r_{ij}(R)\,x^i,
\label{general}
\eeq where  $x=ig^2/2$ is the expansion parameter. The quantities
$\alpha_{ij}(C)$, or geometrical factors, are combinations
of path integrals along the loop $C$,  and the $r_{ij}$
are  traces of products of generators of the Lie algebra associated to the
gauge  group $G$. The  index $i$ corresponds to the order in perturbation
theory, and 
$j$ labels independent contributions at a given order, $d_i$  being the
number  of these at order $i$. In (\ref{general}) ${\rm dim}\,R$ denotes
the dimension of the representation $R$. Notice the convention
$\alpha_{01}(C) = 1$. For a given order in perturbation theory,
$\{r_{ij}\}_{\{j= 1 \dots d_i\}}$ represents a basis of independent group
factors. 

The quantities $\alpha_{ij}(C)$ in (\ref{general}) are Vassiliev invariants 
of
order $i$ \cite{bilin}. Our goal is to compute them in the temporal gauge. 
Their
form in one of the more studied covariant gauges, the Landau gauge, involves
path and multidimensional space integrals. In the light-cone gauge its form 
is
simpler, but one is forced to multiply the resulting perturbative series
expansion by a factor to obtain true invariants. We will discover that  
there must be, in the
temporal gauge, also a multiplicative factor to render the terms 
of
the perturbative series expansion invariant. We will obtain some conditions 
that
this factor must satisfy and we will present an ansatz for it possessing the
same structure as the one present in the light-cone gauge. With this problem
around and the arbitrary distribution $f$ present in the propagator
(\ref{prop}), one would not expect that it is possible to obtain concrete
combinatorial expressions for the Vassiliev invariants. However, it turns 
out
that the structure of the perturbative series expansion  is so much 
constrainted
that making a natural hypotheses on the form of the multiplicative factor  
we
are able to achieve our goal.

The structure of the gauge propagator (\ref{prop}) indicates that in the
temporal gauge we must consider loops in three-space, which do not make the
argument of the delta functions vanish along a finite segment. This fact
restricts us to knot configurations which possess a regular projection
into the plane $x_1, x_2$. This does not imply any loss of generality since 
any
knot can be continuously deformed  to one of that type.

Given a regular projection $\k$  of a knot $K$  onto the $x_1,x_2$ plane we 
will
construct first the perturbative series expansion of the vacuum expectation
value of the corresponding Wilson loop using only the first term of the
propagator (\ref{prop}). At order $i$ the terms in the series involve $i$
propagators and so, in considering only the first term of (\ref{prop}), we 
are
missing terms with all powers of $f(x_1,x_2)$ from 1 to $i$. The term with 
$f^i$
does not depend on $x_0$ and therefore it only contains information on the
shadow of the knot projection $\k$ on the plane $x_1,x_2$, \ie\ it does not
contain information on the signs of the crossings. Terms with lower powers 
of
$f$ contain some information on the crossings, but they vanish after 
considering 
the signed sums involved in the invariants associated to singular knots.
Vassiliev invariants for singular knots are defined as an iterated sum of 
the
differences when a singular point is resolved by an overcrossing and an
undercrossing. Terms with $f^j$ vanish for singular knots with $i-j+1$
singularities. Thus what we are missing by considering at order $i$ only the
first term in the propagator (\ref{prop}) is something that vanishes on 
knots
with $i$ singularities. Of course, being an invariant of  finite type, the 
whole
term vanishes for knots with more than $i$ singularities. The main goal of 
this
paper is to conjecture that the full invariant can be reconstructed from the 
quantities that result from a consideration of only the first term in  
(\ref{prop}).  We will
call those quantities  {\it kernels} of Vassiliev invariants. We will show
explicitly  the reconstruction from the kernels  at orders two, three
and four.

In order to understand the role played by the first term of (\ref{prop}) in 
the
perturbative series expansion, let us analyse in detail the second-order 
contribution. The quantity to be computed is of the form: 
\beq \int_{v<w} dv dw
\dot x^m(v) \dot x^n(w) \varepsilon_{mn} {\hbox{\rm sign}} (x_0(v)-x_0(w))
\delta(x_1(v)-x_1(w)) \delta(x_2(v)-x_2(w)). \label{dinas}
\eeq
Two types of contributions can be encountered. There are other contributions 
when 
$v$ and
$w$ get close to each other. These contributions are related to framing and 
they
can be analysed as in \cite{lcone}, giving the standard framing factor. 
Since
they are simple to control, we will not consider them here. There are
contributions when  $x_1(v)=x_1(w)$ and $x_2(v)=x_2(w)$, but $v\neq w$. 
These
situations correspond precisely to the crossings. Let us suppose that the 
knot
projection $\k$ has $n$ crossings labelled by $j$, $j=1,\dots,n$. At a 
crossing
$j$ the parameters $v$ and $w$ take values $v=s_j$ and $w=t_j$, with 
$s_j\neq
t_j$. The delta functions present in (\ref{dinas}) can be evaluated very 
easily.
Equation (\ref{dinas}) becomes: 
\beq \sum_{j=1}^n \int_{v<w} dv dw {\dot x^m(v) \dot
x^n(w) \varepsilon_{mn}\over |\dot x^m(v) \dot x^n(w) \varepsilon_{mn}|}
{\hbox{\rm sign}} (x_0(v)-x_0(w)) \delta(v-s_j) \delta(w-t_j), 
\label{dinasdos}
\eeq which is precisely a sum of the crossing signs $\epsilon_j$ at the
crossings $j=1,\dots,n$: \beq \sum_{j=1}^n \epsilon_j. \label{dinastres} 
\eeq

The structure of the computation of (\ref{dinas}) clearly generalizes. 
Whenever
a term containing the first part of the propagator (\ref{prop}) appears in 
the
perturbative series expansion, it can be traded by crossing signs. In 
general,
one obtains powers of crossing signs multiplying quantities which depend 
only 
on
the shadow of the regular projection. This proves that, as stated above, an
order-$i$ contribution with $j$ powers of $f$ (and therefore $i-j$ powers of
crossing signs) vanishes for singular knots with $i-j+1$ singularities.

\begin{figure}
\centerline{\hskip.4in \epsffile{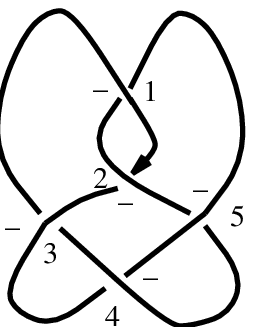}}
\caption{Example of a knot projection.}
\label{labels}
\end{figure}

The argument leading to (\ref{dinastres}) generalizes and allows us to write 
the general form of the perturbative
series expansion when only the first term in (\ref{prop}) is taken into 
account.
Let $K$ be a knot whose regular projection $\k$ presents $n$ crossings with
crossing signs $\epsilon_j$, $j=1,\dots,n$. The resulting perturbative 
series
expansion turns out to be: 
\bear {\cal N}(\k) &=& \sum_{k=0}^\infty\Bigg(
\sum_{i_1 < \cdots <i_k} \epsilon_{i_1} \cdots \epsilon_{i_k}  {\cal T}
(i_1,\dots,i_k) \nonumber \\ & & + {1\over (2!)^2} \sum_{\sigma \in P_2 
\atop
{j\neq i_1,\dots, i_{k-2} \atop i_1 < \cdots < i_{k-2}} } \epsilon_{j}^2
\epsilon_{i_1} \cdots \epsilon_{i_{k-2}} {\cal T} 
(j,\sigma;i_1,\dots,i_{k-2})
\nonumber \\ & & + {1\over (3!)^2} \sum_{\sigma \in P_3 \atop {j\neq 
i_1,\dots,
i_{k-3} \atop i_1 < \cdots < i_{k-3} }} \epsilon_{j}^3 \epsilon_{i_1} \cdots
\epsilon_{i_{k-3}} {\cal T} (j,\sigma;i_1,\dots,i_{k-3}) \nonumber \\ & &
\nonumber \\ & & {\hskip2cm} \cdots \nonumber \\
& & \nonumber \\
& & + {1\over (m!)^2} \sum_{\sigma \in P_m \atop {j\neq i_1,\dots, i_{k-m} 
\atop
i_1 < \cdots < i_{k-m} }} \epsilon_{j}^m \epsilon_{i_1} \cdots
\epsilon_{i_{k-m}} {\cal T} (j,\sigma;i_1,\dots,i_{k-m}) \nonumber \\
& & \nonumber \\
& & {\hskip2cm} \cdots \nonumber \\
& & \nonumber \\
& & + {1\over (k!)^2} \sum_{\sigma \in P_k \atop j} \epsilon_{j}^k  {\cal T} 
(j,\sigma) \Bigg)
\label{nucleos}
\eear
Several comments are in order relative to this expression. The first term 
comes
from the contribution in which all the propagators are attached to different
crossings. The second when two propagators are attached to the same 
crossing and the rest to different crossings, and so
on. Since we are dealing with an ordered integral, a correction must be 
included when there are more than one propagator at a crossing. This 
correction accounts for the fact that we do not have full integrations over 
the delta functions and corresponds to the factor $1/(m!)^2$. In
(\ref{nucleos}),
$P_m$ denotes the permutation group. When several propagators coincide at a
crossing one must include all the permutations in which they can be
arranged. The factors 
${\cal T} (j,\sigma;i_1,\dots,i_{k-m})$ are group factors, and they are 
computed in the
following way: given a set of crossings $i_1, i_2,\dots, i_{i-m}$ and a
permutation $\sigma\in P_m$, the corresponding group factor ${\cal T}
(j,\sigma;i_1,\dots,i_{k-m})$  is the result of taking a trace over the 
product
of group generators which is obtained after assigning a group generator to 
each of the crossings $i_1, i_2,\dots, i_{i-m}$, and $m$ generators to 
crossing $j$,
and travelling along the knot starting from a base point. The first time  
one encounters $j$ a product of $m$ group generators is introdiced; the 
second
time the product is similar, but with the indices rearranged according to 
the
permutation $\sigma$.

\begin{figure}
\centerline{\hskip.4in \epsffile{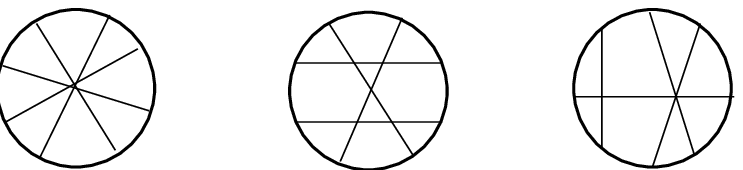}}
\caption{Group factors.}
\label{exgroup}
\end{figure}

In order to clarify the content of (\ref{nucleos}) we will  work out an 
example.
Let us consider the knot shown in fig. \ref{labels} and let us concentrate 
on 
the 
fourth
order term ($k=4$) containing some permutation  $\sigma \in P_2$: 
\beq {1\over
(2!)^2} \sum_{ {j\neq i_1,i_2} \atop i_1 < i_2} \epsilon_j^2 \epsilon_{i_1}
\epsilon_{i_2} {\cal T}(j,\sigma;i_1,i_2).
\label{ejem}
\eeq
Examples of the group factors entering this expression are:
\bear
{\cal T}(1,\sigma;2,3) &=& \tr ( T^{b_1}T^{b_2}T^{a_1}T^{a_2}
T^{\sigma(b_1)}T^{\sigma(b_2)}T^{a_1}T^{a_2}), \nonumber \\
{\cal T}(1,\sigma;3,5) &=& \tr ( T^{b_1}T^{b_2}T^{a_1}T^{a_2}
T^{\sigma(b_1)}T^{\sigma(b_2)}T^{a_2}T^{a_1}), \nonumber \\
{\cal T}(3,\sigma;1,4) &=& \tr ( T^{b_1}T^{b_2}T^{a_1}T^{a_2}
T^{a_1}T^{\sigma(b_1)}T^{\sigma(b_2)}T^{a_2}), 
\label{masejem}
\eear
where we have used the labels specified in fig. \ref{labels}. Group factors 
can 
be
represented by chord diagrams. For example if one chooses $\sigma=(12)$ the
three chord diagrams corresponding to the group factors in (\ref{masejem}) 
are
the ones pictured in fig. \ref{exgroup}.

The terms of the series (\ref{nucleos}) are not knot invariants. Besides the
knot $K$, they clearly depend on the knot projection  chosen. However, at 
order
$k$ they are knot invariants modulo terms that vanish when an order-$k$ 
signed
sum is considered. We will call the terms appearing in (\ref{nucleos}) at 
each 
order 
in
perturbation theory {\it kernels} of the Vassiliev invariants of order $k$. 
We
conjecture that these kernels contain enough information to allow 
reconstruction of the full Vassiliev invariants at each order. In the next 
section we will
present the reconstruction procedure for all primitive Vassiliev invariants 
up
to order four.

\vfill
\newpage

\section{Reconstruction of Vassiliev invariants up to order four}
\setcounter{equation}{0}

\subsection{Outline of the calculation procedure}

In this section we will obtain  combinatorial formulae for the  primitive
Vassiliev invariants up to order four using the kernels (\ref{nucleos}) 
presented in 
the
previous section. In order  to be able to reconstruct the invariant we need 
to
exploit as much as possible  all the information contained in the 
perturbative
series expansion of a Wilson loop in Chern-Simons theory. The crucial 
ingredient
is, as we observed above, that all the dependence on the crossings comes
from the first term of the propagator (\ref{prop}). Contributions not 
involving
that part are crossing independent, \ie\ they would not  distinguish between
over- or undercrossings, and consequently neither between the diagram of a 
knot
projection $\k$ and its standard ascending diagram $\alpha ( {\k} )$. Recall
that the ascending diagram of a knot projection is defined as the diagram
obtained by switching, when travelling along the knot from the base point, 
all the
undercrossings to overcrossings. There is a straightforward consequence of 
this
fact that will help us in  symplifying  our calculations.  Under the action 
of
an inversion of space, a Vassiliev invariant of even order  does not vary,
while one of odd order change sign. This means that all the  signature
contributions should be of even order  in the former and odd order in the  
later. As 
we
claim that all these contributions come from the first term of the 
propagator
(\ref{prop}), it follows that integrals with  an odd number of powers of the
function $f$ in (\ref{prop}) will not contribute to the invariant. 

Many of the ingredients entering the reconstruction procedure of the full
invariants rely on the use of the factorization theorem in Chern-Simons 
theory 
proved in \cite{factor}. According to this theorem, once a canonical basis 
is
chosen for the group factors, any  non-primitive Vassiliev invariant of a 
given
order can be written in terms of invariants of lower orders. Using this 
theorem
we will obtain a series of relations involving unknown integrals, which will
allow their solution to be such that a combinatorial formula for the
Vassiliev invariants will be obtained.

As stated above, we will have  to deal also with a Kontsevich-type global 
factor. We will assume that this factor can be written as the invariant of 
the
unknot raised to some exponent that depends on some features of the knot
projection under consideration. This will modify the perturbative series at
every even  order. We will obtain a series of consistency relations for the
exponent, which admit simple solutions.

As crucial as the calculation itself is  finding a convenient way to deal 
with 
the integrals appearing in the perturbative series expansion. On the one 
hand,
we would like to write them as explicitly as possible so that, when we know  
how they behave, relations appearing  at a given order can be used in higher
orders; on the other hand, we would like to describe them in a compact way 
so that the notation does not become too clumsy. We will introduce a 
notation that we think satisfies these two conditions.

\begin{figure}
\centerline{\hskip.4in \epsffile{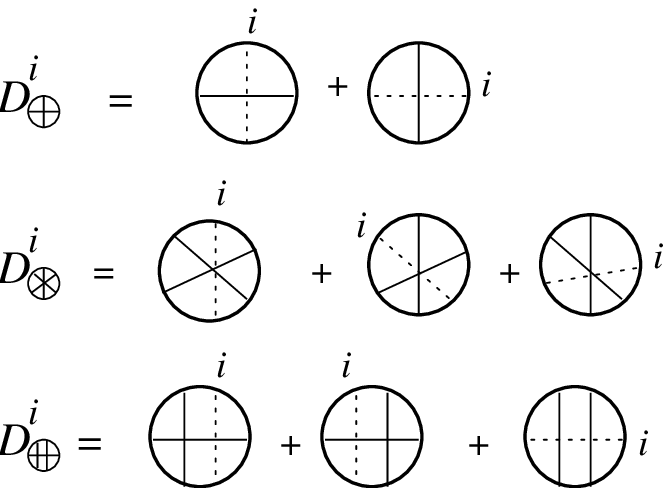}}
\caption{Examples of $D^i$ integrals.}
\label{dintegral}
\end{figure}

We will basically denote an integral made out of the $f$-dependent part of 
the 
propagator (\ref{prop}) by a capital $D$ and a subindex which will actually 
be
the Feynman diagram it comes from. Our calculations, though,  require a more
subtle labelling. Given a Feynman diagram, each chord in it usually 
represents
the propagator of the theory. Our propagator (\ref{prop}) contains  two 
pieces:
the explicit one, which leads to the signatures of the crossings, and the
$f$-dependent one. A Feynman integral will be a sum over all the possible 
ways
of identifying the chords with each of them.   So for a given Feynman 
diagram 
we
will  end up with different types of $D$  integrals, depending on how many
$f$-terms they contain. When  all the propagators in the Feynman integral 
are of this kind,  we will  simply denote it by $D_{\bigcirc}$, where the 
subindex
represents the  corresponding Feynman diagram (the void circle stands for 
any
diagram). If  only one chord  stands for the signature-dependent part, its
evaluation will simply result in a crossing sign, $\epsilon_i$, plus a
restriction of the original integration domain. We will say that the chord
standing for this factor is attached to the $i$-th crossing, meaning that 
the
ordered integration domain of the other chords of the $D$ integral is now
limited by the position of that crossing. We will write down the resulting
integral as: 
\begin{equation}  \epsilon_i D^i_{\bigcirc} \, , \label{noma}
\end{equation} 
with the superindex of $D$ denoting that one of the chords in the  diagram 
is
attached to the $i$-th crossing. This time $D$ is in fact a  sum over all 
the
possible choices of placing the signature-dependent part of the propagator
(\ref{prop}) in the $i$-th crossing (and of course a sum over the 
permutations
of the given diagram is understood everywhere). Some examples are shown in 
fig.
\ref{dintegral}. There the  integrals are represented directly by their 
Feynman
diagrams, with a dashed  line standing for a signature-dependent term of the
propagator, attached to a crossing $i$, and a continuous line representing 
the
$f$-dependent term. 

A more involved case arises when the integrand contains  two 
signature-dependent terms of the propagator (\ref{prop}). In this case we 
will distinguish three
subcases: 

- when both are attached to the same crossing the integral will then be
written  as: 
\begin{equation} 
 {1 \over 4} \epsilon_i^2 D^{ii}_{\bigcirc} \, ,
\label{nomb}
\end{equation}

\begin{figure}
\centerline{\hskip.4in \epsffile{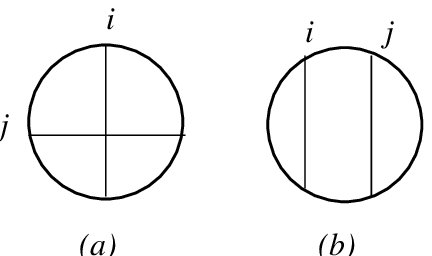}}
\caption{Possible configurations of two crossings.}
\label{twocross}
\end{figure}

\begin{figure}
\centerline{\hskip.4in \epsffile{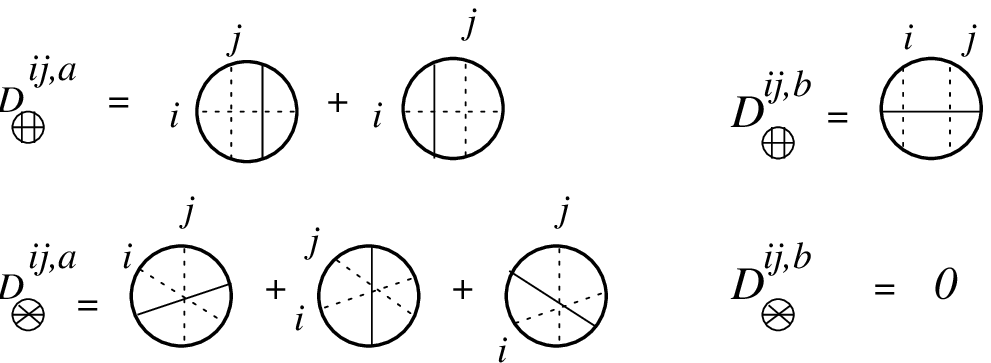}}
\caption{Examples of $D^{ij}$ integrals.}
\label{dijintegral}
\end{figure}

- when the crossings are different and, while travelling along the knot 
path, 
their labels follow the pattern in fig. \ref{twocross}$(a)$; then the 
integral will be 
denoted as:
\begin{equation} 
\epsilon_i \epsilon_j D^{ij,a}_{\bigcirc} \, ,
\label{nomc}
\end{equation}

- and when they are as in figure  \ref{twocross}$(b)$:
\begin{equation} 
\epsilon_i \epsilon_j D^{ij,b}_{\bigcirc} \, .
\label{nomd}
\end{equation}
We will denote by ${\cal C}_a$ the set of all pairs of crossings like those 
in 
\ref{twocross}$(a)$, and by ${\cal C}_b$ the pairs like in fig.
\ref{twocross}$(b)$. Examples of these cases are drawn in fig. 
\ref{dijintegral}. As we are dealing with invariants up to order four, we do 
not
need to handle the case where three or  more signature-dependent terms of 
the
propagator (but not all) are fixed to crossings. When the contribution does 
not
contain $f$-dependent terms, the Feynman integral may be read from the 
kernels
(\ref{nucleos}). We  will denote by $E_{\bigcirc}$ the sum of terms in
(\ref{nucleos}) corresponding to  the diagram specified in its subindex,
encoded in that formula in the form of the group factor ${\cal T}$.

In order to organize the perturbative series expansion, we have to make a 
choice
of basis for the group factors entering the Feynman diagrams. Once this is 
done,
the coefficients of the basis elements will be built out of a sum of Feynman
integrals, with the aproppriate factors. We will  denote these sums by
$S^E_{\bigcirc}$ when the terms involve only the  signature-dependent part 
of
the propagator, and by  $S^D_{\bigcirc}$ when they involve  the 
$f$-dependent
part. Indexes in the capital $D$ will have the same meaning as above. This 
time,
the diagram will stand for the independent group  factor as obtained 
following
the group Feynman rules given below,  in fig. \ref{groupr}.

Additional  notation is needed to write explicit combinatorial  formulae for 
the
invariants. These involve the so-called {\it crossing  numbers} \cite{hirs},
which build up the signature contributions in every  $E_{\bigcirc}$ 
integral. 
We will use the notation introduced in  \cite{hirs} for some of these
functions, as well as new ones. The key  ingredient of that notation (see
\cite{hirs} or \cite{hirsdos} for a more detailed explanation) is the 
following
definition of the signature function:

Let $\pi : S^1 \longrightarrow {\bf R}^2$ be the projection, into the $x_1,
x_2$ plane of a knot diagram $\k$. Let $s_i \in  S^1$, $i \in {\cal I}$, be 
the
pre-images of the $n$ crossings  in $\k$, with  ${\cal I} = \{ 1, \dots ,2 
n\}$ 
the index set of the labellings of the  crossings. Then, following 
\cite{hirs},
we define:
\beq  
\epsilon (i,j) = \left\{ \begin{array}{ll} \epsilon(\pi(s_i)) & \mbox{ if 
$\pi(s_i) = \pi(s_j)$  and  $i \neq j$} \\
0 & \mbox{ otherwise} 
 \end{array}
\right. 
\label{signatura}
\eeq

\begin{figure}
\centerline{\hskip.4in \epsffile{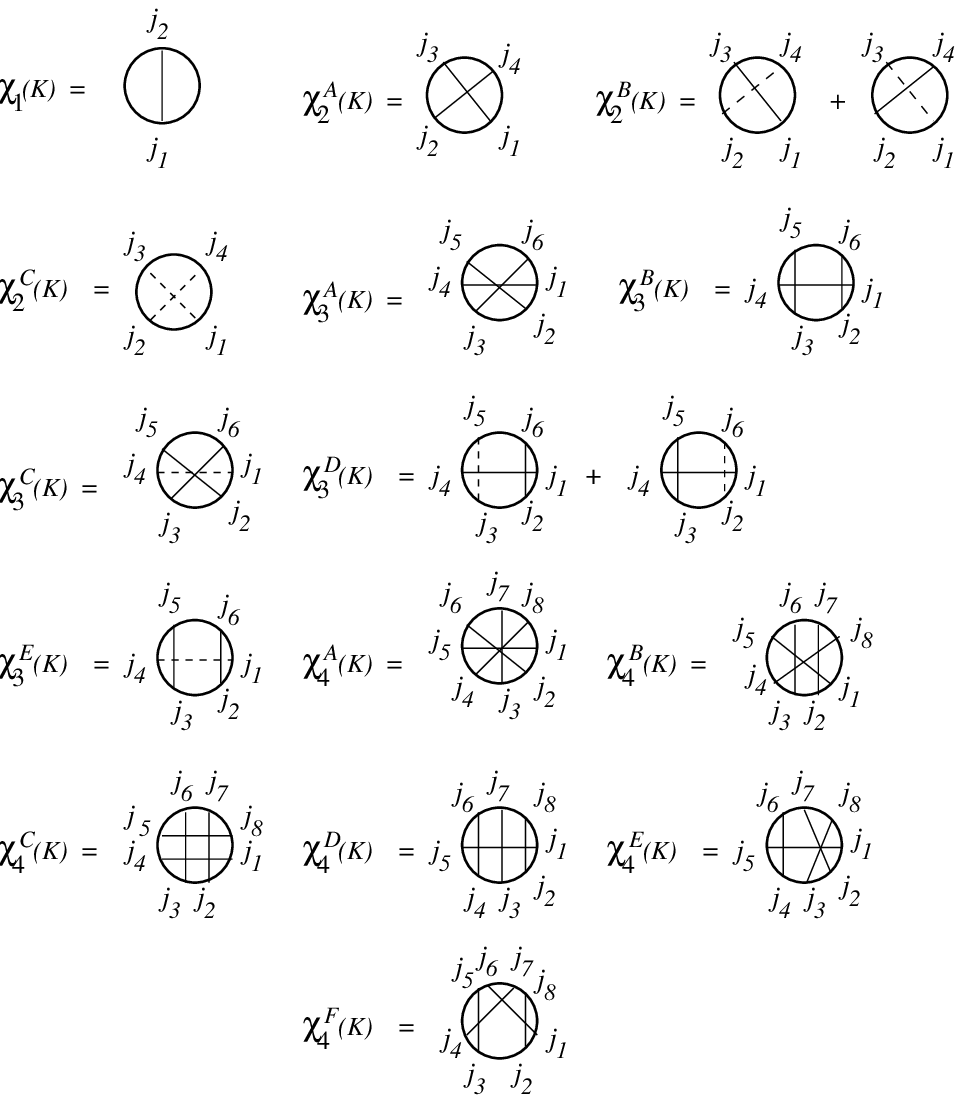}}
\caption{Diagrammatic expression for crossing numbers.}
\label{crosnumber}
\end{figure}

This function is such that  whenever two labellings happen to label the same
crossing, it gives its signature, and if they do not,  it returns zero. With 
the
help of this function one defines quantities involving powers of the 
signatures.
The following is a list of all the definitions of these quantities required
in our computations (notice that our notation for the order-two functions is
slightly different  from that in \cite{hirs}). The sums are taken over all
possible ways of  choosing the labellings in the index set $\cal I$ within 
the
given ordering.  In fig. \ref{crosnumber} we present a diagrammatic notation 
for
these definitions. There, the solid  lines stand for the signature function, 
and
the dashed ones for its square.  The diagram tells us in a straightforward 
way
the ordering that the labels must  follow when we travel along the knot, and
thus which collection of crossings  would contribute to a given crossing 
number.
Only one representative is chosen from those entering each sum. The others
can be obtained very simply from the representative performing a rotation of 
the
diagram while keeping the labels fixed. The list of functions that will be
needed below is the following: 
\bear \chi_1 (\k) &=& \sum_{j_1>j_2} \epsilon
(j_1,j_2) \label{crossnum} \\ \chi_2^A (\k) &=& \sum_{j_1>j_2>j_3>j_4} 
\epsilon
(j_1,j_3) \epsilon  (j_2,j_4)  \nonumber \\ \chi_2 ^B(\k) &=&
\sum_{j_1>j_2>j_3>j_4} \big[ \epsilon (j_1,j_3)^2 \epsilon  (j_2,j_4) + 
\epsilon (j_1,j_3)  \epsilon (j_2,j_4)^2 \big] \nonumber \\ \chi_2^C (\k) 
&=&
\sum_{j_1>j_2>j_3>j_4} \epsilon (j_1,j_3)^2 \epsilon  (j_2,j_4)^2  \nonumber 
\\
\chi_3 ^A(\k) &=& \sum_{j_1> \cdots>j_6}  \epsilon (j_1,j_4) \epsilon 
(j_2,j_5) \epsilon (j_3,j_6)  \nonumber \\
\chi_3 ^B(\k) &=& \sum_{j_1>\cdots>j_6} \big[ \epsilon (j_1,j_3) \epsilon 
(j_2,j_5) \epsilon (j_4,j_6 ) +  \epsilon (j_1,j_4) 
\epsilon (j_2,j_6) \epsilon (j_3,j_5 ) \nonumber \\
&+& \epsilon (j_1,j_5) 
\epsilon (j_2,j_4 ) \epsilon (j_3,j_6) \big]
\nonumber \\
\chi_3 ^C(\k) &=& \sum_{j_1>\cdots>j_6} \big[ \epsilon (j_1,j_4)^2 \epsilon 
(j_2,j_5) \epsilon (j_3,j_6)   +  \epsilon (j_1,j_4) \epsilon (j_2,j_5)^2 
\epsilon (j_3,j_6)   \nonumber \\
&+& \epsilon (j_1,j_4) \epsilon (j_2,j_5) \epsilon (j_3,j_6)^2   \big]
\nonumber \\
\chi_3 ^D(\k) &=& \sum_{j_1>\cdots>j_6} \big[ \epsilon (j_1,j_3)^2 \epsilon 
(j_2,j_5) \epsilon (j_4,j_6 ) +  
\epsilon (j_1,j_3) \epsilon (j_2,j_5) \epsilon (j_4,j_6 )^2 \nonumber \\ 
&+& \epsilon (j_1,j_4) \epsilon (j_2,j_6)^2 \epsilon (j_3,j_5 ) +
\epsilon (j_1,j_4) \epsilon (j_2,j_6) \epsilon (j_3,j_5 )^2 \nonumber \\
&+& \epsilon (j_1,j_5)^2 \epsilon (j_2,j_4 ) \epsilon (j_3,j_6) +
\epsilon (j_1,j_5) \epsilon (j_2,j_4 )^2 \epsilon (j_3,j_6)
 \big]
\nonumber \\
\chi_3 ^E(\k) &=& \sum_{j_1>\cdots >j_6} \big[ \epsilon (j_1,j_3) \epsilon 
(j_2,j_5)^2 \epsilon (j_4,j_6 ) +  \epsilon (j_1,j_4)^2 
\epsilon (j_2,j_6) \epsilon (j_3,j_5 ) \nonumber \\
&+& \epsilon (j_1,j_5) 
  \epsilon (j_2,j_4 ) \epsilon (j_3,j_6)^2 \big] \nonumber \\
\chi_4 ^A(\k) &=& \sum_{j_1>\cdots>j_8}  \epsilon (j_1,j_5) \epsilon 
(j_2,j_6) \epsilon (j_3,j_7) \epsilon (j_4,j_8) \nonumber \\
\chi_4 ^B(\k) &=& \sum_{j_1>\cdots>j_8} \big[ \epsilon (j_1,j_5) \epsilon 
(j_2,j_7) \epsilon (j_3,j_6) \epsilon (j_4,j_8) \nonumber \\
&+&  
\epsilon (j_1,j_6) \epsilon (j_2,j_5) \epsilon (j_3,j_7) \epsilon (j_4,j_8)
+ \epsilon (j_1,j_4) \epsilon (j_2,j_6) \epsilon (j_3,j_7) \epsilon 
(j_5,j_8) \nonumber \\
&+&  
\epsilon (j_1,j_5) \epsilon (j_2,j_6) \epsilon (j_3,j_8) \epsilon (j_4,j_7) 
\big]
\nonumber \\
\, \nonumber \\
\chi_4 ^C(\k) &=& \sum_{j_1>\cdots>j_8} \big[ \epsilon (j_1,j_6) \epsilon 
(j_2,j_5) \epsilon (j_3,j_8) \epsilon (j_4,j_7) \nonumber \\
&+&  
\epsilon (j_1,j_4) \epsilon (j_2,j_7) \epsilon (j_3,j_6) \epsilon (j_5,j_8)
\big]
\nonumber \\
\, \nonumber \\
\chi_4 ^D(\k) &=& \sum_{j_1>\cdots>j_8} \big[ \epsilon (j_1,j_7) \epsilon 
(j_2,j_6) \epsilon (j_3,j_5) \epsilon (j_4,j_8) \nonumber \\
&+&  
\epsilon (j_1,j_5) \epsilon (j_2,j_4) \epsilon (j_3,j_7) \epsilon (j_6,j_8)
+ \epsilon (j_1,j_3) \epsilon (j_2,j_6) \epsilon (j_4,j_8) \epsilon 
(j_5,j_7) \nonumber \\
&+&  
\epsilon (j_1,j_5) \epsilon (j_2,j_8) \epsilon (j_3,j_7) \epsilon (j_4,j_6) 
\big]
\nonumber \\
\, \nonumber \\
\chi_4 ^E(\k) &=& \sum_{j_1>\cdots>j_8} \big[ \epsilon (j_1,j_6) \epsilon 
(j_2,j_7) \epsilon (j_3,j_5) \epsilon (j_4,j_8) \nonumber \\
&+&  
\epsilon (j_1,j_6) \epsilon (j_2,j_4) \epsilon (j_3,j_7) \epsilon (j_5,j_8)
+ \epsilon (j_1,j_3) \epsilon (j_2,j_6) \epsilon (j_4,j_7) \epsilon 
(j_5,j_8) \nonumber \\
&+&  
\epsilon (j_1,j_5) \epsilon (j_2,j_8) \epsilon (j_3,j_6) \epsilon (j_4,j_7) 
+ \epsilon (j_1,j_7) \epsilon (j_2,j_5) \epsilon (j_3,j_6) \epsilon 
(j_4,j_8) \nonumber \\
&+&  
\epsilon (j_1,j_4) \epsilon (j_2,j_5) \epsilon (j_3,j_7) \epsilon (j_6,j_8)
+ \epsilon (j_1,j_4) \epsilon (j_2,j_6) \epsilon (j_3,j_8) \epsilon 
(j_5,j_7) \nonumber \\
&+&  
\epsilon (j_1,j_5) \epsilon (j_2,j_7) \epsilon (j_3,j_8) \epsilon (j_4,j_6) 
\big]
\nonumber \\
\, \nonumber \\
\chi_4 ^F(\k) &=& \sum_{j_1>\cdots>j_8} \big[ \epsilon (j_1,j_4) \epsilon 
(j_2,j_8) \epsilon (j_3,j_6) \epsilon (j_5,j_7) \nonumber \\
&+&  
\epsilon (j_1,j_7) \epsilon (j_2,j_5) \epsilon (j_3,j_8) \epsilon (j_4,j_6)
+ \epsilon (j_1,j_4) \epsilon (j_2,j_7) \epsilon (j_3,j_5) \epsilon 
(j_6,j_8) \nonumber \\
&+&  
\epsilon (j_1,j_6) \epsilon (j_2,j_4) \epsilon (j_3,j_8) \epsilon (j_5,j_7) 
+ \epsilon (j_1,j_3) \epsilon (j_2,j_7) \epsilon (j_4,j_6) \epsilon 
(j_5,j_8) \nonumber \\
&+&  
\epsilon (j_1,j_6) \epsilon (j_2,j_8) \epsilon (j_3,j_5) \epsilon (j_4,j_7)
+ \epsilon (j_1,j_7) \epsilon (j_2,j_4) \epsilon (j_3,j_6) \epsilon 
(j_5,j_8) \nonumber \\
&+&  
\epsilon (j_1,j_3) \epsilon (j_2,j_5) \epsilon (j_4,j_7) \epsilon (j_6,j_8) 
\big]
\nonumber 
\eear

\subsection{Vassiliev invariants of order two and three}

In this subsection we will present the reconstruction procedure to obtain a
combinatorial expression for each of the primitive Vassiliev invariants at 
orders
two and three. We will obtain the same combinatorial expressions as the ones
computed in \cite{hirs,hirsdos},  working in a covariant gauge.

As we argued above we will assume that the perturbative series expansion 
emerging
in the temporal gauge must be accompanied by a global factor which involves 
the
topological invariant for the unknot to some power. The topological 
invariant
vacuum expectation value of the Wilson line corresponding to the knot $K$ 
has
the form 
\beq \langle W({K},G)\rangle = \langle W({\k},G)\rangle_{{\rm temp}}
\times  \langle W(U,G)\rangle^{b(\k)}, \label{global} 
\eeq
where, as before, $\k$ denotes the regular projection into the $x_1,x_2$ 
plane
chosen, and $b(\k)$ is an unknown function. As in the case of the 
light-cone,
gauge we will assume that this function depends only on the shadow 
corresponding
to the projection $\k$ of the knot $K$. In other words, the quantity $b(\k)$ 
is
insensitive to crossing changes in $\k$.

We will denote the perturbative series expansion of $W({K},G)$ by: 
\beq
{1 \over d} \langle W({K},G)\rangle = 1 + \sum_{i=1}^{\infty} v_i(K) x^i,
\label{expansiona}
\eeq
where $v_i(K)$ stands for the combination of Vassiliev invariants appearing 
at order $i$, while that of $W({\k},G)_{{\rm temp}}$ denoted by:
\beq
{1 \over d} \langle W({\k},G)\rangle_{{\rm temp}} = 1 + \sum_{i=1}^{\infty} 
\hat v_i(\k) 
x^i.
\label{expansionb}
\eeq
 The quantities $\hat v_i(\k)$ do not need to be topological invariants. 
Actually, as explicitly shown in its labelling, they depend on the 
projection 
$\k$ of the knot $K$. In (\ref{expansiona}) and (\ref{expansionb}), $d=$
dim $R$, the dimension of the representation carried by the Wilson loop.

\begin{figure}
\centerline{\hskip.4in \epsffile{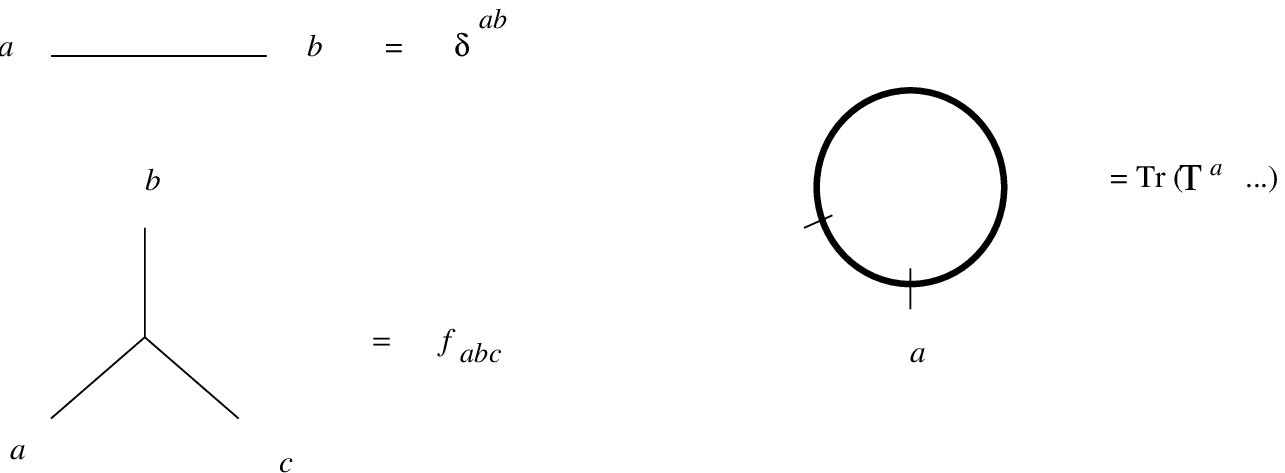}}
\caption{Group factor Feynman rules.}
\label{groupr}
\end{figure}

\begin{figure}
\centerline{\hskip.4in \epsffile{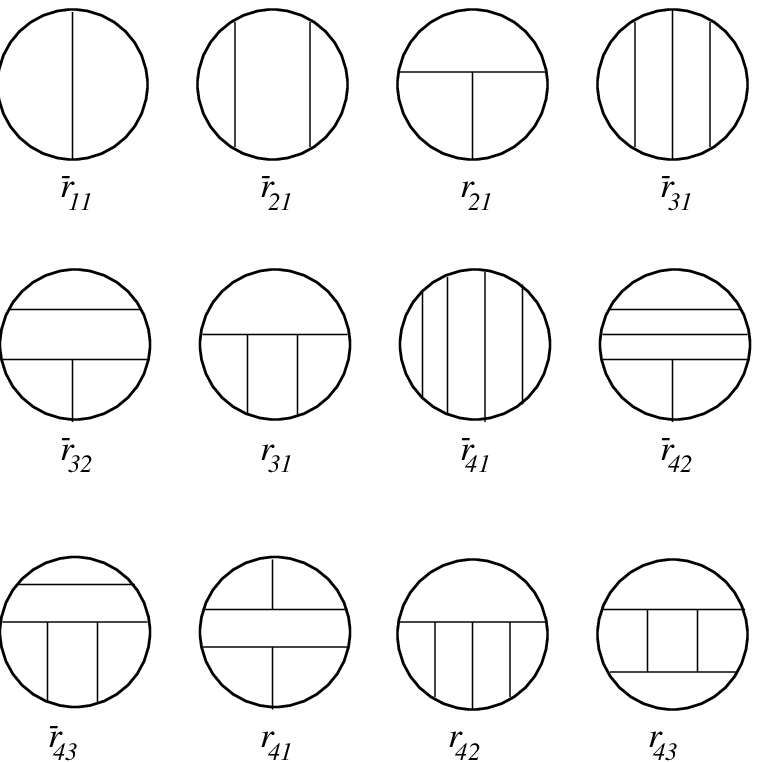}}
\caption{Choice of canonical basis up to order four.}
\label{canonical}
\end{figure}

\begin{figure}
\centerline{\hskip.4in \epsffile{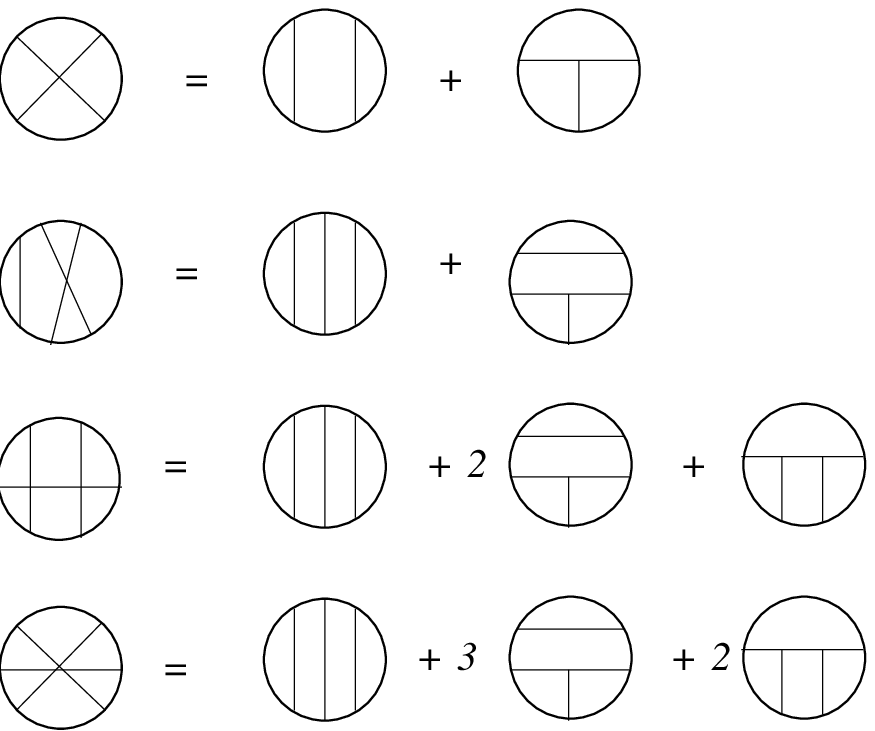}}
\caption{Expansion of chord diagrams in the canonical basis: orders two and 
three.}
\label{chor23}
\end{figure}

\begin{figure}
\centerline{\hskip.4in \epsffile{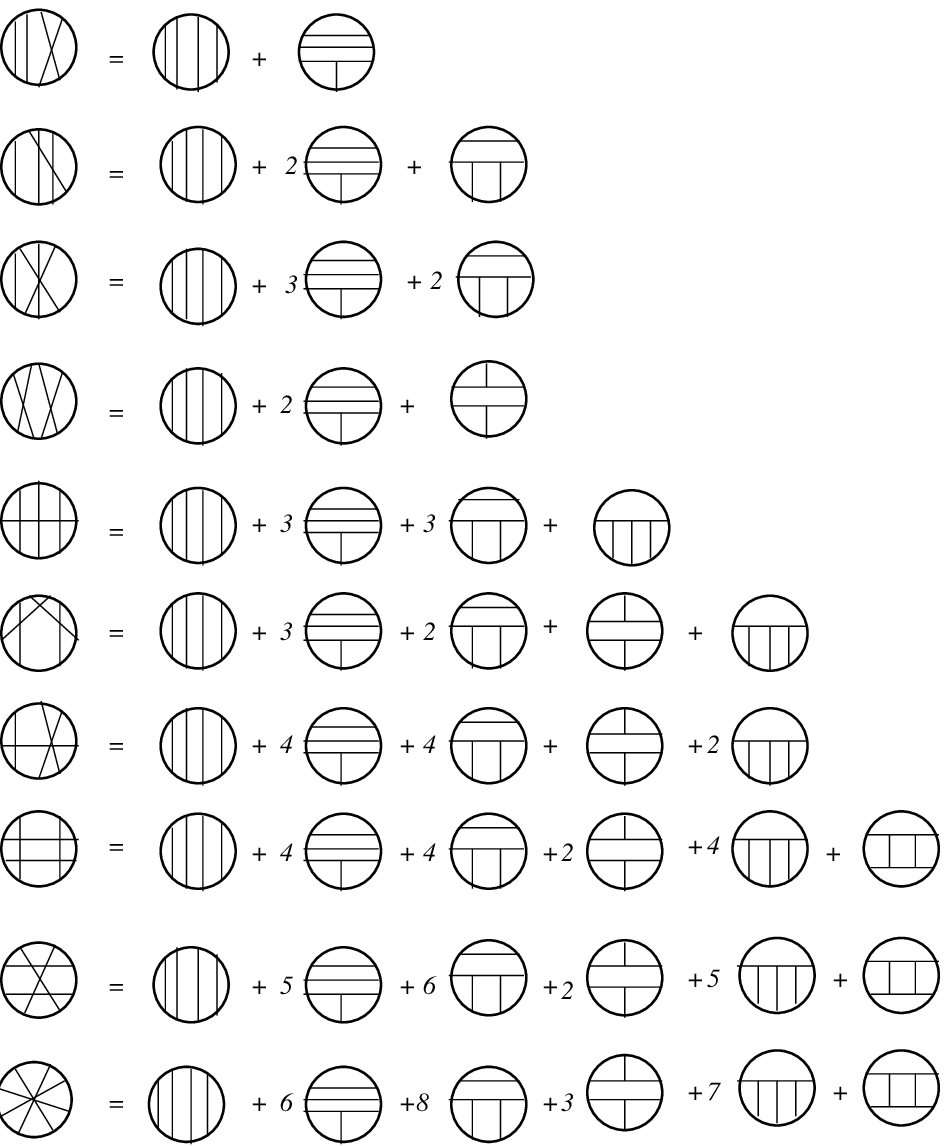}}
\caption{Expansion of chord diagrams in the canonical basis: order four.}
\label{chords}
\end{figure}

As explained in detail in \cite{alla}, and summarized in eq. 
(\ref{general}), 
to
obtain  universal Vassiliev  invariants (just depending on the knot class, 
and
not on the chosen gauge  group) we first express the contribution from a 
given
diagram in the  perturbative series as a sum of products of two factors,
geometrical and  group factors;  then we choose a  basis for the 
independent
group  factors. The coefficients of the basis elements will turn out to be
Vassiliev  invariants. In order to obtain the primitive invariants, and also 
the
relations holding  for the non-primitive ones, there is a preferred family 
of 
bases called  canonical \cite{factor}. Our choice of basis will be the same 
as
in  \cite{alla}, but here we will refer to it using  diagrams. In  fig.
\ref{groupr} we have drawn the Feynman rules needed to build up a group  
factor
out of  its diagram. Our choice of canonical basis is depicted in fig.
\ref{canonical}. Notice that we are including diagrams with isolated chords 
or
collapsible propagators. The reason for this is that their inclusion 
provides
useful information when working in the non-trivial vertical framing.  
Instead 
of
factorizing them out as in \cite{alla}, we will keep them in our analysis.
This implies that the number of elements in the basis at a given order will 
increase
with respect  to ref. \cite{alla}.  The expressions of all the chord 
diagrams
in terms of the elements of the canonical basis have been collected in figs.
\ref{chor23} and \ref{chords}.

The perturbative series expansions entering  (\ref{global}) get some
modifications relative to their form in (\ref{general}). We will write them 
in
the form: 
\bear
 {1 \over d} \langle W({K},G)\rangle &=& 1 + \sum_{i=1}^{\infty} 
\sum_{j=1}^{d_i} 
\alpha_{ij}(K) 
r_{ij}(G) x^i  + \sum_{i=1}^{\infty} \sum_{j=1}^{\tilde d_i} \gamma_{ij}(K) 
\tilde r_{ij}(G) x^i ,\nonumber \\
{1 \over d} \langle W({\k},G)\rangle_{{\rm temp}} &=& 1 + 
\sum_{i=1}^{\infty} 
\sum_{j=1}^{d_i} \hat \alpha_{ij}(\k)  r_{ij}(G) x^i + \sum_{i=1}^{\infty} 
\sum_{j=1}^{\tilde d_i} \hat \gamma_{ij}(\k) 
\tilde r_{ij}(G) x^i. \nonumber \\
\label{expansionc}
\eear
Notice that we have split the perturbative series into two sums. In the 
first
sum the group factors,  and their corresponding coefficients, are exactly 
those 
appearing in (\ref{general}), while in the second sum they are all the 
non-primitive elements coming from diagrams with collapsible propagators.  
The
quantities $r_{ij}(G)$ and $\tilde r_{ij}(G)$ denote the respective group
factors, while $d_i$ and $\tilde d_i$ are the dimension of their basis at  
order
$i$. As for the geometrical factors, 
 $\alpha_{ij}(K)$  and $\gamma_{ij}(K)$ denote the Vassiliev invariants, 
primitive or not, we are looking for, while $\hat\alpha_{ij}(\k)$  and
$\hat\gamma_{ij}(\k)$ are just the  geometrical coefficients in the 
canonical
basis of the perturbative  Chern-Simons theory in the temporal gauge.

  Our strategy is the following. First we will analyse the behaviour of the  
unknown integrals entering 
 $\hat\alpha_{ij}(\k)$ and $\hat\gamma_{ij}(\k)$; then we will build the 
whole
invariant, taking into account the corresponding global term as dictated by
(\ref{global}). Since, as shown in \cite{factor}, the perturbative series
expansion of the vacuum expectation value of the Wilson loop  exponentiates 
in
terms of the primitive basis elements, we have the following simple relation
among primitives: 
\beq \alpha_{ij}(K) = \hat\alpha_{ij}(\k) + b(\k)\,
\alpha_{ij}(U).
 \label{globcorrec} \eeq

Let us begin with the analysis of $\hat v_i(\k)$ in (\ref{expansionb}). At 
first
order we  have no correction term (recall we are using the vertical 
framing), and the temporal gauge series  provides the full regular 
invariant: 
\begin{equation}
v_1(K)= \hat v_1 ( {\cal K} ) =\Big( \eea + \lld\dda \Big) \times  \ffaa.
\label{ordone}
\end{equation}
From the expression (\ref{nucleos}) for the kernels we easily find, 
extracting
the $k=1$ contribution: 
\begin{equation}
 \eea = \sum_{i=1}^n \epsilon_i,
\label{linka}
\end{equation}
where $n$ is the number of crossings in $\k$. This corresponds to the 
linking
number in the vertical framing, which is known to be the correct answer for
$v_1(K)$. Thus we must have 
\beq \lld\dda = 0,
\label{relat}
\eeq 
which agrees with our general arguments, showing that contributions with an 
odd
number of $f$-dependent terms vanish.

At order two, the series expansion of (\ref{expansionb}) can be expressed 
as:
\beq
 \hat v_2 (\k) =  \Big( \eeb + \lld \ddb \Big) \times \ffab +
     \Big( \eec + \lld \ddc \Big) \times \ffac.
\label{ordtwoa}
\eeq
Notice that we have not included terms of the form $\sum\limits_{i=1}^n
\epsilon_i D^i_{\bigcirc}$, since they have an odd number of $f$-dependent 
terms,
and should not contribute.  In terms of the group factors of the chosen
canonical basis,  the last expression takes the form: 
\bear
 \hat v_2 (\k) &=& \hat\gamma_{21}(\k) \times \ffab +
\hat\alpha_{21} (\k) \times \ffah \nonumber \\ 
&=& \Big( \eeb + \eec + \lld \ddb + \lld \ddc \Big) \times \ffab \nonumber 
\\
&+& \Big( \eec + \lld \ddc \Big) \times \ffah . 
\label{ordtwob}
\eear
We can easily compute from the expression (\ref{nucleos}) for the kernels, 
the
two signature-dependent terms entering this expression. One finds: 
\bear
\eeb &=& {1\over 4}n(\k) + \sum_{j_1>j_2>j_3>j_4}(\epsilon(j_1,j_2)
\epsilon(j_3,j_4)
+\epsilon(j_1,j_4)\epsilon(j_2,j_3)), \nonumber \\
\, \nonumber \\
\eec &=& {1\over 4} n(\k) + \sum_{j_1>j_2>j_3>j_4}\epsilon(j_1,j_3)
\epsilon(j_2,j_4),
\label{lola}
\eear
where $n(\k)$ is the total number of crossings of the knot projection $\k$. 
These
give the following contribution to the sum entering the first group factor 
in
(\ref{ordtwoa}): 
\beq \lls^{\,E} {\hskip -5pt \ddb} \equiv \eeb + \eec = {1
\over 2} 
   \bigg( \sum_{i=1}^n \epsilon_i  \bigg)^2.
\label{ordtwoc}
\eeq
According to the factorization theorem \cite{factor} this is the whole
non-primitive regular invariant of order two, $\gamma_{21}={1\over 2}(\sum
\epsilon_i)^2$. Thus, we conclude that the order-two $D$ integrals must 
satisfy:
\beq \lls^{\,D} {\hskip -5pt \ddb} \equiv \lld\ddb + \lld \ddc = 0.
\label{relata} \eeq 

The second equation in (\ref{lola}) gives us the crossing-dependent part of 
the
primitive element $\hat\alpha_{21}(\k)$, which can alternatively be written,
using the crossing numbers in (\ref{crossnum}), as: 
\beq \eec = {1 \over
4}n \,(\k)  + \chi_2^A(\k). \label{epsila}
\eeq
 Adding the corresponding global factor term from (\ref{globcorrec}) we end 
with
the following expression for the primitive invariant at order two: 
\beq
\alpha_{21}(K) = {1 \over 4}n \,(\k)  + \chi_2^A(\k)   + \lld \ddc (\k) + 
b({\cal K})\; \alpha_{21}(U),
\label{primtwoa}
\eeq
where $\alpha_{21}(U)$ stands for the value of this invariant for the 
unknot.
The function $b({\cal  K})$ is the unknown exponent in the global factor in
(\ref{global}). Using the fact that $D_{\bigcirc}$ and  $b(\k)$  are equal 
in
$\k$  and $\alpha(\k)$, and that the  latter is equivalent under ambient 
isotopy
to the  unknot, we find: 
\beq \lld \ddc (\k) = \alpha_{21}(U)\, [ 1 - b(\k) ] -
\chi_2^A (\alpha(\k)) - {1  \over 4}n \,(\k).
\label{primtwob}
\eeq
The final expression for the invariant is:
\beq
\alpha_{21}(K) = \alpha_{21}(U) +   \chi_2^A(\k) -
\chi_2^A (\alpha(\k)),
\label{primtwoc}
\eeq
which agrees with the formulae given in \cite{hirs} and \cite{lannes}. 
Notice
that  its dependence on $b({\cal K})$ has disappeared, so up to this order 
we 
do  not get any condition on this function. It might be identically zero. It 
is important to remark that the derivation of (\ref{primtwoc}) that we have
presented is much simpler than the one in the covariant gauge obtained in 
\cite{hirs}. This simplicity is rooted in the special features of the 
temporal
gauge that permit to have the compact expression (\ref{nucleos}) for the
kernels, which are the essential building blocks of the combinatorial
expressions for Vassiliev invariants. These features will become more 
prominent
in the  third-order analysis to which we now turn.

At order three, the expression of the perturbative series of 
(\ref{expansionb}) takes the form:
\bear
&& \hat v_3(\k) = \hat\gamma_{31}(\k) \times \ffae + \hat\gamma_{32}(\k) 
\times \ffai +
\hat\alpha_{31}(\k) \times \ffaj \nonumber \\
\, \nonumber \\
 &=& \bigg( \lls^{\,E}{\hskip -5pt\dde} + \sum_{i} \epsilon_i 
\lls^{\,Di}{\hskip -5pt\dde} \bigg)
	    \times \ffae + \bigg(  \lls^{\,E}{\hskip -5pt\ddi}  + \sum_{i} 
\epsilon_i  
\lls^{\,D^i}{\hskip -5pt\ddi}  \bigg)
     \times \ffai \nonumber \\
 &+& \bigg(  \lls^{\,E}{\hskip -5pt\ddj}  + \sum_{i} \epsilon_i \lls^{\,Di}
{\hskip -5pt\ddj} \bigg)
     \times \ffaj ,
\label{ordthreea}
\eear
where we have made the choice of canonical basis shown in fig. 
\ref{canonical}.
In order to write down  the sums $S^E_{\bigcirc}$ and $S^D_{\bigcirc}$ in  
terms
of their Feynman integrals, one has to take into account the change of  
basis
described in fig. \ref{chor23}. Given a basis element, its sum will be built 
up 
with all  the chord diagrams whose expansion in the canonical basis contains
that  element, each multiplied by the corresponding coefficient.

At this order there are three independent group factors, but only one is 
primitive. The  factorization theorem  
 provides  a sufficiently large number of relations between the $D$  
integrals,
so that we will be able to solve for the primitive invariant. The Feynman 
integrals proportional to $\epsilon^0$ or $\epsilon^2$ times some $D$  
integral
are not written in (\ref{ordthreea}) since, as we argued before, they do not 
contribute. Recall that the integrals  $D^i$ are built out of two  
$f$-dependent
terms of the propagator (\ref{prop}). The third factor, which is a
signature-dependent one, leads to the sign $\epsilon_i$ and a restriction in
the  integration domain. Thus, we may expect them to be related to the 
order-two
independent  integral $\lld\ddc$. As we show below, this is indeed the case.
With the help of the factorization theorem we will find relations for the 
non-primitive diagrams.  Our task is to use these relations to find 
expressions
for the  unknown integrals  in the primitive factor $\hat\alpha_{31}$ in 
terms 
of
the  order-two integrals evaluated in the whole knot or in some closed piece 
of
it.

Similarly to the case of lower order, the computation of the signature
contributions is easily obtained from the kernels (\ref{nucleos}). For the 
case
$k=3$ in (\ref{nucleos}) and the first group factor in (\ref{ordthreea})  
one
finds: 
\bear  \lls^{\,E}{\hskip -5pt\dde} &\equiv& \eee + \eed + \eef + \eeg
\nonumber \\ &=& {1 \over 3} \big( \sum_{i} \epsilon_i \big)^3 = \gamma_{31}
(K), \label{linktres} \eear
where we have used the factorization theorem \cite{factor}. The other term
associated to the  group factor under consideration must therefore vanish: 
\beq
\lls^{\,D{i}}{\hskip -5pt\dde} \equiv \lldi\dde +  \lldi\ddd +
\lldi\ddf + \lldi\ddg = 0 \;\;\;\;\; \forall \, i.
\label{relatb}
\eeq
We thus end with a non-trivial relation for the $D_{\bigcirc}$ integrals of 
order three: they sum up to zero. Notice that it is the same kind of 
relation 
that we found  in (\ref{relata}) at order two. 

To the second group factor in (\ref{ordthreea})  is associated  the other
non-primitive  factor
$\hat\gamma_{32}(\k)$:
\beq
\hat\gamma_{32} (\k) = \bigg(  \lls^{\,E}{\hskip -5pt\ddi}  + \sum_{i} 
\epsilon_i  
\lls^{\,D^i}{\hskip -5pt\ddi}  \bigg),
\label{nonprima}
\eeq
whose relation with the corresponding regular invariant, following
(\ref{global}),  is:
\beq
\gamma_{32}(K) = \hat\gamma_{32}(\k) + b(\k) \, \alpha_{21}(U) \, 
\sum_i\epsilon_i.
\label{nonprimb}
\eeq
 Due to the Chern-Simons factorization theorem \cite{factor}, the invariant 
must
fulfil the relation: 
\beq
\gamma_{32}(K) = \alpha_{21}(K) \, \sum_i \epsilon_i .
\label{nonprimc}
\eeq
These last two equations trivially imply that the following relation must 
hold:
\beq
\hat\gamma_{32}(\k) + b(\k) \, \alpha_{21}(U) \, \sum_i\epsilon_i
 = \alpha_{21}(K) \, \sum_i \epsilon_i.
\label{iguales}
\eeq

This equation will provide important relations to solve  the unknown
quantities in (\ref{nonprima}). The strategy to obtain them is the 
following.
First we extract the signature-dependent part of (\ref{nonprima}), using  
the
kernels (\ref{nucleos}); then one substitutes (\ref{primtwoa}) and 
 (\ref{nonprimb}) into (\ref{iguales}).  The signature contributions in
(\ref{nonprima}) turn out to be,  using the definition of the signature 
function
given in (\ref{signatura}) and the crossing numbers (\ref{crossnum}): 
\bear
\lls^{\,E}{\hskip -5pt\ddi} &\equiv& \eed + 2\, \eef + 3 \, \eeg 
\label{signatres} \\
& & \nonumber \\
&=& \Big( \sum_{j_1 > \cdots > j_6} \epsilon(j_1, j_2) \epsilon(j_3, j_5) 
\epsilon(j_4, j_6) + \; \hbox{\rm c. p.}\Big) \; + 2\, \chi_3^B + 3 \, 
\chi_3^A,
\nonumber
\eear
where c. p. stands for the five inequivalent cyclic permutations of the 
indices.
The right-hand side of (\ref{signatres}) factorizes as : 
\beq
\lls^{\,E}{\hskip -5pt\ddi} = \eec \, \sum_{i} \epsilon_i 
\label{equaltres}
\eeq
where $\eec$ is given in (\ref{lola}).
Substituting this result into (\ref{iguales}), we find  a relation which 
involves
three of the four  $D^i$ integrals present at this order: 
\beq
\lls^{\,Di}{\hskip -5pt\ddi} \equiv \lldi\ddd + 2 \lldi\ddf + 
3 \lldi\ddg = \lld \ddc.
\label{relatc}
\eeq
Notice that the left-hand side of (\ref{relatc}) depends on the crossing 
$i$,
while the right-hand side does not, \ie\ the precise combination of $D$
integrals in the left, whose domain of integration in principle depends on 
the
crossing $i$, is in fact equal to an order-two $D$ integral evaluated in the
whole knot. 

\begin{figure}
\centerline{\hskip.4in \epsffile{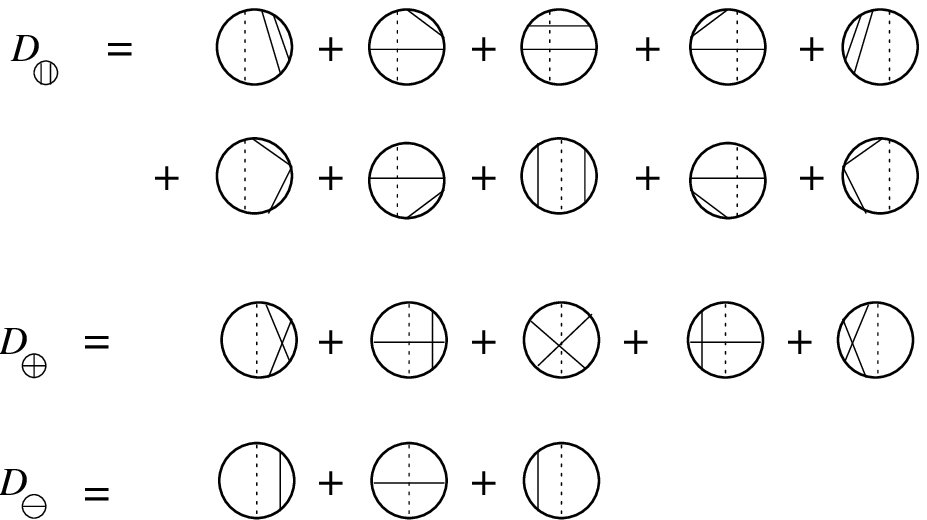}}
\caption{Splitting of $D$ integrals.}
\label{spliting}
\end{figure}

Equations (\ref{relatb}) and (\ref{relatc}) can also be proved  without 
making
use of the factorization theorem. It is worthwhile to describe how this is 
so, since
it provides some insight that will be useful later.
 The left-hand side of both equations is a sum over Feynman  integrals, with 
one
propagator fixed at a crossing and the other two running  over the two 
regions
in which that crossing divides the path of the knot in a way  consistent 
with
the corresponding diagram. To better understand the argument, let us first 
think
that we want to compute the  integral $\lld \ddc$, or  $\lld \ddb$, 
splitting 
the knot path into two regions from a selected  crossing $i$, so one runs 
over
the parametric interval $(s_{i_-}, s_{i_+})$,  and the other over $(s_{i_+},
s_{i_-})$.   
The integration region can be
decomposed in a sum of partial contributions as depicted in fig. 
\ref{spliting},
where the dashed line represents the crossing $i$. The linearity of the 
integral
guarantees that the sum of all these partial contributions leads to the full 
$\lld \ddc$, or $\lld \ddb$. The order-two integrals $D^i$ appearing in
(\ref{relatb})  can be organized in a similar way, leading to: 
\beq \lldi\dde + 
\lldi\ddd + \lldi\ddf + \lldi\ddg =  \lld\ddb + \lld \ddc, \eeq
so (\ref{relatb}) is a consequence of (\ref{relata}).
For (\ref{relatc}) the splitting procedure is a bit more elaborated.  The 
$D$
integrals entering in its left-hand side can be read from fig. 
\ref{spliting}:
given some $D^i_\bigcirc$, all the diagrams in  fig. \ref{spliting} whose 
three
chords build the Feynman diagram in the subscript (no matter whether they 
are dashed or
not) contribute to it. We can see that some cancellations occur between 
them.
For instance, when the first three integrals in the second column of fig.
\ref{spliting} (the first two entering $\lldi\ddd$ and the third 
$\lldi\ddf$)
are summed up, they factorize to the first integral times the second in the
last equality of fig. \ref{spliting}. This factorization can be written as: 
\beq 
\lld\dda(\k _{i_+} )  \times \lld\dda (\k _{i_-},\k _{i_+} ),
\label{primerfac} 
\eeq 
where the  notation used is the following: $\k _{i_+}$ and $\k _{i_-}$ are 
the two components obtained from the original knot when the $i$ crossing is
removed;  $\lld\dda(\k _{i_+} )$ stands for the integral of the 
$f$-dependent
part of the propagator over the component $\k _{i_+}$, while $\lld\dda (\k
_{i_-},\k _{i_+} )$ represents the integral of the $f$-part of the 
propagator
with one of its end-points running over $\k _{i_-}$ and the other over $\k
_{i_+}$. Recall that the diagram in the subscript denotes the way that the
propagators are attached to the knot. The first factor in (\ref{primerfac}) 
is
zero because of (\ref{relat}), as it is the evaluation of an odd number of
$f$-dependent parts of the propagator (\ref{prop}) over the knot  $\k 
_{i_+}$.
Similar arguments hold for the first three integrals in the  fourth column 
of
fig. \ref{spliting}, which again sum up to zero. Also, the sum of the second 
and
third integrals in the third column of fig. \ref{spliting} factorize to: 
\beq {1
\over 2} \Bigg( \lld \dda (\k _{i_-},\k _{i_+} ) \Bigg)^2 . 
\label{segunfac}
\eeq 
But this integral can be seen to be zero by using the
decomposition of the first-order integral in the last equality of fig.
\ref{spliting}, and (\ref{relat}). One can now see that the remaining 
integrals
in (\ref{relatc}) build the decomposition of $\lld \ddc$ shown in fig.
\ref{spliting}, so (\ref{relatc}) is proved.

The splitting argument  will again lead to a relation for  the unknown 
integrals entering  the  primitive diagram of eq. (\ref{ordthreea}). These 
are: 
\beq
\lls^{\,Di}{\hskip -5pt\ddj} \equiv \lldi\ddf + 2 \lldi\ddg .
\label{descuatro}
\eeq

All the integrals entering (\ref{descuatro}) can  again be read from fig.
\ref{spliting}: just add up those diagrams whose chords build the subscripts 
in
(\ref{descuatro}) with the aproppriate factors. A factorization of the type
explained in (\ref{segunfac}) occurs, and the remaining terms build up the
decomposition of $\lld \ddc$, except for the first and last terms in the 
second
equality of fig. \ref{spliting}, leading to the following result: 
\beq
\lls^{\,Di}{\hskip -5pt\ddj} = \lld \ddc (\k  )- \lld \ddc (\k _{i_+} )- 
\lld
\ddc  (\k _{i_-} ) . \label{relatd} 
\eeq 
Thus, we have achieved our goal: we have expressed the unknown integrals in 
the
primitive factor at order three in terms of an  known  integrals of order 
two.
Given a crossing $i$, the $D$ integrals of order three in the left-hand side 
of
(\ref{relatd}) (made out of two $f$-dependent parts of propagator 
(\ref{prop}))
can be expressed as a combination of the order-two integral $\lld \ddc$,
evaluated in  $\k$ and in the two components  $\k _{i_+}$ and $\k _{i_-}$ in
which the original knot projection is divided when the $i$ crossing is 
removed.
Notice that now both sides of (\ref{relatd}) depend on the crossing $i$.

Equation  (\ref{relatd}) can also be obtained by using the factorization 
theorem
and the relations  (\ref{relatb}) and  (\ref{relatc}). Taking into account 
that
\beq
\lls^{\,Di}{\hskip -5pt\ddi} - \lls^{\,Di}{\hskip -5pt\ddj} = 
 \lldi\ddd + \lldi\ddf +  \lldi\ddg = - \lldi\dde,
\label{aunno}
\eeq
where, in the last equality, we have made use of (\ref{relatb}), and using
(\ref{relatc}) and
 (\ref{descuatro}), one easily finds: 
\beq
\lls^{\,Di}{\hskip -5pt\ddj} = \lld \ddc + \lldi\dde .
\label{casi}
\eeq
The last term of this equation can be written as:
\bear
\lldi\dde (\k) &=& \lld\ddb (\k _{i_+} )  +  \lld\ddb (\k _{i_-} ) \nonumber 
\\
&+& \lld\dda(\k _{i_+} )  \times \lld\dda (\k _{i_-} ) .
\label{bonita}
\eear
Taking into account that  the third and fourth terms of 
this
equation vanishes because of (\ref{relat}) and that, using (\ref{relata}),  
the first two can be
substituted by the corresponding order-two $D$-terms, one finds that
(\ref{relatd}) holds. 

From the procedure that we have developed to obtain (\ref{relatd}),  we 
learn
that the factorization theorem provides enough relations to express all the
unknown parts of the primitive invariants in terms of the lower-order 
integral
$\lld \ddc$. Alternatively to the use of the factorization theorem, we also 
possess
a splitting procedure which leads to the same results and that sheds some 
light
on the origin of the relations involved.

In order to write down the order-three primitive invariant explicitly, we 
need
to  compute in detail the signature contributions. These are easily obtained
by using the general formula for the kernels in (\ref{nucleos}). The 
resulting
contribution can be written in a compact form, using the crossing functions
(\ref{crossnum}): 
\bear 
\lls^{\,E}{\hskip -5pt\ddj}  &\equiv&  \eef + 2 \, \eeg
\nonumber \\  &=& {1 \over 9} \chi_1(\k) + 
{3 \over 4} \chi_2^B (\k) + \chi_3^B
(\k) + 2 \chi_3^A (\k). \label{sigthree}
\eear
Using (\ref{primtwob}), (\ref{relatd}), (\ref{sigthree}), and the following 
two relations,
\bear
 \chi_2^B (\k)  &=&  \sum_i \epsilon_i \, n(\k _{i_+}, \k _{i_-} ), 
\label{chiene}
 \\
 n(\k _{i_+}, \k _{i_-} ) + 1 &=& n(\k  ) - n(\k _{i_+} ) - n( \k _{i_-} ),
\label{enes}
\eear
where $n(\k _{i_+}, \k _{i_-} )$ stands for the number of crossings between 
the
two components $\k _{i_+}$ and $\k _{i_-}$, we end with the formula (recall
that, according to (\ref{global}), there is no contribution from the global
factor at odd orders): 
\bear \alpha_{31} (K) &=&  \Big[ {1 \over 9} - {1 \over
4} - \alpha_{21} (U) \,  \Big] \, \chi_1(\k) + {1 \over 2} \chi_2^B (\k) +
\chi_3^B (\k) + 2 \chi_3^A  (\k)
\nonumber \\
&-& \sum_i \epsilon_i \Big[ \chi_2^A (\alpha(\k)) - \chi_2^A (\alpha(\k 
_{i_+})) -
\chi_2^A (\alpha(\k _{i_-})) \Big] \nonumber \\
&-& \alpha_{21} (U) \sum_i \epsilon_i \, \big[ b (\k) - 
b (\k _{i_+}) - b (\k _{i_-}) \big].
\label{primthree}
\eear
Contrary to the order-two result (\ref{primtwoc})  we now find an expression
 depending explicitly on the unknown function $b$. Invariance of 
$\alpha_{31}
(K)$ provides a relation for the terms involving $b$ in (\ref{primthree}), 
which,
in turn, leads to a $b$-independent expression. The simplest way to achieve 
this
is to consider a knot $K$ with two projections which differ by a 
Reidemeister
move of type I. It is easy to find that the value of $\chi_1$ varies in 
one
unit while all the other terms, except the last one involving the function 
$b$ 
in
(\ref{primthree}), remain invariant (this will be shown in full detail in 
the
next section). Thus, the contribution from the last term in 
(\ref{primthree})
must cancel the one from the first term. This implies that the unknown 
function
$b$ must satisfy: 
\beq b (\k) - b (\k _{i_+}) - b (\k _{i_-}) = x,
\label{incoga} \eeq 
where $x$ is a constant such that: 
 \beq {1 \over 9} - {1 \over 4} -
\alpha_{21} (U) - \alpha_{21} (U)\, x = 0. \label{incogb} \eeq Throughout 
this
paper we will use the following normalization for the unknot:  
\beq \alpha_{21}
(U) = -{1 \over 6}, \label{unknota}
\eeq
which implies:
\beq
x = -{1 \over 6}.
\label{incogc}
\eeq

Taking into account (\ref{incoga}) and (\ref{incogc}), expression 
(\ref{primthree}) for the order-three primitive Vassiliev invariant turns 
out to be: 
\bear
\alpha_{31} (K) &=& {1 \over 2} \chi_2^B (\k) + \chi_3^B (\k) + 2 \chi_3^A 
(\k)
\nonumber \\
&-& \sum_i \, \epsilon_i \Big[ \chi_2^A (\alpha(\k)) - \chi_2^A (\alpha(\k 
_{i_+})) - \chi_2^A (\alpha(\k _{i_-})) \Big] .
\label{primthreeb}
\eear
This combinatorial formula is the same as the one obtained in  
\cite{hirsdos},
using Chern-Simons gauge theory in a covariant gauge and in \cite{lannes} 
using
other methods. As compared to the calculation  in the covariant gauge, our
computation is much simpler. It is very unlikely that with the 
covariant-gauge
methods utilized in \cite{hirsdos} one could obtain combinatorial 
expressions
for higher-order invariants. It turns out that our procedure goes beyond and 
can
be implemented at higher order. We will show in the next section how this is
achieved at order four, obtaining combinatorial formulae for the two 
primitive
Vassiliev invariants present at that order.

\subsection{Vassiliev invariants of order four}

\hskip .25cm 

In this section we will apply our reconstruction procedure to compute the 
two
combinatorial expressions for the two primitive invariants at  order four.
Using  our diagrammatic notation  for the group factors and the choice of 
basis
shown in fig. \ref{canonical}, the perturbative series expansion in the 
temporal 
gauge  at this order takes the form: 
\bear
 \hat v_4(\k) &=& \hat\gamma_{41}(\k) \times \ffak + \hat\gamma_{42}(\k) 
\times 
\ffau + 
\hat\gamma_{43}(\k) \times \ffay + \hat\alpha_{41}(\k) \times \ffav    
\nonumber 
\\
\, \nonumber \\
&+& \hat\alpha_{42}(\k) \times \ffaw + \hat\alpha_{43}(\k) \times \ffax, 
\label{ordcuatro}
\eear
which, after writing down the geometrical contributions  more explicitly,
making use of the notation introduced above to separate the $D$ integrals 
from
the $E$ integrals, becomes: 
\bear
 \hat v_4(\k) = \hskip 9cm &\,& \nonumber \\
\; \nonumber \\
  \bigg( \lls^{\,E}{\hskip -5pt\ddk} + {1 \over 4} \sum_{i} \epsilon_i^2 
\lls^{\,Dii}{\hskip -5pt\ddk}  +
\sum_{i>j} \epsilon_i \epsilon_j \lls^{\,Dij}{\hskip -5pt\ddk}  +
 \lls^{\,D}{\hskip -5pt\ddk}  \bigg) \times \ffak &+& \nonumber \\ 
\bigg( \lls^{\,E}{\hskip -5pt\ddu} 
+ {1 \over 4} \sum_{i} \epsilon_i^2 \lls^{\,Dii}{\hskip -5pt\ddu}  +
\sum_{i>j} \epsilon_i \epsilon_j \lls^{\,Dij}{\hskip -5pt\ddu}  +
 \lls^{\,D}{\hskip -5pt\ddu}  \bigg) \times \ffau &+& \nonumber \\
\bigg( \lls^{\,E}{\hskip -5pt\ddy} + {1 \over 4} \sum_{i} \epsilon_i^2 
\lls^{\,Dii}{\hskip -5pt\ddy}  +
\sum_{i>j} \epsilon_i \epsilon_j \lls^{\,Dij}{\hskip -5pt\ddy}  +
 \lls^{\,D}{\hskip -5pt\ddy}  \bigg) \times \ffay &+& \nonumber \\
\bigg( \lls^{\,E}{\hskip -5pt\ddv} + {1 \over 4} \sum_{i} \epsilon_i^2 
\lls^{\,Dii}{\hskip -5pt\ddv}  +
\sum_{i>j} \epsilon_i \epsilon_j \lls^{\,Dij}{\hskip -5pt\ddv}  +
 \lls^{\,D}{\hskip -5pt\ddv}  \bigg) \times \ffav &+& \nonumber \\ 
\bigg( \lls^{\,E}{\hskip -5pt\ddw} + {1 \over 4} \sum_{i} \epsilon_i^2 
\lls^{\,Dii}{\hskip -5pt\ddw}  +
\sum_{i>j} \epsilon_i \epsilon_j \lls^{\,Dij}{\hskip -5pt\ddw}  +
 \lls^{\,D}{\hskip -5pt\ddw}  \bigg) \times \ffaw &+& \nonumber \\
 \bigg( \lls^{\,E}{\hskip -5pt\ddx} + {1 \over 4} \sum_{i} \epsilon_i^2 
\lls^{\,Dii}{\hskip -5pt\ddx}  +
\sum_{i>j} \epsilon_i \epsilon_j \lls^{\,Dij}{\hskip -5pt\ddx}  +
 \lls^{\,D}{\hskip -5pt\ddx}  \bigg) \times \ffax.
\label{ordfoura}
\eear
Notice that in this expression we have not included the  Feynman integrals
proportional to $\epsilon$ and $\epsilon^3$, as they do not contribute. Also 
it
is worthwhile to point out that $D^{ii}_\bigcirc$ denotes an integral where 
two
chords for the signature-dependent part of (\ref{prop}) are attached to the 
same crossing $i$, while $D^{ij}_\bigcirc$ corresponds to one in which the 
two
chords  are attached to two different  crossings. In the latter, there are 
in
fact two different sums: one for $i,j \in {\cal C}_a$, and another for $i,j 
\in
{\cal C}_b$, where ${\cal C}_a$ and ${\cal C}_b$ are the sets which entered 
in
(\ref{nomc}) and (\ref{nomd}). All these integrals are built out of products 
of
two $f$-terms, while the ones of $D_\bigcirc$ contain four $f$-terms.

At  order four there are six independent group factors, but only two are 
primitive. The factorization theorem will allow us to obtain ways of 
relating
all the $D$-integrals in terms of the second-order one $\lld\ddc$. This will
lead to an expression for the ones associated to the primitive group 
factors,
$\lls^{\,Dij}{\hskip -5pt\ddw}$ and $\lls^{\,Dij}{\hskip -5pt\ddx}$, similar 
to
that obtained at third order in (\ref{relatd}).

As in previous orders, one easily finds, with the aid of the kernels
(\ref{nucleos}) and of the factorization theorem, that the sum over all the
signature contributions coming from the propagator (\ref{prop}), which are 
contained
in $\lls^{\,E}{\hskip -5pt\ddk}$,  builds up the whole regular invariant: 
\bear
\lls^{\,E}{\hskip -5pt\ddk} &\equiv&  \eek + \eel + \eem + \eez + \een + 
\eeo 
\nonumber \\ &+&   \eep + \eeqq + \eer + \ees + \eet \nonumber \\
&=& {1 \over 4} \big( \sum_i \epsilon_i 
\big)^4 = \gamma_{41}(K).
\label{linkfour}
\eear
This implies that the  rest of the coefficients associated to that group 
factor
vanish: 
\bear
\lls^{\,Dij}{\hskip -5pt\ddk} &\equiv&
\lldij\ddk + \lldij\ddl + \lldij\ddm + \lldij\ddz + \lldij\ddn 
 \nonumber \\ 
\, \nonumber \\ 
 &+&  \lldij\ddo + \lldij\ddp + \lldij\ddq + \lldij\ddr + \lldij\dds  
\nonumber \\
&+& \lldij\ddt = 0 \;\; \forall i,j\, , 
\,  \nonumber \\
\, \nonumber \\
\lls^{\,D}{\hskip -5pt\ddk} &\equiv&
\lld\ddk + \lld\ddl + \lld\ddm + \lld\ddz + \lld\ddn + \lld\ddo  
\nonumber \\ 
\, \nonumber \\ 
 &+&  \lld\ddp + \lld\ddq + \lld\ddr + \lld\dds + \lld\ddt = 0.
\label{relate}
\eear

The next non-primitive factor, $\hat\gamma_{42}(\k)$, has the form:
\beq
\hat\gamma_{42}(\k) = \lls^{\,E}{\hskip -5pt\ddu} 
+ {1 \over 4} \sum_{i} \epsilon_i^2 \lls^{\,Dii}{\hskip -5pt\ddu}  +
\sum_{i>j} \epsilon_i \epsilon_j \lls^{\,Dij}{\hskip -5pt\ddu}  +
 \lls^{\,D}{\hskip -5pt\ddu},
\label{nprimfoura}
\eeq
and, as follows from (\ref{global}), its relation with the corresponding
invariant is: 
\beq
\gamma_{42}(K) = \hat\gamma_{42}(\k) + {1 \over 2} 
b(\k) \, \alpha_{21}(U) \, \big( \sum_i \epsilon_i  \big)^2.
\label{globale}
\eeq
From the factorization theorem, it follows that this  invariant  factorizes 
as:
\beq
\gamma_{42}(K)={1 \over 2} \big( \sum_i \epsilon_i  \big)^2 \, 
\alpha_{21}(K).
\label{factc}
\eeq
 Following the procedure used at order three, \ie\ computing the
signature-dependent part of $\hat\gamma_{42}(\k)$ with the aid of the 
kernels 
in
(\ref{nucleos}), and comparing these last two equations, we find that the
signature contributions match, 
\beq \lls^{\,E}{\hskip -5pt\ddu} =
{1 \over 2} \big( \sum_i \epsilon_i  \big)^2 \,  \eec, \label{facuatro}
\eeq
 so the $D$ integrals have to fulfil the following relations:
\bear
\lls^{\,Dii}{\hskip -5pt\ddu} &\equiv& \lldii\ddl + 2 \lldii\ddm + 3 
\lldii\ddz
+  2 \lldii\ddn + 3 \lldii\ddo 
\nonumber \\
\,\nonumber \\
&+& 3 \lldii\ddp + 4 \lldii\ddq + 4 \lldii\ddr + 5 \lldii\dds + 6 \lldii\ddt
\nonumber \\
\, \nonumber \\
&=& 2 \lld\ddc \;\;\;\;\;\; \forall i\, ,
\label{relatf}
\eear
\bear
\lls^{\,Dij}{\hskip -5pt\ddu} &\equiv& 
\lldij\ddl + 2 \lldij\ddm + 3 \lldij\ddz
+  2 \lldij\ddn + 3 \lldij\ddo 
\nonumber \\
\,\nonumber \\
&+& 3 \lldij\ddp + 4 \lldij\ddq + 4 \lldij\ddr + 5 \lldij\dds + 6 \lldij\ddt
\nonumber \\ 
\, \nonumber \\ &=&
\lld\ddc \;\;\;\;\;\; \forall i \neq j \, ,
\label{relatg}
\eear
\bear
\lls^{\,D}{\hskip -5pt\ddu} &\equiv& 
\lld\ddl + 2 \lld\ddm + 3 \lld\ddz
+  2 \lld\ddn + 3 \lld\ddo
\nonumber \\
\, \nonumber \\
&+& 3 \lld\ddp + 4 \lld\ddq + 4 \lld\ddr + 5 \lld\dds + 6 \lld\ddt 
\nonumber \\
\, \nonumber \\ &=& 0.
\label{relath}
\eear
Recall that the coefficients multiplying the $D$-integrals come from the 
choice
of basis that we have made. They can be computed with the aid of fig.
\ref{chords}. Notice that in principle the left-hand side of (\ref{relatf}) 
and
(\ref{relatg})  could depend upon the pair of crossings chosen. The
factorization  theorem, however, implies that this is  not the case. 
Actually,
these relations are even more remarkable. In (\ref{relatf}) and 
(\ref{relatg}),
we are  dealing with $D$ integrals where two  of the propagators are placed 
in
the same crossing $(D^{ii})$, in two  different crossings belonging to 
${\cal
C}_a$ $(D^{ij,a})$, or in another two  in  ${\cal C}_b$ $(D^{ij,b})$. A 
given
pair of chords in a given Feynman  diagram will fulfil only one of the 
last two
conditions, as is easily seen  from their picture. So in fact eq. 
(\ref{relatg}) is not one but two  different relations. It is also 
worthwhile to
point out that factorization  provides also a check for  the kernels
(\ref{nucleos}): the computation of the signature contributions  encoded in 
the
symbol $S^{\,E}{\hskip -5pt\ddu}$, done   with the aid of (\ref{nucleos}), 
has
to match  that coming from the factorized expression of the invariant given 
in 
(\ref{factc}). Equation (\ref{facuatro}) shows that this is indeed the case.

For the other  non-primitive factors one proceeds similarly. For 
$\hat\gamma_{43}(\k)$ we have: 
\bear 
\hat\gamma_{43}(\k) &=& \lls^{\,E}{\hskip -5pt\ddy} + {1 \over 4} \sum_{i} 
\epsilon_i^2 \lls^{\,Dii}{\hskip -5pt\ddy}  +
\sum_{i>j} \epsilon_i \epsilon_j \lls^{\,Dij}{\hskip -5pt\ddy}  +
 \lls^{\,D}{\hskip -5pt\ddy}, \\
\gamma_{43}(K) &=& \hat\gamma_{43}(K), \label{papa}
\\ \, \nonumber \\ 
\gamma_{43}(K) &=& \sum_i \epsilon_i \, \alpha_{31}(K).
\label{globalf}
\eear
Comparing (\ref{papa}) with (\ref{globalf}), and making use of 
(\ref{nucleos}) 
to check that the signature contributions in $\gamma_{43}(K)$ match those of 
the
right-hand side of (\ref{globalf}), we find the following relations for the
$D$ integrals: 
\bear \lls^{\,Dii}{\hskip -5pt\ddy} &\equiv& \lldii\ddm +  2
\lldii\ddz +   + 2 \lldii\ddp + 3 \lldii\ddo   + 4 \lldii\ddq 
 \nonumber \\
\, \nonumber \\
&+&   4 \lldii\ddr +  6 \lldii\dds + 8 \lldii\ddt \nonumber \\
\nonumber \\
&=& 4 \lls^{\,Di}{\hskip -5pt\ddj} 
=4 \big[\lld \ddc (\k  )- \lld \ddc (\k _{i_+} )- \lld \ddc 
 (\k _{i_-} ) \big] , \nonumber \\
\label{relatm}
\eear
\bear
\lls^{\,Dij}{\hskip -5pt\ddy} &\equiv& 
\lldij\ddm +  2 \lldij\ddz +  
+ 2 \lldij\ddp + 3 \lldij\ddo   + 4 \lldij\ddq  
 \nonumber \\
\, \nonumber \\
&+& 4 \lldij\ddr +  6 \lldij\dds + 8 \lldij\ddt \nonumber \\
\, \nonumber \\
&=&  \lls^{\,Di}{\hskip -5pt\ddj} 
= \lld \ddc (\k  )- \lld \ddc (\k _{i_+} )- \lld \ddc 
 (\k _{i_-} ),
\label{relatn}
\eear
\bear
\lls^{\,D}{\hskip -5pt\ddy} &\equiv& \lld\ddm +  2 \lld\ddz
+ 2 \lld\ddp + 3 \lld\ddo   + 4 \lld\ddq + 4 \lld\ddr
 \nonumber \\
\, \nonumber \\
&+&   6 \lld\dds + 8 \lld\ddt = 0,
\label{relato}
\eear
where, for the last step in eqs. (\ref{relatm}) and 
(\ref{relatn}), we have made use of (\ref{relatd}).

For the last non-primitive factor $\hat\alpha_{41}(\k)$, we find:
\bear 
\hat\alpha_{41}(\k) &=& \lls^{\,E}{\hskip -5pt\ddv} + {1 \over 4} \sum_{i} 
\epsilon_i^2 \lls^{\,Dii}{\hskip -5pt\ddv}  +
\sum_{i>j} \epsilon_i \epsilon_j \lls^{\,Dij}{\hskip -5pt\ddv}  +
 \lls^{\,D}{\hskip -5pt\ddv}, \\
\alpha_{41}(K) &=& \hat\alpha_{41}(\k) + b(\k) \, {1 \over 2} \big( 
\alpha_{21}(U) \big)^2,
\\
\alpha_{41}(K) &=& {1 \over 2} \big( \alpha_{21}(K) \big)^2,
\label{globald}
\eear
and the relations obtained for the $D$-integrals turn out to be:
\bear
\lls^{\,Dii}{\hskip -5pt\ddv} &\equiv& \lldii\ddn +  \lldii\ddp +  
\lldii\ddq
+  2 \lldii\ddr  
\nonumber \\
\, \nonumber \\
&+& 2 \lldii\dds  + 3 \lldii\ddt =  \lld\ddc \;\;\; \forall i\, ,
\label{relati}
\eear

\bear
\lls^{\,Dij,a}{\hskip -5pt\ddv} &\equiv& 
\lldia\ddn +  \lldia\ddp +  \lldia\ddq
+  2 \lldia\ddr   
\nonumber \\ 
\, \nonumber \\
&+& 2 \lldia\dds + 3 \lldia\ddt =
\lld\ddc \;\; \forall i,j \in {\cal C}_a \, ,
\label{relatj}
\eear
\bear
\lls^{\,Dij,b}{\hskip -5pt\ddv} &\equiv& 
\lldib\ddn +  \lldib\ddp +  \lldib\ddq
+  2 \lldib\ddr   
\nonumber \\ 
\, \nonumber \\
 &+& 2 \lldib\dds + 3 \lldib\ddt = 0
\;\; \forall i,j \in {\cal C}_b \, ,
\label{relatk}
\eear
\bear
\lls^{\,D}{\hskip -5pt\ddv} &\equiv& 
\lld\ddn +  \lld\ddp +  \lld\ddq
+  2 \lld\ddr + 2 \lld\dds + 3 \lld\ddt
\nonumber \\ 
\, \nonumber \\
&=& {1 \over 2} \bigg( \lld\ddc \bigg)^2.
\label{relatl}
\eear
Notice that in this case we have now  found different relations for
the $D^{ij}_\bigcirc$ integrals depending on the relative position between 
the
crossing labels.

\begin{figure}
\centerline{\hskip.4in \epsffile{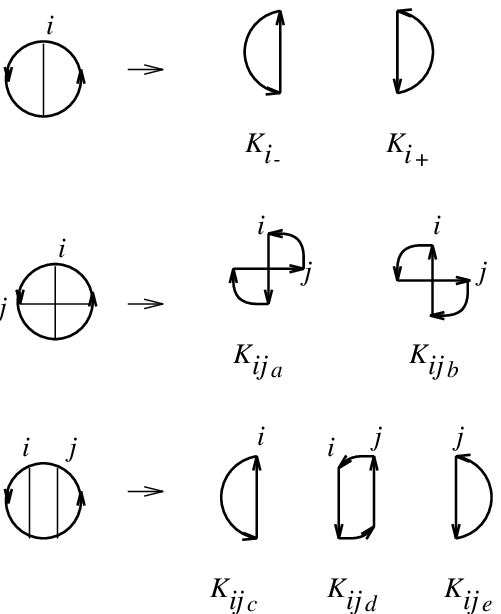}}
\caption{Dividing the knot in other knots.}
\label{subknots}
\end{figure}

We have obtained a series of relations which will be used in the  
determination
of the unknown terms in the primitive factors. As in order  three, a 
fundamental
step to carry out the computation is the expression of the integral 
$\lld\ddc$
in terms of all the integrals appearing when we split its  integration 
domain 
in
the pieces defined by two selected crossings $i$ and  $j$. Now there are two 
ways of doing this, depending on which are the  relative positions of the
crossings labels. The notation  we will use when a closed path is split
after removing two crossings is shown in fig. \ref{subknots}. Notice  that,
when the crossings are alternating,  orientation has to be reversed in  
two
of the four segments the knot is divided  into, so as to have actual  {\it 
closed} paths.  

Using all the previous relations for the $D$ integrals at this order,  and
applying a splitting procedure analogous to the one at order three, one 
finds 
the following expressions  for the unknown integrals present in the two 
primitive
factors:  
\bear
\lls^{\,Dii}{\hskip -5pt\ddw} &\equiv&
\lldii\ddp + \lldii\ddo + 2 \lldii\ddq + 4 \lldii\ddr \nonumber \\
\, \nonumber \\
&+& 5 \lldii\dds + 7 \lldii\ddt \nonumber \\
\, \nonumber \\ 
&=& 3 \, \bigg[ \lld\ddc (\k) -  \lld\ddc (\k_{i_+}) - \lld\ddc 
(\k_{i_-}) \bigg], 
\label{primfoura}
\eear

\bear
\lls^{\,Dij,a}{\hskip -5pt\ddw} &\equiv&
\lldia\ddp + \lldia\ddo + 2 \lldia\ddq + 4 \lldia\ddr \nonumber \\
\, \nonumber \\
&+& 5 \lldia\dds + 7 \lldia\ddt \nonumber \\
\, \nonumber \\ 
&=& 3  \lld\ddc (\k) \nonumber \\
\, \nonumber \\ 
 &-&  2\, \bigg[ \lld\ddc (\k_{i_+}) + \lld\ddc (\k_{i_-})  + 
\lld\ddc (\k_{j_+}) + \lld\ddc (\k_{j_-}) \bigg] \nonumber \\
\, \nonumber \\
&+& \lld\ddc (\k_{ij_a}) + 
\lld\ddc (\k_{ij_b}),
\label{primfourb}
\eear

\bear
\lls^{\,Dij,b}{\hskip -5pt\ddw} &\equiv&
\lldib\ddp + \lldib\ddo + 2 \lldib\ddq + 4 \lldib\ddr \nonumber \\
\, \nonumber \\
&+& 5 \lldib\dds + 7 \lldib\ddt \nonumber \\
\, \nonumber \\ 
&=&  \lld\ddc (\k) \nonumber \\
\, \nonumber \\ 
 &-&   \bigg[ \lld\ddc (\k_{i_+}) + \lld\ddc (\k_{i_-})  + 
\lld\ddc (\k_{j_+}) + \lld\ddc (\k_{j_-}) \bigg]  \nonumber \\
\, \nonumber \\
&+& \lld\ddc (\k_{ij_c}) + 
\lld\ddc (\k_{ij_d}) + \lld\ddc (\k_{ij_e}),
\label{primfourc}
\eear

\bear
\lls^{\,Dii}{\hskip -5pt\ddx} &\equiv&
\lldii\ddr + \lldii\dds +  \lldii\ddt = 0, \\
\, \nonumber \\
\lls^{\,Dij,b}{\hskip -5pt\ddx} &\equiv&
\lldib\ddr + \lldib\dds +  \lldib\ddt = 0, 
\label{primfoure}
\eear

\bear
\lls^{\,Dij,a}{\hskip -5pt\ddx} &\equiv&
\lldia\ddr + \lldia\dds +  \lldia\ddt \nonumber \\
\, \nonumber \\
&=& \lld\ddc (\k) \nonumber \\
\, \nonumber \\ 
 &-&  \bigg[ \lld\ddc (\k_{i_+}) + \lld\ddc (\k_{i_-})  + 
\lld\ddc (\k_{j_+}) + \lld\ddc (\k_{j_-}) \, \bigg] \nonumber \\
\, \nonumber \\
&+& \lld\ddc (\k_{ij_a}) + 
\lld\ddc (\k_{ij_b}).
\label{primfourf}
\eear
As before, the relations we have found depend on how the two crossings are 
related. Notice that all the possible ways of dividing the knot into other 
closed knots when one or two  crossings are removed appear. The alternating 
case
is a bit subtle  because, to form closed paths, 
orientation in some  segments has to be reversed, and so the integrals in 
the left-hand side of
(\ref{primfourb}) and  (\ref{primfourf})  have to appear with the 
aproppriate sign. 
                                                         
Our aim is now to calculate $\hat\alpha_{42}(\k)$ and $\hat\alpha_{43}(\k)$, 
the primitive factors at this order. It follows from (\ref{ordfoura})
that they have the form:
\beq
\hat\alpha_{42}(\k) =\lls^{\,E}{\hskip -5pt\ddw} + {1 \over 4} \sum_{i} 
\epsilon_i^2 \lls^{\,Dii}{\hskip -5pt\ddw}  +
\sum_{i>j} \epsilon_i \epsilon_j \lls^{\,Dij}{\hskip -5pt\ddw}  +
 \lls^{\,D}{\hskip -5pt\ddw}, 
\label{hola}
\eeq
\beq
\hat\alpha_{43}(\k) =  \lls^{\,E}{\hskip -5pt\ddx} + {1 \over 4} \sum_{i} 
\epsilon_i^2 \lls^{\,Dii}{\hskip -5pt\ddx}  +
\sum_{i>j} \epsilon_i \epsilon_j \lls^{\,Dij}{\hskip -5pt\ddx}  +
 \lls^{\,D}{\hskip -5pt\ddx}.
\eeq
The signature contributions appearing in them can be computed from
 the kernels (\ref{nucleos}). Using the crossing numbers notation introduced 
in
 eq. (\ref{crossnum}), we find: 
\bear
\lls^{\,E}{\hskip -5pt\ddw} &\equiv& 
\eep + \eeo + 2 \eeqq + 4 \eer \nonumber \\
\, \nonumber \\
&+& 5 \ees + 7 \eet \nonumber \\
\, \nonumber \\
&=& {38 \over (4!)^2 } n(\k) + {44 \over (3!)^2 } \chi_2^A(\k) + {21 \over 
16} \chi_2^C(\k) + 3  \chi_3^C(\k) \nonumber \\
\, \nonumber \\
&+& {3 \over 4} \chi_3^E(\k) + 
{9 \over 4} \chi_3^D(\k)
+ 7 \chi_4^A(\k) + 5  \chi_4^B(\k) + 4 \chi_4^C(\k) \nonumber \\
\, \nonumber \\
&+&
 \chi_4^D(\k) + 2  \chi_4^E(\k) +  \chi_4^F(\k),
\label{primfourg}
\eear

\bear
\lls^{\,E}{\hskip -5pt\ddx} &\equiv& 
\eer + \ees +  \eet \nonumber \\
\, \nonumber \\
&=& {5 \over (4!)^2 } n(\k) + {1 \over 3! } \chi_2^A(\k) + {1 \over 4} 
\chi_2^C(\k) + {1 \over 2} \chi_3^C(\k) \nonumber \\
\, \nonumber \\
&+& {1 \over 2} \chi_3^E(\k) 
+  \chi_4^A(\k) +  \chi_4^B(\k) +  \chi_4^C(\k).
\label{primfourh}
\eear

Making use of all these formulae and adding the corresponding global terms 
coming from (\ref{global}) we obtain the following expressions for the 
primitive invariants at order four:
\bear
&\alpha_{42}(K) &= {38 \over (4!)^2 } n(\k) + {44 \over (3!)^2 } 
\chi_2^A(\k) + {21 \over 16} \chi_2^C(\k) + 3  \chi_3^C(\k) \nonumber \\
\, \nonumber \\
&+& {3 \over 4} \chi_3^D(\k) + 
{9 \over 4} \chi_3^E(\k)
+ 7 \chi_4^A(\k) + 5  \chi_4^B(\k) + 4 \chi_4^C(\k) \nonumber \\
\, \nonumber \\
&+&
 \chi_4^D(\k) + 2  \chi_4^E(\k) +  \chi_4^F(\k) \nonumber \\
\, \nonumber \\
&+& {3 \over 4} \sum_i \epsilon_i^2  \Big[ \lld\ddc (\k) -  \lld\ddc 
(\k_{i_+}) 
- \lld\ddc (\k_{i_-}) \Big] \nonumber \\
&+& \sum_{i,j \in {\cal C}_a } \epsilon_i \epsilon_j \bigg\{
3  \lld\ddc (\k) -  2 \Big[ \lld\ddc (\k_{i_+}) + \lld\ddc (\k_{i_-}) 
\nonumber \\ \, \nonumber \\
  &+& 
\lld\ddc (\k_{j_+}) + \lld\ddc (\k_{j_-}) \Big] + \lld\ddc (\k_{ij_a}) + 
\lld\ddc (\k_{ij_b}) \bigg\} \nonumber \\
\, \nonumber \\
&+& \sum_{i,j \in {\cal C}_b} \epsilon_i \epsilon_j  \bigg\{ \lld\ddc (\k)
 -   \Big[ \lld\ddc (\k_{i_+}) + \lld\ddc (\k_{i_-})  + 
\lld\ddc (\k_{j_+})  \nonumber \\
\, \nonumber \\
&+&
\lld\ddc (\k_{j_-}) \Big] + \lld\ddc (\k_{ij_c}) + 
\lld\ddc (\k_{ij_d}) + \lld\ddc (\k_{ij_e}) \bigg\} \nonumber \\
\, \nonumber \\ 
&+& \lld\ddw(\k) + b(\k)\,\alpha_{42}(U),
\label{primfouri}
\eear

\bear
&\alpha_{43}(K)&= {5 \over (4!)^2 } n(\k) + {1 \over 3! } \chi_2^A(\k) + {1 
\over 4} \chi_2^C(\k) + {1 \over 2} \chi_3^C(\k) \nonumber \\
&+& {1 \over 2} \chi_3^E(\k) 
+  \chi_4^A(\k) +  \chi_4^B(\k) +  \chi_4^C(\k) \nonumber \\
&+& \sum_{i,j \in {\cal C}_a } \epsilon_i \epsilon_j \bigg\{
\lld\ddc (\k) 
 -  \Big[ \lld\ddc (\k_{i_+}) + \lld\ddc (\k_{i_-})  + 
\lld\ddc (\k_{j_+}) \nonumber \\
\, \nonumber \\ 
&+& \lld\ddc (\k_{j_-}) \, \Big] + \lld\ddc (\k_{ij_a}) + 
\lld\ddc (\k_{ij_b}) \bigg\} \nonumber \\
\, \nonumber \\ 
&+& \lld\ddx(\k) + b(\k)\,\alpha_{43}(U).
\label{primfourj}
\eear

To get rid of the $D$-integrals appearing in this expression  one has  to 
first
replace $\lld\ddc$ by  (\ref{primtwob}), and then take
into account the fact that all $D$ integrals,  as well as the unknown 
function
$b(\k)$, only depend on the shadow of the  knot. 
In order to simplify the equations, we will use in (\ref{primfouri}) and
(\ref{primfourj}) the following set of relations:
\bear
\chi_3^E(\k) &=& \sum_{i>j \in {\cal C}_b} \epsilon_i \epsilon_j (\k) \, 
\Big[ n (\k) -   n (\k_{i_+}) - n (\k_{i_-})  \nonumber \\
- n (\k_{j_+}) &-&  n (\k_{j_-}) 
  + n (\k_{ij_c}) + 
n (\k_{ij_d}) + n (\k_{ij_e}) \Big], \nonumber 
\eear
\bear
 6 \, \chi_2^A(\k) + 4 \, \chi_3^C(\k) + \chi_3^D(\k) &=& 
\sum_{i>j \in {\cal C}_a } \epsilon_i \epsilon_j(\k)   \,
\Big[ 3 \, n (\k) -  2 \, n (\k_{i_+}) -2 \, n (\k_{i_-}) \nonumber \\
- 2 \, n (\k_{j_+}) &-&  2\, n (\k_{j_-})  
+ n (\k_{ij_a}) + 
n (\k_{ij_b}) \Big], \nonumber \\
2 \, \chi_2^A(\k) + 2 \, \chi_3^C(\k) &=&
\sum_{i>j \in {\cal C}_a }  \epsilon_i \epsilon_j(\k) 
 \, \Big[ n (\k) -    n (\k_{i_+}) - n (\k_{i_-}) \nonumber \\
- n (\k_{j_+}) &-&  n (\k_{j_-}) + n (\k_{ij_a}) + 
n (\k_{ij_b}) \Big].
\label{doschiene}
\eear
These relations are analogous to  (\ref{chiene}) and their use will make 
(\ref{primfouri}) and (\ref{primfourj}) independent of the function $n(\k)$.
Evaluating  (\ref{primfouri}) and (\ref{primfourj}) for the ascending 
diagram
$\alpha(\k)$, a projection of the unknot, one obtains expressions for
the order-four integrals  $\lld\ddw$ and $\lld\ddx$. Substituting them back 
into
(\ref{primfouri}) and (\ref{primfourj}), using the  normalization for the
unknot  invariant at order two given in (\ref{unknota}) and the relations
(\ref{doschiene}), one
obtains the following expressions:

\bear
&& {\hskip -1.5cm} \alpha_{42}(K) = \alpha_{42}(U) + {2 \over 9 }\, \big[ 
\chi_2^A(\k) - 
\chi_2^A(\alpha(\k)) \big] + 2 \, \big[ \chi_3^C(\k) - \chi_3^C (\alpha(\k)) 
\big]  \nonumber \\
&& {\hskip -1.5cm} + \, {1 \over 2} \, \big[ \chi_3^D(\k) - 
\chi_3^D(\alpha(\k)) 
\big] + 
2 \, \big[ \chi_3^E(\k) - \chi_3^E(\alpha(\k)) \big] +  7 \, \big[ 
\chi_4^A(\k) 
- \chi_4^A(\alpha(\k)) \big] \nonumber \\
\, \nonumber \\
&&{\hskip -1.5cm} + \,
5 \, \big[ \chi_4^B(\k) - \chi_4^B(\alpha(\k)) \big]  +  4 \, \big[ 
\chi_4^C(\k) 
- \chi_4^C (\alpha(\k)) \big] 
+ \chi_4^D(\k) - \chi_4^D(\alpha(\k)) \nonumber \\
\, \nonumber \\
&& {\hskip -1.5cm} +\, 2 \, \big[ \chi_4^E(\k) - \chi_4^E(\alpha(\k)) \big] 
 +  \chi_4^F(\k) - \chi_4^F(\alpha(\k)) \nonumber \\
\, \nonumber \\
&& {\hskip -1.5cm} -\,  \sum_{i>j \in {\cal C}_a }\big[ \epsilon_i 
\epsilon_j(\k) - \epsilon_i 
\epsilon_j(\alpha(\k)) \big] \, \bigg\{ \,
3  \chi_2^A(\alpha(\k)) - \,  2 \Big[ \chi_2^A(\alpha(\k_{i_+})) + 
\chi_2^A(\alpha (\k_{i_-})) \nonumber \\
\, \nonumber \\
&& {\hskip -1.5cm}  
+ \; \chi_2^A(\alpha(\k_{j_+})) + \chi_2^A(\alpha (\k_{j_-})) \Big] + 
\chi_2^A(\alpha (\k_{ij_a})) + 
\chi_2^A(\alpha (\k_{ij_b})) \bigg\} \nonumber \\
\, \nonumber \\
&& {\hskip -1.5cm}- \, \sum_{i>j \in {\cal C}_b} \big[ \epsilon_i 
\epsilon_j(\k) 
- \epsilon_i 
\epsilon_j(\alpha(\k)) \big] \,  \bigg\{ \,
 \chi_2^A(\alpha(\k)) -  \chi_2^A(\alpha(\k_{i_+})) - \chi_2^A(\alpha 
(\k_{i_-})) \nonumber \\
\, \nonumber \\
 && {\hskip -1.5cm} 
- \, \chi_2^A(\alpha(\k_{j_+})) - \chi_2^A(\alpha (\k_{j_-})) + 
 \chi_2^A(\alpha (\k_{ij_c})) + 
\chi_2^A(\alpha (\k_{ij_d})) + \chi_2^A(\alpha (\k_{ij_e})) \bigg\} 
\nonumber 
\\
\, \nonumber \\
&& {\hskip -1.5cm} + \, {1 \over 6} \sum_{i>j \in {\cal C}_a }\big[ 
\epsilon_i 
\epsilon_j(\k) - 
\epsilon_i \epsilon_j(\alpha(\k)) \big] \, \Big\{
3  b (\k) -  2  b (\k_{i_+}) -2 b (\k_{i_-}) \nonumber \\
&& {\hskip -1.5cm} - \, 2 b (\k_{j_+}) -2 b (\k_{j_-})  + b (\k_{ij_a}) + 
b (\k_{ij_b}) \Big\} \nonumber \\
&& {\hskip -1.5cm} + \, {1 \over 6} \sum_{i>j \in {\cal C}_b} \big[ 
\epsilon_i 
\epsilon_j(\k) - 
\epsilon_i \epsilon_j(\alpha(\k)) \big] \,  \Big\{  b (\k) -   b (\k_{i_+}) 
- 
b (\k_{i_-}) \nonumber \\
&& {\hskip -1.5cm} - \, b (\k_{j_+}) - b (\k_{j_-}) + b (\k_{ij_c}) + 
b (\k_{ij_d}) + b (\k_{ij_e}) \Big\},
\label{primfourk}
\eear

\bear
&& {\hskip -1.5cm} \alpha_{43}(K)= \alpha_{43}(U) - {1 \over 6 }\,  \big[ 
\chi_2^A(\k) - \chi_2^A(\alpha(\k)) \big] +  {1 \over 2} \, \big[ 
\chi_3^E(\k) 
- \chi_3^E(\alpha(\k)) \big]
\nonumber \\
\, \nonumber \\
&& {\hskip -1.5cm}
+ \,  \chi_4^A(\k) - \chi_4^A(\alpha(\k))  +   \chi_4^B(\k) - 
\chi_4^B(\alpha(\k))  +   \chi_4^C(\k) - 
\chi_4^C(\alpha(\k)) \nonumber \\
\, \nonumber \\
&& {\hskip -1.5cm} - \, \sum_{i>j \in {\cal C}_a } \big[ \epsilon_i 
\epsilon_j(\k) - \epsilon_i 
\epsilon_j(\alpha(\k)) \big] \, 
\bigg\{ \,
 \chi_2^A(\alpha(\k)) -    \chi_2^A(\alpha(\k_{i_+})) - \chi_2^A(\alpha 
(\k_{i_-}))  \nonumber \\
\, \nonumber \\
&& {\hskip -1.5cm} 
- \, \chi_2^A(\alpha(\k_{j_+})) - \chi_2^A(\alpha (\k_{j_-})) + 
\chi_2^A(\alpha 
(\k_{ij_a)})) + 
\chi_2^A(\alpha (\k_{ij_b})) \bigg\} \nonumber \\
&& {\hskip -1.5cm} + \, {1 \over 6} \sum_{i>j \in {\cal C}_a } \big[ 
\epsilon_i 
\epsilon_j(\k) - 
\epsilon_i \epsilon_j(\alpha(\k)) \big] \, 
\Big\{
 b (\k) -   b (\k_{i_+}) - b (\k_{i_-}) \nonumber \\
&& {\hskip -1.5cm} - \, b (\k_{j_+}) - b (\k_{j_-})  
+ b (\k_{ij_a}) + 
b (\k_{ij_b}) \Big\}.
\label{primfourl}
\eear
Notice that some coefficients of the crossing functions have changed because
there are new contributions coming from the use of (\ref{primtwob}) and 
(\ref{doschiene}). Also, the terms depending only on $\alpha(\k)$ (like 
$n(\k)$ or $\chi_2^C$ ) disappear after  substituting back $\lld\ddw$ and 
$\lld\ddx$.

The expressions (\ref{primfourk}) and (\ref{primfourl}) contain
 sums involving the function $b(\k)$ evaluated in different closed paths.
In analogy with order three, we will require the factors of these sums to be
constants. Actually, as we argue below, this is the only possibility for 
(\ref{primfourk}) and (\ref{primfourl}) to be invariants.
Making use of the constraint imposed  at order three (see (\ref{incoga})) we
define the  following: 
\bear
2x &-& [ b(\k) - b(\k_{ij_c}) - b(\k_{ij_d}) - b(\k_{ij_e}) ] = y, \nonumber 
\\
2x &-& [ b(\k) - b(\k_{ij_a}) - b(\k_{ij_b})  ] = z, \nonumber \\
4x &-& [ b(\k) - b(\k_{ij_a}) - b(\k_{ij_b}) ] = t, 
\label{incogd}
\eear
where $y$, $z$ and $t$ are the constants  and $x=- {1 \over 6}$ 
 follows from (\ref{incogc}).
Recall that the labels of $\k$ refer to the closed paths in which the 
original
knot is split  when  two of its crossings are removed (see fig.
\ref{subknots}). Consistency  between the 
last two equations leads to $t= 2 x + z$. A solution for the  other two 
can be obtained by using  the values of $\alpha_{42}$ and 
$\alpha_{43}$  for some non-trivial knot (for example, for the trefoil knot 
$T$:
 $\alpha_{42}(T)=62/3 + 1/360$ and  $\alpha_{43}(T)=10/3 - 1/360$). They turn 
out to be: 
\beq
y=z=0.
\label{incoge}
\eeq

The final combinatorial expressions for the two order-four primitive 
Vassiliev invariants are:

\bear
&& {\hskip -1.5cm} \alpha_{42}(K) = \alpha_{42}(U) + {1 \over 6 }\, \big[ 
\chi_2^A(\k) - 
\chi_2^A(\alpha(\k)) \big] + 2 \, \big[ \chi_3^C(\k) - \chi_3^C (\alpha(\k)) 
\big]  \nonumber \\
&& {\hskip -1.5cm} + \, {1 \over 2} \, \big[ \chi_3^D(\k) - 
\chi_3^D(\alpha(\k)) 
\big] + 
2 \, \big[ \chi_3^E(\k) - \chi_3^E(\alpha(\k)) \big] +  7 \, \big[ 
\chi_4^A(\k) 
- \chi_4^A(\alpha(\k)) \big] \nonumber \\
\, \nonumber \\
&&{\hskip -1.5cm} + \,
5 \, \big[ \chi_4^B(\k) - \chi_4^B(\alpha(\k)) \big]  +  4 \, \big[ 
\chi_4^C(\k) 
- \chi_4^C (\alpha(\k)) \big] 
+ \chi_4^D(\k) - \chi_4^D(\alpha(\k)) \nonumber \\
\, \nonumber \\
&& {\hskip -1.5cm} +\, 2 \, \big[ \chi_4^E(\k) - \chi_4^E(\alpha(\k)) \big] 
 +  \chi_4^F(\k) - \chi_4^F(\alpha(\k)) \nonumber \\
\, \nonumber \\
&& {\hskip -1.5cm} -\,  \sum_{i>j \in {\cal C}_a }\big[ \epsilon_i 
\epsilon_j(\k) - \epsilon_i 
\epsilon_j(\alpha(\k)) \big] \, \bigg\{ \,
3  \chi_2^A(\alpha(\k)) - \,  2 \Big[ \chi_2^A(\alpha(\k_{i_+})) + 
\chi_2^A(\alpha (\k_{i_-})) \nonumber \\
\, \nonumber \\
&& {\hskip -1.5cm}  
+ \; \chi_2^A(\alpha(\k_{j_+})) + \chi_2^A(\alpha (\k_{j_-})) \Big] + 
\chi_2^A(\alpha (\k_{ij_a})) + 
\chi_2^A(\alpha (\k_{ij_b})) \bigg\} \nonumber \\
\, \nonumber \\
&& {\hskip -1.5cm}- \, \sum_{i>j \in {\cal C}_b} \big[ \epsilon_i 
\epsilon_j(\k) 
- \epsilon_i 
\epsilon_j(\alpha(\k)) \big] \,  \bigg\{ \,
 \chi_2^A(\alpha(\k)) -  \chi_2^A(\alpha(\k_{i_+})) - \chi_2^A(\alpha 
(\k_{i_-})) \nonumber \\
\, \nonumber \\
 && {\hskip -1.5cm} 
- \, \chi_2^A(\alpha(\k_{j_+})) - \chi_2^A(\alpha (\k_{j_-})) + 
 \chi_2^A(\alpha (\k_{ij_c})) + 
\chi_2^A(\alpha (\k_{ij_d})) + \chi_2^A(\alpha (\k_{ij_e})) \bigg\} 
\nonumber 
\\
\, 
\label{primfourm}
\eear

\bear
&& {\hskip -1.5cm} \alpha_{43}(K)= \alpha_{43}(U) - {1 \over 6 }\,  \big[ 
\chi_2^A(\k) - \chi_2^A(\alpha(\k)) \big] +  {1 \over 2} \, \big[ 
\chi_3^E(\k) 
- \chi_3^E(\alpha(\k)) \big]
\nonumber \\
\, \nonumber \\
&& {\hskip -1.5cm}
+ \,  \chi_4^A(\k) - \chi_4^A(\alpha(\k))  +   \chi_4^B(\k) - 
\chi_4^B(\alpha(\k))  +   \chi_4^C(\k) - 
\chi_4^C(\alpha(\k)) \nonumber \\
\, \nonumber \\
&& {\hskip -1.5cm} - \, \sum_{i>j \in {\cal C}_a } \big[ \epsilon_i 
\epsilon_j(\k) - \epsilon_i 
\epsilon_j(\alpha(\k)) \big] \, 
\bigg\{ \,
 \chi_2^A(\alpha(\k)) -    \chi_2^A(\alpha(\k_{i_+})) - \chi_2^A(\alpha 
(\k_{i_-}))  \nonumber \\
\, \nonumber \\
&& {\hskip -1.5cm} 
- \, \chi_2^A(\alpha(\k_{j_+})) - \chi_2^A(\alpha (\k_{j_-})) + 
\chi_2^A(\alpha 
(\k_{ij_a)})) + 
\chi_2^A(\alpha (\k_{ij_b})) \bigg\}.
\label{primfourn}
\eear
Again, the coefficient of $\chi_2^A$ in (\ref{primfourm}) has changed 
because 
of the contribution coming from the last equation in (\ref{incogd}). Note 
that 
both formulae have the same structure: a sum over some crossing numbers 
evaluated in $\k$ minus the same sum evaluated in the ascending diagram
$\alpha(\k)$. In addition, there are residual sums involving some 
combination of
the functions  $\chi_2^A$  evaluated in the different pieces the knot is 
divided
into when two crossings  are selected. In $\alpha_{42}(K)$ we have two of 
these
sums: one for all the  pairs of crossings belonging to ${\cal C}_a$, and 
another
for those in ${\cal  C}_b$. In $\alpha_{43}(K)$, however, only the former 
set
contributes. This,  together with the fact that there is a larger number of
order-four crossing  numbers appearing in (\ref{primfourm}), makes the
expression for  $\alpha_{42}(K)$ more complicated. 

There is another important 
comment to be made: the term in both invariants  proportional to 
$\chi_2^A(\k) - \chi_2^A(\alpha(\k))$ is in fact a Vassiliev invariant by 
itself, that of order two with the unknot normalized to zero. So the rest of 
the sum also has to be  a topological invariant (as we prove in the next
section). Then,  the value of the coefficient of 
$\chi_2^A(\k) - \chi_2^A(\alpha(\k))$ does not affect the 
topological invariance of our formulae. 
The last two constraints for the function $b(\k)$ in (\ref{incogd}) only 
affect
that term, so that  any other values for $t$ and $z$, as long as they are 
constants,
would not spoil the  topological invariance of our formulae. With our 
present
knowledge, the only way to fix $t$ and $z$ is to compare our expression for 
 $\alpha_{42}(K)$ and  $\alpha_{43}(K)$ to a known one for some non-trivial
knot, as we did to get (\ref{incoge}). Of course, this would not be 
necessary
if we had an independent argument to obtain the function $b(\k)$. To fix the
constant $y$, however, there is no need to make explicit comparisons: it
follows from invariance, as was the case of $x$ at order three. Indeed, 
under
the  first Reidemeister move, the variation of the sum that multiplies $y$,
\beq 
\sum_{i>j \in
{\cal C}_b}  \epsilon_i \epsilon_j(\k), 
\eeq
is proportional to the writhe, $\chi_1(\k)$, while the rest of the terms 
remain
invariant (this will be explicitly shown in the next section). Thus $y=0$ is
the only solution.

We have implemented the combinatorial expressions
(\ref{primfourm}) and (\ref{primfourn}), as well as the known ones,
(\ref{primtwoc}) and (\ref{primthreeb}) into  a
Mathematica algorithm. In the tables 1 and 2 of the appendix we present a
list of the values of the four primitive invariants $\alpha_{21}$,
$\alpha_{31}$,
$\alpha_{42}$  and
$\alpha_{43}$ for all prime knots up to nine crossings. Actually, the
values presented in those tables are $\alpha_{21}$, $\alpha_{31}$,
$\alpha_{42}$  and
$\alpha_{43}$ once their value for the unknot has been substracted.
The values for the new combinatorial
expressions for
$\alpha_{42}$ and
$\alpha_{43}$ agree for all knots for which those
quantities are known \cite{alla,torusknots,simon}.

The constraints that we have obtained for the function
$b (\k)$ can be summarized in the following equations:
 \bear
& & b (\k) - b (\k _{i_+}) - b (\k _{i_-}) = - {1 \over 6}, \nonumber \\
& & b(\k) - b(\k_{ij_a}) - b(\k_{ij_b})  = - {1 \over 3},  \nonumber\\
& & b(\k) - b(\k_{ij_c}) - b(\k_{ij_d}) - b(\k_{ij_e}) = - {1 \over 3}. 
\label{lulu}
\eear
These constraints on $b$ have a very simple solution.
 Let us consider a representative of $\k$ which is a Morse
function in both the $x$ and the $y$ directions. Certainly this can always 
be
done without lost of generality for any projection $\k$. This representative
has well defined  numbers  of
critical points in both the $x$ and the $y$ directions. Let us denote these
numbers by $n_x$ and $n_y$ respectively. A solution of the equations
(\ref{lulu}) is: 
\beq
b(\k) = {1\over 12} (n_x+n_y).
\label{hipotesis}
\eeq
To prove that this is indeed a solution, let us consider the three possible
splittings of $\k$ contained in eq. (\ref{lulu}), which are represented in 
fig.
\ref{subknots}. Under the first splitting
we find that $n_x+n_y \rightarrow n_x+n_y+2$, \ie\ that the number of 
extrema is
increased by 1 in each of the resulting components. Under the second 
splitting,
the number of extrema is increased by 2 in each of the resulting components, 
and
therefore $n_x+n_y \rightarrow n_x+n_y+4$. Finally, in the third splitting
$n_x+n_y \rightarrow n_x+n_y+4$, since one component increases by 2 while in
the other two it increases by 1. Thus (\ref{hipotesis}) satisfies the 
relations
(\ref{lulu}). Notice that the ansatz (\ref{hipotesis}) is symmetric under 
the
interchange of $x$ and $y$. This is consistent with the rotational 
invariance 
on 
the plane normal to the time direction present in the temporal gauge.
We conjecture that (\ref{hipotesis}) is the correct form of the function 
$b(\k)$ 
when a representative of $\k$, which is a Morse function in both the $x$ and 
the 
$y$ directions, is chosen.

\vfill
\newpage

\section{Invariance under Reidemeister moves}
\setcounter{equation}{0}

\begin{figure}
\centerline{\hskip.4in \epsffile{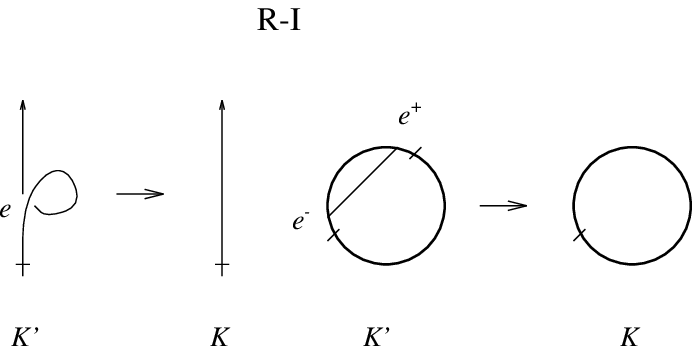}}
\caption{ Reidemeister I .}
\label{reidI}
\end{figure}

\begin{figure}
\centerline{\hskip.4in \epsffile{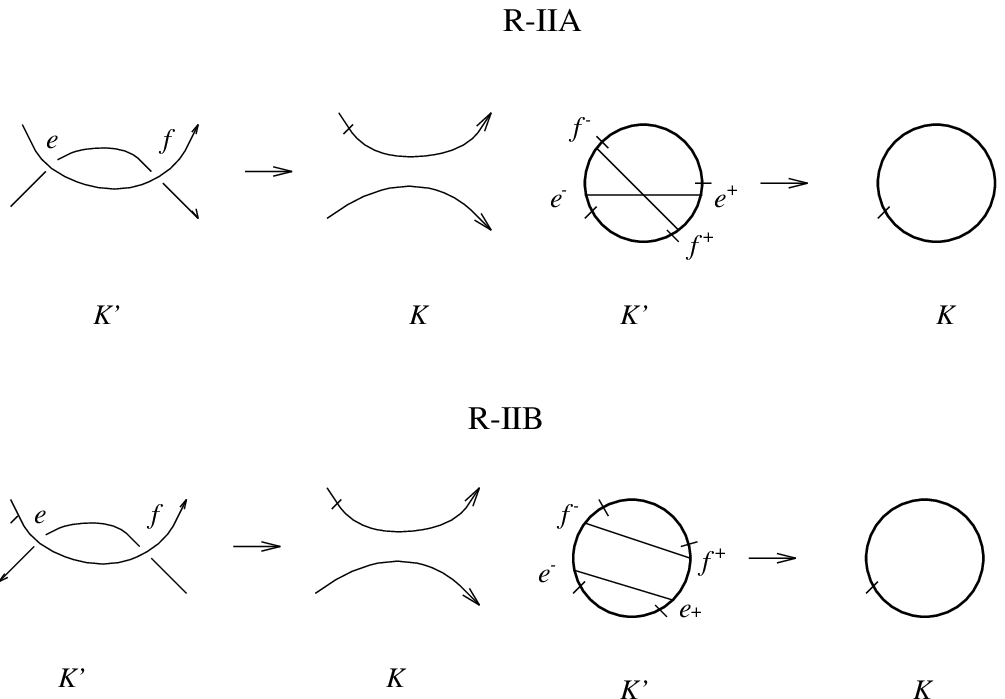}}
\caption{ Reidemeister II.}
\label{reidII}
\end{figure}

\begin{figure}
\centerline{\hskip.4in \epsffile{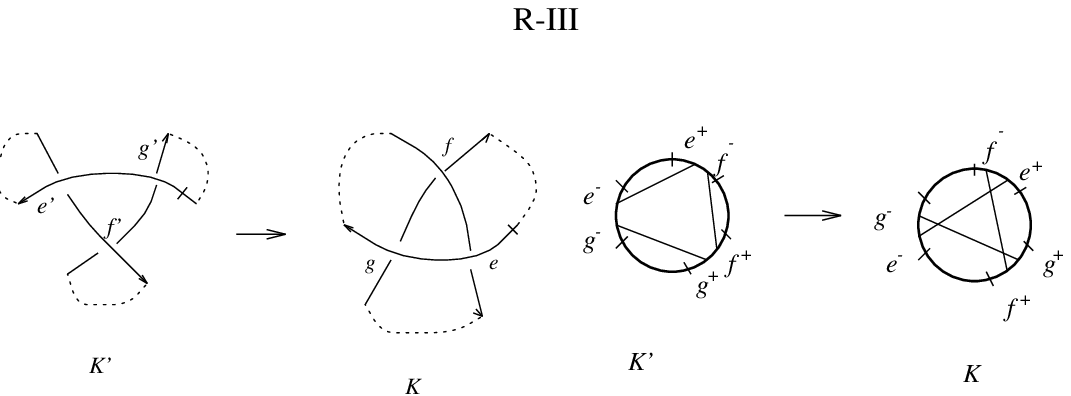}}
\caption{ Reidemeister III.}
\label{reidIII}
\end{figure}

In this section we will prove that the  combinatorial expressions for the 
two 
primitive invariants of order four, (\ref{primfourm}) and (\ref{primfourn}),
are actually topological invariants, showing that
they are invariant under the three Reidemeister moves. To do so, we have to 
know 
how the crossing functions behave under these moves. For some of them this 
has 
been done in \cite{hirs,hirsdos}. For the others, we will work out the form
of their variations   proceeding in an analogous way. 

The Reidemeister moves are depicted in figs. \ref{reidI}, \ref{reidII},
\ref{reidIII}.  For concreteness, we have made a choice of base points and
orientations, as well  as a choice of joining the three curves in R-III, but 
it
should be clear that these  choices do not affect our proof. They have also 
been
taken such that the  signature of the crossings involved in the moves does not 
vary
when we change from  $\k$ to $\alpha(\k)$ or from $\k'$ to $\alpha(\k ')$. 
As
the number of different  functions is quite large, we are going to provide 
details
only for one case, $\chi_4^A$,  and give a list of the variations for  the  
rest
of the crossing functions. 

 It is easier to understand the procedure to obtain the variations of the 
crossing functions  using diagrams. Recall that in fig. \ref{crosnumber} we gave 
a
 diagrammatic  definition of  the crossing functions in which a circle
represented the ordered set of crossing  labels on $\k$, and a series of 
chords
joining the labels $i$ and $j$ represented the signature  function 
$\epsilon(i,j)$.
From those diagrams one can  immediately figure out that, given a selected  
group
of crossings, only those that  follow the pattern given by the diagram 
contribute
to the corresponding function.  In figs. \ref{reidI}, \ref{reidII},
\ref{reidIII} we have drawn   similar diagrams to represent the three 
Reidemeister moves, which are labelled  R-I, R-IIA and R-IIB, and R-III. 
In
these diagrams, the circle representing $\k$ is divided into  sections, some 
of
them affected by the move and some not. Only the labels affected by the move 
are
depicted and a chord is pictured as joining them. All other crossings do not 
change
under the moves.

Under R-I, the function $\chi_4^A$ does not vary:
\beq
\chi_4^A (\k') - \chi_4^A (\k) = 0.
\label{ruchica}
\eeq
The $e$ crossing does not contribute to any term in $\chi_4^A$, as can be 
seen 
from the symbolic expression of the move in fig. \ref{reidI}: there is no 
way 
to
choose $e$ and other  three crossings so that they reproduce the 
diagrammatic 
expression of this  function, because all the others lay in the region 
outside
$(e^-, e^+)$. The  same argument holds for every other crossing function
appearing in  eqs. (\ref{primfourm}) and (\ref{primfourn}) for
$\alpha_{42}(K)$ and $\alpha_{43}(K)$, because none has a  diagrammatic
expression with isolated chords.

 Under a move of type  R-IIA we find:
\beq
\chi_4^A (\k') - \chi_4^A (\k) = \epsilon_e \, \epsilon_f 
 \sum_{ i_1, i_2 \in (f^-, e^+) \atop
       i_3, i_4 \in (f^+, e^-)} \epsilon(i_1,i_3) 
\epsilon(i_2,i_4),
\label{rdachica}
\eeq
where the crossings $e$ and $f$ are such
that $\epsilon_e = -  \epsilon_f$. In this expression and throughout this 
section, the crossing labels $i_1,\dots,i_n$
fulfil the natural order: $i_1 < i_2 < \cdots < i_n$.
In the variation (\ref{rdachica}) there is a potential term  proportional
to $\epsilon_e + \epsilon_f$  but it does not  contribute because
its coefficient turns out to be zero. The  sum on the right-hand side of 
(\ref{rdachica}) can be expressed in diagrammatic form.
The regions to which the labels are  attached 
are the  regions not affected by the move. The signature functions fix
how to draw the chords. These two chords, together with the ones 
corresponding to $e$ and $f$, build the diagram associated to $\chi_4^A$. 

Under R-IIB we find, 
\beq
\chi_4^A (\k') - \chi_4^A (\k) = 0.
\label{rdbchica}
\eeq
As in the previous case the crossings $e$ and $f$ are such that $\epsilon_e 
= - 
\epsilon_f$, and the terms linear in $e$ and $f$ cancel. This time  no 
quadratic
contribution is left, because there is no way to choose $e$,  $f$ and two 
other
crossings so as to end up with the chord diagram corresponding to 
$\chi_4^A$.

Under R-III the three crossings involved in the move are such that 
$\epsilon_f 
= 
\epsilon_g = - \epsilon_e$. Their signature values do not vary from $\k'$  
to
$\k$. The variation turns out to be: 
\bear
&\,& \chi_4^A (\k') - \chi_4^A (\k) =
- \, \epsilon_f \, \epsilon_g {\hskip -.25cm \sum_{ i_1, i_2 \in (f^+, e^-) 
\atop 
       i_3, i_4 \in (e^+, g^+)  } }  {\hskip -.25cm \epsilon(i_1,i_3) 
\epsilon(i_2,i_4)}
  \nonumber \\  
  \, \nonumber \\   
&-& \epsilon_e \, \epsilon_g 
{\hskip -.25cm  \sum_{ i_1, i_2 \in (f^+, e^-) \atop
       i_3, i_4 \in (g^-, f^-)   }} {\hskip -.25cm   \epsilon(i_1,i_3) 
\epsilon(i_2,i_4) }
- \epsilon_f \, \epsilon_e 
 {\hskip -.25cm \sum_{ i_1, i_2 \in (g^-, f^-) \atop 
       i_3, i_4 \in (e^+, g^+) }} {\hskip -.25cm   \epsilon(i_1,i_3) 
\epsilon(i_2,i_4)}.
\label{rtchica}
\eear

In order to make the resulting expressions simpler, from now on we will 
assume
that $\epsilon_f=1$ in any of the moves. Certainly this does not imply a 
loss
of generality. We now present the lists of variations of the crossing
functions. Under R-IIA one finds:
 \bear
&\,& \chi_2^A (\k') - \chi_2^A (\k) = - 1  \nonumber \\
\, \nonumber \\
&\,& \chi_3^C (\k') - \chi_3^C (\k) = -{\hskip -.25cm  \sum_{ i_1  
\in (f^+, e^-) \atop 
       i_2 \in (f^-, e^+) }} {\hskip -.25cm  \epsilon^2 (i_1,i_2)  }   
+ 2 {\hskip -.25cm  \sum_{ i_1, i_2 \in (f^+, e^-) \atop 
       i_3, i_4 \in (f^-, e^+) }} {\hskip -.25cm  \epsilon(i_1,i_3) 
\epsilon(i_2,i_4)}, \nonumber \\
 \, \nonumber \\
&\,& \chi_3^D (\k') - \chi_3^D (\k) = 2 {\hskip -.25cm  
\sum_{ i_1, i_2 ,i_3 \in (f^+, e^-) \atop i_4 \in (f^-, e^+) }} {\hskip 
-.25cm 
\epsilon(i_1,i_3) \epsilon(i_2,i_4)}
 + 2 {\hskip -.25cm   \sum_{ i_1 \in (f^+, e^-) \atop
 i_2 ,i_3, i_4 \in (f^-, e^+) }} {\hskip -.25cm   \epsilon(i_1,i_3) 
\epsilon(i_2,i_4)}, \nonumber \\
 \, \nonumber \\
&\,& \chi_3^E (\k') - \chi_3^E (\k) =  2 {\hskip -.25cm  
\sum_{ i_1, i_2 \in (f^+, e^-) \atop i_3, i_4 \in (f^-, e^+)} } {\hskip 
-.25cm 
\epsilon(i_2,i_3)  \epsilon(i_1,i_4)}, \nonumber \\
\, \nonumber \\
&\,& \chi_4^B (\k') - \chi_4^B (\k) = - {\hskip -.25cm  
\sum_{ i_1, i_2 \in (f^+, e^-) \atop 
i_3, i_4 \in (f^-, e^+) }} {\hskip -.25cm   
\epsilon(i_2,i_3) \epsilon(i_1,i_4)}, \nonumber \\
\, \nonumber \\
&\,& \chi_4^C (\k') - \chi_4^C (\k) = 0, \nonumber \\
\, \nonumber \\
&\,& \chi_4^D (\k') - \chi_4^D (\k) = 0, \nonumber \\
\, \nonumber \\
&\,& \chi_4^E (\k') - \chi_4^E (\k) = - {\hskip -.25cm \sum_{ i_1, 
i_2 ,i_3 \in (f^+, e^-) \atop 
   i_4 \in (f^-, e^+) } } {\hskip -.25cm \epsilon(i_1,i_3) 
\epsilon(i_2,i_4)} \;
-   {\hskip -.25cm  \sum_{ i_1 \in (f^+, e^-) \atop
 i_2 ,i_3, i_4 \in (f^-, e^+) } } {\hskip -.25cm \epsilon(i_1,i_3) 
\epsilon(i_2,i_4)}, \nonumber \\
 \, \nonumber \\
&\,& \chi_4^F (\k') - \chi_4^F (\k) = 0.
\label{todosrda}
\eear 

Under R-IIB the variations turn out to be:
\bear
&\,& \chi_2^A (\k') - \chi_2^A (\k) = 0,  \nonumber \\
\, \nonumber \\
&\,& \chi_3^C (\k') - \chi_3^C (\k) =  2 {\hskip -.25cm \sum_{
i_1, i_2 \in (f^+, e^-) \atop 
  i_3, i_4 \in (f^-, e^+) } } {\hskip -.25cm  \epsilon(i_1,i_3) 
\epsilon(i_2,i_4)}, \nonumber \\
 \, \nonumber \\
&\,& \chi_3^D (\k') - \chi_3^D (\k) = 2 {\hskip -.5cm  \sum_{ 
i_1, i_2 ,i_3 \in (f^+, e^-) \atop i_4 \in (f^-, e^+)}} {\hskip -.25cm  
\epsilon(i_1,i_3) \epsilon(i_2,i_4)} \;
 + 2 {\hskip -.5cm   \sum_{ i_1 \in (f^+, e^-) \atop
 i_2 ,i_3, i_4 \in (f^-, e^+)}  } {\hskip -.25cm \epsilon(i_1,i_3) 
\epsilon(i_2,i_4)}, \nonumber \\
 \, \nonumber \\
&\,& \chi_3^E (\k') - \chi_3^E (\k) = - {\hskip -.25cm \sum_{ i_1 
\in (f^+, e^-) \atop
       i_2 \in (f^-, e^+)} } {\hskip -.25cm   \epsilon^2 (i_1,i_2)  } \;   
+ 2 {\hskip -.25cm   \sum_{ i_1, i_2 \in (f^+, e^-) \atop
i_3, i_4 \in (f^-, e^+)   } } {\hskip -.25cm  \epsilon(i_2,i_3) 
\epsilon(i_1,i_4)}, \nonumber \\
\, \nonumber \\
&\,& \chi_4^B (\k') - \chi_4^B (\k) = - {\hskip -.25cm  \sum_{ i_1, 
i_2 \in (f^+, e^-) \atop 
i_3, i_4 \in (f^-, e^+)   }}{\hskip -.25cm   \epsilon(i_1,i_3) 
\epsilon(i_2,i_4)}, \nonumber \\
\, \nonumber \\
&\,& \chi_4^C (\k') - \chi_4^C (\k) =  - {\hskip -.25cm  \sum_{ 
i_1, i_2 \in (f^+, e^-) \atop
i_3, i_4 \in (f^-, e^+)   }} {\hskip -.25cm   \epsilon(i_2,i_3) 
\epsilon(i_1,i_4)}, \nonumber \\
\, \nonumber \\
&\,& \chi_4^D (\k') - \chi_4^D (\k) = - {\hskip -.6cm  \sum_{ i_1, 
i_2 ,i_3 \in (f^+, e^-) \atop 
   i_4 \in (f^-, e^+)   } }{\hskip -.5cm  \epsilon(i_1,i_3) 
\epsilon(i_2,i_4)} \;
-   {\hskip -.25cm  \sum_{ i_1 \in (f^+, e^-) \atop
 i_2 ,i_3, i_4 \in (f^-, e^+)   }} {\hskip -.25cm   \epsilon(i_1,i_3) 
\epsilon(i_2,i_4)}, \nonumber \\
\, \nonumber \\
&\,& \chi_4^E (\k') - \chi_4^E (\k) = 0, \nonumber \\
\, \nonumber \\
&\,& \chi_4^F (\k') - \chi_4^F (\k) = 0.
\label{todosrdb}
\eear

Under R-III moves we find  the  variations of the crossing numbers 
collected
below. For simplicity, we have not written explicitly the terms where the
signature function is squared (these  terms appear in  $\chi_3^C$, 
$\chi_3^D$
and $\chi_3^E$). The reason is that, whatever they might be, they are 
trivially 
cancelled when we compute $\alpha_{42}(K') - \alpha_{42}(K)$, or 
$\alpha_{43}(K') - \alpha_{43}(K)$, because they  always appear in terms of 
the
form $\chi(\k) - \chi(\alpha(\k))$ and  $\epsilon(i,j)^2$ has the same value 
in
both $\k$ and $\alpha(\k)$ for any  $i,j$. We will generally denote these
contributions by $F(\epsilon^2)$. The variations under R-III moves  are:
\bear
 \chi_2^A (\k') - \chi_2^A (\k) &=& 1  \nonumber \\
\, \nonumber \\
 \chi_3^C (\k') - \chi_3^C (\k) &=&  1 - \; 2  
{\hskip -.25cm \sum_{ 
i_1 \in (e^+, g^+) \atop 
        i_2 \in (f^+, e^-)   }} {\hskip -.25cm  \epsilon(i_1,i_2) 
 + F(\epsilon^2)}, \nonumber \\
 \, \nonumber \\
 \chi_3^D (\k') - \chi_3^D (\k) &=& 0 + F(\epsilon^2), \nonumber \\
 \, \nonumber \\
 \chi_3^E (\k') - \chi_3^E (\k) &=&  2 {\hskip -.25cm  \sum_{ 
i_1 \in (e^+, g^+) \atop 
       i_2 \in (f^+, e^-)   }} {\hskip -.25cm  \epsilon (i_1,i_2)}     
- \; 2 {\hskip -.25cm  \sum_{ i_1 \in (g^-, f^-) \atop 
i_2 \in (f^+, e^-) } }   
{\hskip -.25cm \epsilon(i_1,i_2)}
- \; 2 {\hskip -.25cm  \sum_{ i_1 \in (g^-, f^-) \atop
i_2 \in (e^+, g^+)   } }
{\hskip -.25cm \epsilon(i_1,i_2)}, \nonumber 
\eear

\bear
 \chi_4^B (\k') - \chi_4^B (\k) &=& 
- {\hskip -.25cm \sum_{ i_1, 
  i_2 \in (e^+, g^+) \atop 
  i_3, i_4 \in (f^+, e^-)   }} {\hskip -.25cm \big[ \epsilon(i_1,i_4) 
   \epsilon(i_2,i_3) - \epsilon(i_1,i_3) \epsilon(i_2,i_4) - 
   \epsilon (i_1,i_3) \big] } \nonumber \\
\, \nonumber \\
&+& { \hskip -.25cm \sum_{ i_1, i_2 \in (g^-, f^-) \atop 
   i_3, i_4 \in (e^+, g^+)   }} {\hskip -.25cm \big[ \epsilon(i_1,i_4) 
    \epsilon(i_2,i_3) - \epsilon(i_1,i_3) \epsilon(i_2,i_4)+ 
\epsilon (i_1,i_3) \big] }
\nonumber \\
\, \nonumber \\
&+& {\hskip -.25cm \sum_{ i_1, 
    i_2 \in (g^-, f^-) \atop 
    i_3, i_4 \in (e^+, g^+)   }} {\hskip -.25cm \big[  \epsilon(i_1,i_4) 
\epsilon(i_2,i_3) - \epsilon(i_1,i_3) \epsilon(i_2,i_4) + \epsilon (i_1,i_3) 
\big] } \nonumber \\
\, \nonumber \\
&+&  2 {\hskip -.5cm  \sum_{ i_1, 
  i_2 \in (g^-, f^-) \atop 
   i_3 \in (e^+, g^+), i_4 \in (f^+, e^-)   }  } {\hskip -.5cm
\epsilon(i_1,i_3) \epsilon(i_2,i_4) }\nonumber \\
\, \nonumber \\
\, \nonumber \\
 \chi_4^C (\k') - \chi_4^C (\k) &=& - {\hskip -.9cm \sum_{ i_1, 
  i_2 \in (g^-, f^-) \atop i_3 \in (e^+, g^+), i_4 \in (f^+, e^-)   }}
   {\hskip -.6cm \epsilon(i_1,i_3) \epsilon(i_2,i_4) } +
{\hskip -.9cm  \sum_{ i_2, 
  i_3 \in (e^+, g^+) \atop  i_1 \in (g^-, f^-) , i_4 \in (f^+, e^-)   }}
   {\hskip -.5cm \epsilon(i_1,i_3) \epsilon(i_2,i_4) }   \nonumber \\
   \, \nonumber \\ 
&+&  {\hskip -.7cm \sum_{ i_3, 
  i_4 \in (f^+, e^-) \atop i_1 \in (g^-, f^-) , i_2 \in (e^+, g^+)    }}
   {\hskip -.25cm \epsilon(i_1,i_3) \epsilon(i_2,i_4) } -
  {\hskip -.25cm \sum_{ i_1, i_2 \in (g^-, f^-) \atop
   i_3, i_4 \in (e^+, g^+)   }} {\hskip -.25cm  \epsilon(i_1,i_4) 
    \epsilon(i_2,i_3) }  \nonumber \\
\, \nonumber \\
&+& {\hskip -.25cm \sum_{ i_1, 
  i_2 \in (e^+, g^+) \atop
   i_3, i_4 \in (f^+, e^-)   }  } {\hskip -.25cm \epsilon(i_1,i_4) 
    \epsilon(i_2,i_3)} -  {\hskip -.25cm\sum_{ i_1, i_2 \in (g^-, f^-) \atop 
   i_3, i_4 \in (f^+, e^-)   }} {\hskip -.25cm  \epsilon(i_1,i_4) 
    \epsilon(i_2,i_3) }\nonumber \\ 
  \, \nonumber \\ 
  \, \nonumber \\ 
 \chi_4^D (\k') - \chi_4^D (\k) &=& 2 {\hskip -.25cm  \sum_{ 
i_1, i_2 \in (e^+, g^+) \atop i_3, i_4 \in (f^+, e^-)   } }
{\hskip -.25cm \epsilon(i_1,i_4) \epsilon(i_2,i_3)}
+  {\hskip -.25cm\sum_{ i_1, i_2, i_3 \in (e^+, g^+) \atop  i_4 \in 
(f^+, e^-)   } } {\hskip -.25cm
 \epsilon(i_1,i_3) \epsilon(i_2,i_4)} \nonumber \\ 
\, \nonumber \\ 
&+& 
 {\hskip -.5cm \sum_{ i_1 \in (e^+, g^+) \atop  i_2, i_3 ,i_4 \in 
(f^+, e^-)   } }{\hskip -.5cm
 \epsilon(i_1,i_3) \epsilon(i_2,i_4) }
- {\hskip -.25cm \sum_{ i_1, i_2, i_3 \in (g^-, f^-) \atop  i_4 \in 
(e^+, g^+)   } } {\hskip -.25cm 
 \epsilon(i_1,i_3) \epsilon(i_2,i_4) } \nonumber \\ 
\, \nonumber \\ &-&
{\hskip -.5cm \sum_{ i_1 \in (g^-, f^-) \atop  i_2, i_3 ,i_4 \in 
(e^+, g^+)   } } {\hskip -.5cm
 \epsilon(i_1,i_3) \epsilon(i_2,i_4) } 
- {\hskip -.25cm \sum_{ i_1, i_2, i_3 \in (g^-, f^-) \atop i_4 \in 
(f^+, e^-)   } }  {\hskip -.25cm
 \epsilon(i_1,i_3) \epsilon(i_2,i_4)} \nonumber \\ 
\, \nonumber \\  &-&
 {\hskip -.25cm \sum_{ i_1 \in (g^-, f^-) \atop  i_2, i_3 ,i_4 \in 
(f^+, e^-)   } } {\hskip -.25cm
 \epsilon(i_1,i_3) \epsilon(i_2,i_4)}, \nonumber \\
\, \nonumber \\
\, \nonumber \\ 
\chi_4^E (\k') - \chi_4^E (\k) &=& - 2 {\hskip -.5cm \sum_{ 
i_1, i_2 \in (g^-, f^-) \atop i_3 \in (e^+, g^+), i_4 \in (f^+, e^-)   
} } {\hskip -.5cm \big[ \epsilon(i_1,i_3) \epsilon(i_2,i_4) - 
\epsilon(i_1,i_4) 
\epsilon(i_2,i_3) \big] } \nonumber \\
\, \nonumber \\ 
 &+& \; 2 {\hskip -.25cm \sum_{ i_1, i_2 \in (e^+, g^+) \atop i_3 , 
i_4 \in (f^+, e^-)   } } {\hskip -.25cm  \epsilon(i_1,i_3) 
\epsilon(i_2,i_4)} 
 - {\hskip -.25cm \sum_{ i_1, i_2, i_3 \in (e^+, g^+) \atop  i_4 \in 
(f^+, e^-)   } } {\hskip -.25cm
 \epsilon(i_1,i_3) \epsilon(i_2,i_4) }\nonumber \\ 
\, \nonumber \\  &-& 
{\hskip -.25cm \sum_{ i_1 \in (e^+, g^+) \atop  i_2, i_3 ,i_4 \in 
(f^+, e^-)   } } {\hskip -.25cm
 \epsilon(i_1,i_3) \epsilon(i_2,i_4) } 
+ {\hskip -.25cm \sum_{ i_1, i_2, i_3 \in (g^-, f^-) \atop i_4 \in 
(e^+, g^+)   }} {\hskip -.25cm 
 \epsilon(i_1,i_3) \epsilon(i_2,i_4)}\nonumber \\ 
\, \nonumber \\  &+&
{\hskip -.25cm \sum_{ i_1 \in (g^-, f^-) \atop  i_2, i_3 ,i_4 \in 
(e^+, g^+)   } } {\hskip -.25cm
 \epsilon(i_1,i_3) \epsilon(i_2,i_4)} 
+ {\hskip -.25cm \sum_{ i_1, i_2, i_3 \in (g^-, f^-) \atop  i_4 \in 
(f^+, e^-)   }} {\hskip -.25cm 
 \epsilon(i_1,i_3) \epsilon(i_2,i_4)} \nonumber \\ 
\, \nonumber \\ &+&
 {\hskip -.25cm \sum_{ i_1 \in (g^-, f^-) \atop  i_2, i_3 ,i_4 \in 
(f^+, e^-)   }}  {\hskip -.25cm 
 \epsilon(i_1,i_3) \epsilon(i_2,i_4)}, \nonumber \\
\, \nonumber \\
\, \nonumber \\ 
 \chi_4^F (\k') - \chi_4^F (\k) &=& {\hskip -.5cm \sum_{ i_1, 
  i_2 \in (g^-, f^-) \atop i_3 \in (e^+, g^+), i_4 \in (f^+, e^-)   }}
   {\hskip -.5cm \big[ \epsilon(i_1,i_3) \epsilon(i_2,i_4) - 3 \,
   \epsilon(i_1,i_4) \epsilon(i_2,i_3) \big] } \nonumber \\
   \, \nonumber \\ 
   &+& {\hskip -.5cm \sum_{ i_2, 
  i_3 \in (e^+, g^+)\atop i_1 \in (g^-, f^-) , i_4 \in (f^+, e^-)   }}
   {\hskip -.25cm \big[ \epsilon(i_1,i_4) \epsilon(i_2,i_3) - 
\epsilon(i_1,i_3) 
\epsilon(i_2,i_4) \big]}  \nonumber \\
  \, \nonumber \\  
    &+& {\hskip -.5cm \sum_{ i_3, 
  i_4 \in (f^+, e^-) \atop i_1 \in (g^-, f^-) , i_2 \in (e^+, g^+)    }}
  {\hskip -.25cm \big[ \epsilon(i_1,i_4) \epsilon(i_2,i_3) - 
\epsilon(i_1,i_3) 
\epsilon(i_2,i_4) \big]} 
 \nonumber \\
  \, \label{todosrt} \\  
&+& \;  2 {\hskip -.5cm \sum_{ i_1, i_2, i_3 \in (e^+, g^+) \atop  
i_4 \in (f^+, e^-)   } }
  {\hskip -.5cm \epsilon(i_1,i_3) \epsilon(i_2,i_4) } + \;  2 {\hskip -.5cm
 \sum_{ i_1 \in (e^+, g^+) \atop  i_2, i_3 ,i_4 \in 
(f^+, e^-)   }} 
 {\hskip -.5cm \epsilon(i_1,i_3) \epsilon(i_2,i_4)} 
\nonumber
\eear

\begin{figure}
\centerline{\hskip.4in \epsffile{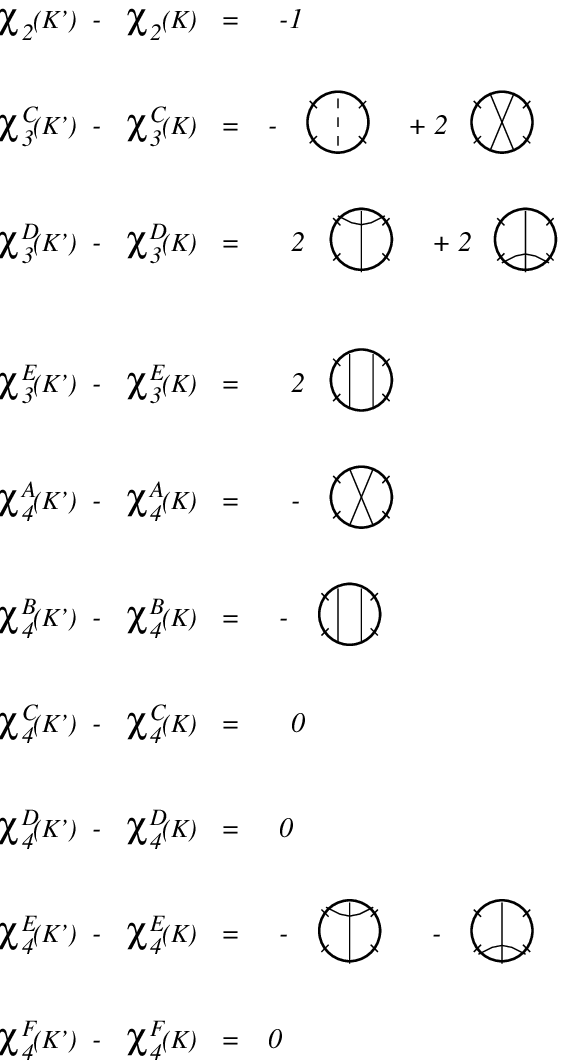}}
\caption{Behaviour of crossing functions under Reidemeister IIA.}
\label{crosIIA}
\end{figure}

\begin{figure}
\centerline{\hskip.4in \epsffile{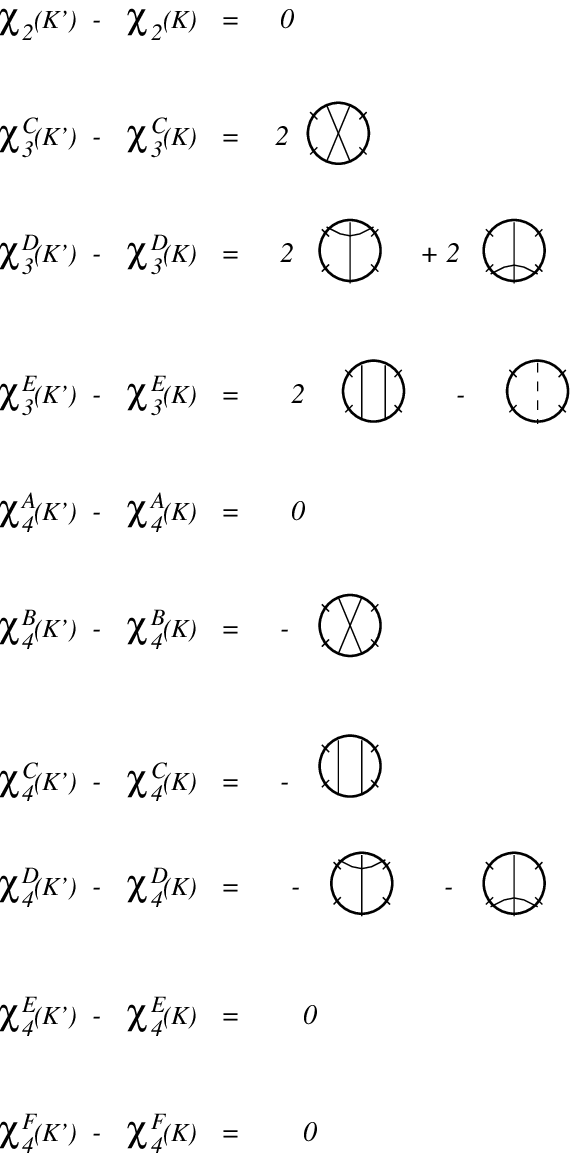}}
\caption{Behaviour of crossing functions under Reidemeister IIB.}
\label{crosIIB}
\end{figure}

\begin{figure}
\centerline{\hskip.4in \epsffile{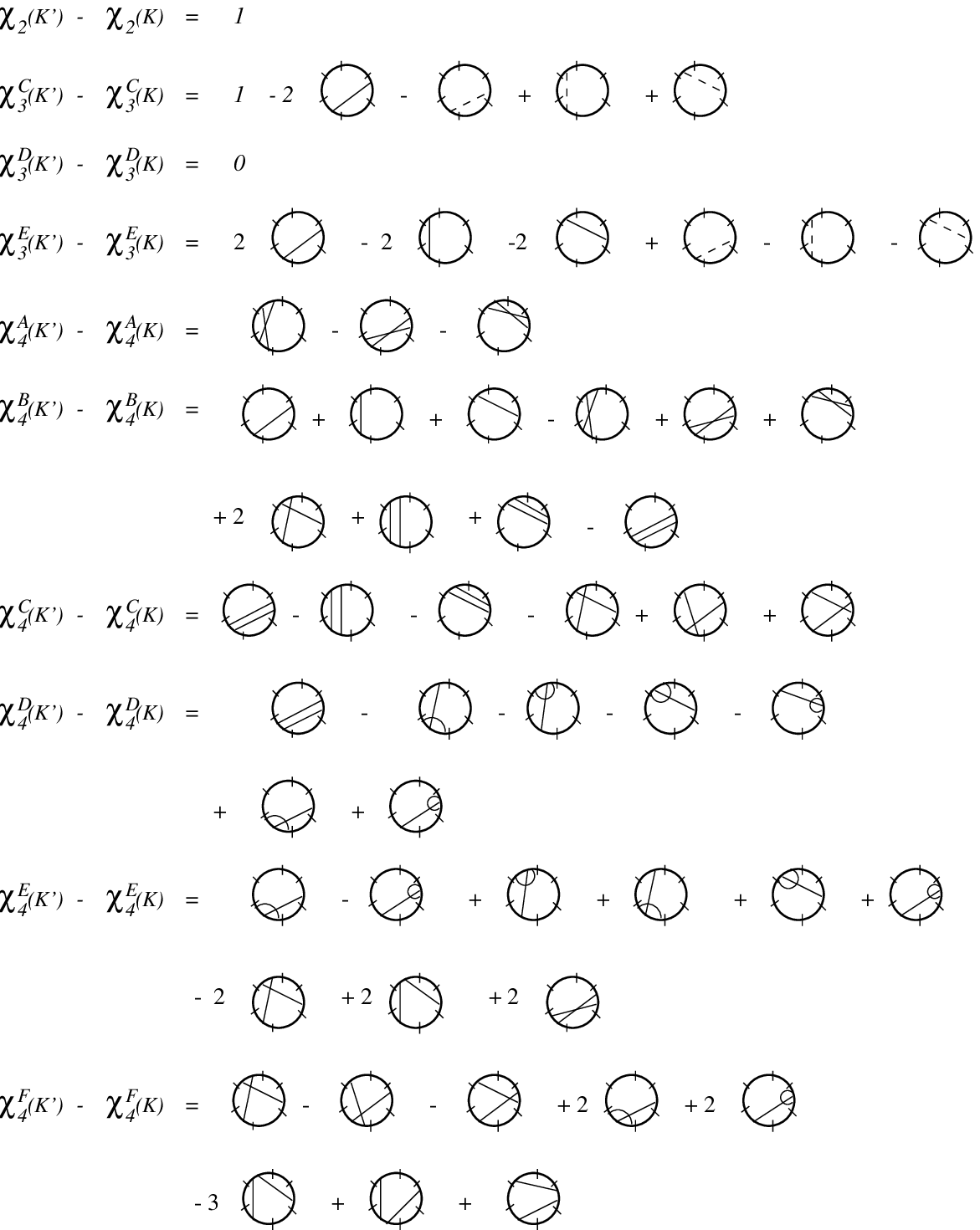}}
\caption{Behaviour of crossing functions under Reidemeister III.}
\label{crosIII}
\end{figure}

All the variations in the previous equations possess simple diagrammatic
expressions. They have been depicted in  figs.
\ref{crosIIA}, \ref{crosIIB} and \ref{crosIII}.

Applying the formulae for the variations to the expressions 
(\ref{primfourm})
and (\ref{primfourn}) for $\alpha_{42}$ and $\alpha_{43}$, we obtain the 
transformation under the moves of all the terms containing only crossing 
numbers. 
The behaviour of the terms containing sums over ${\cal C}_a$ and ${\cal
C}_b$ has to be studied separately. One has to analyse the three different
cases: both  crossings $i$ and $j$ belong  to the  set affected by the
moves, only one of them, or none. Recall that these sums are essentially 
made
out of the crossing number $\chi_2^A$ evaluated in the ascending diagram of
different closed pieces of the knot. In order to clarify the analysis, let us
reproduce those terms here: 
\bear 
{\hskip -.4cm} I_1(\k) &=&  \sum_{i>j \in {\cal C}_a } \big[
\epsilon_i \epsilon_j(\k) - \epsilon_i  \epsilon_j(\alpha(\k)) \big] \,  
\bigg\{ \, \chi_2^A(\alpha(\k)) \nonumber \\
\, \nonumber \\
&-&   \chi_2^A(\alpha(\k_{i_+})) - \chi_2^A(\alpha (\k_{i_-})) 
- \chi_2^A(\alpha(\k_{j_+})) - \chi_2^A(\alpha (\k_{j_-})) \nonumber \\
\, \nonumber \\ 
&+&  \chi_2^A(\alpha (\k_{ij_a}))  + 
\chi_2^A(\alpha (\k_{ij_b})) \bigg\}, \label{iuno} \\
\, \nonumber \\
{\hskip -.4cm} I_2 (\k) &=&  \sum_{i>j \in {\cal C}_a }\big[ \epsilon_i 
\epsilon_j(\k) 
- \epsilon_i 
\epsilon_j(\alpha(\k)) \big] \, \bigg\{ \,
 3 \, \chi_2^A(\alpha(\k)) \nonumber \\
\, \nonumber \\ 
&-&  2  \chi_2^A(\alpha(\k_{i_+}))-  2 
\chi_2^A(\alpha (\k_{i_-})) 
-   2 \chi_2^A(\alpha(\k_{j_+})) - 2 \chi_2^A(\alpha (\k_{j_-}))\nonumber \\
\, \nonumber \\ 
&+& \chi_2^A(\alpha (\k_{ij_a})) + 
\chi_2^A(\alpha (\k_{ij_b}) \bigg\}, \label{idos} \\
\, \nonumber \\
{\hskip -.4cm} I_3 (\k) &=& \sum_{i>j \in {\cal C}_b} \big[ \epsilon_i 
\epsilon_j(\k) - 
\epsilon_i 
\epsilon_j(\alpha(\k)) \big] \,  \bigg\{ \,
 \chi_2^A(\alpha(\k)) \nonumber \\
\, \nonumber \\
&-&  \chi_2^A(\alpha(\k_{i_+}))- \chi_2^A(\alpha (\k_{i_-})) 
- \chi_2^A(\alpha(\k_{j_+})) - \chi_2^A(\alpha (\k_{j_-})) \nonumber \\
\, \nonumber \\ &+& 
\chi_2^A(\alpha (\k_{ij_c})) + 
\chi_2^A(\alpha (\k_{ij_d})) + \chi_2^A(\alpha (\k_{ij_e})) \bigg\}. 
\nonumber \\ \,
\label{itres}
\eear
These three expressions possess the same structure, so we will refer
to them in the following compact way, whenever we do not need
to take into account particular details:
\beq I_k(\k) = \sum_{i>j \in c(k)}  \big[ \epsilon_i \epsilon_j(\k) - 
\epsilon_i 
\epsilon_j(\alpha(\k)) \big] \, F^k_{ij} (\k),
\label{compac}
\eeq
with $k=1,2,3$,  $F^k_{ij}$ standing for the combination of functions
entering  a given sum. The superindex in $F^k_{ij}$ denotes that this
combination depends on the sum, and the subindexes that it also depends on 
the
pair of crossings. The set over which the sum is taken is  specified by 
$c(k)$,
where $c(k) = {\cal C}_a$ for $k=1,2$ and $c(k) = {\cal C}_b$ for $k=3$.

Under R-I the variation of all these sums is trivially zero for the same
 reasons as stated above: as the crossing $e$ involved in this move is 
isolated,
there is no other crossing that could give a contribution to any of the
$\chi_2^A$ functions. As we have already seen, the variation of all the 
other
terms in $\alpha_{42}(K)$ and $\alpha_{43}(K)$ also vanishes; then it 
follows
trivially that our formulae (\ref{primfourm}) and (\ref{primfourn}) are
invariant under this move.

Under R-IIA or R-IIB the general behaviour of the sums can be written as:
\bear
I_k(\k') - I_k(\k) &=& \sum_{i>j \neq \{e,f\} \atop i,j \in c(k) }  
\big[ \epsilon_i \epsilon_j(\k) - \epsilon_i 
\epsilon_j(\alpha(\k)) \big] \, \big( F^k_{ij} (\k') - F^k_{ij} (\k) \big) 
\nonumber \\
&+& \sum_{i \neq  \{e,f\} \atop i,e \in c(k) } \epsilon_e 
\big[ \epsilon_i (\k) - \epsilon_i  (\alpha(\k)) \big] \,  F^k_{ie} (\k')
\label{gendos} \\ &+& \sum_{i \neq  \{e,f\} \atop i,f \in c(k) } \epsilon_f
\big[ \epsilon_i (\k) - \epsilon_i  (\alpha(\k)) \big] \,  F^k_{if} (\k'). 
\nonumber \eear
Notice that in this equation there are no additional terms 
proportional to $\epsilon_e \epsilon_f$ because the diagram in fig.
\ref{reidIII} has been chosen so that
\beq
\epsilon_e \epsilon_f (\k) - \epsilon_e \epsilon_f (\alpha(\k)) = 0.
\label{zeroo}
\eeq
Recall that all the crossings $i,j \neq \{e,f\}$ do not change when going 
from
$\k'$ to $\k$. In the last two terms, there is no subtraction of the 
function
$F^k(\k)$ because the crossings $e$ and $f$ are not present in $\k$. 

After these general comments on (\ref{compac}) we will evaluate 
(\ref{gendos})
and then we will find out the behaviour of (\ref{iuno}), (\ref{idos}) and
(\ref{itres}) under the second Reidemeister move. We need to specify the two
labellings on each crossing, as in the case of the crossing numbers, to
distinguish on which section of $\k$ they lay. We will write the signature
function as given in (\ref{signatura}) and the labels will fulfil the 
ordering
$i_1<i_2<i_3<i_4$. Under R-IIA we find:  
\bear I_1(\k') - I_1(\k) &=&
- \sum_{ i_1, i_2 \in (f^-, e^+) \atop i_3, i_4 \in (f^+, e^-)}  
\Big[ \epsilon
(i_1, i_3) \epsilon(i_2, i_4)(\k) - \epsilon (i_1, i_3) \epsilon (i_2, i_4)
(\alpha(\k)) \Big], \nonumber \\ \, \nonumber \\ I_2(\k') - I_2(\k) &=& 
\bigg\{
- 3 { \hskip -.5cm \sum_{ i_1, i_2 \in (f^-, e^+) \atop i_3, i_4 
\in (f^+, e^-)}
} 
  - { \hskip -.25cm \sum_{ i_1, i_2 , i_3 \in (f^-, e^+) \atop  i_4 
\in (f^+, e^-)} } - { \hskip -.5cm \sum_{ i_1 \in (f^-, e^+) 
\atop i_2, i_3, i_4
\in (f^+, e^-)} } \bigg\}\nonumber \\ \, \nonumber \\ &\times& 
\Big[ \epsilon
(i_1, i_3) \epsilon(i_2, i_4)(\k)  - \epsilon (i_1, i_3) \epsilon (i_2, i_4)
(\alpha(\k)) \Big], \nonumber \\ \, \nonumber \\
I_3(\k') - I_3(\k) &=& - \sum_{ i_1, i_2 \in (f^-, e^+) 
\atop i_3, i_4 \in (f^+, e^-)} \Big[ \epsilon (i_1, i_4) 
\epsilon(i_2, i_3)(\k)
- \epsilon (i_1, i_4) \epsilon (i_2, i_3) (\alpha(\k)) \Big], 
\nonumber \\ \, 
\label{idosa}
\eear
and, under R-IIB:
\bear
I_1(\k') - I_1(\k) &=& - \sum_{ i_1, i_2 \in (f^-, f^+) \atop i_3, i_4 
\in (e^+, e^-)} \Big[ \epsilon (i_1, i_3) \epsilon(i_2, i_4)(\k) - \epsilon
(i_1, i_3) \epsilon (i_2, i_4) (\alpha(\k)) \Big], \nonumber \\ \, \nonumber 
\\
I_2(\k') - I_2(\k) &=& -  \,\sum_{ i_1, i_2 \in (f^-, f^+) \atop i_3, i_4 
\in (e^+, e^-)} \Big[ \epsilon (i_1, i_3) \epsilon(i_2, i_4)(\k) - \epsilon
(i_1, i_3) \epsilon (i_2, i_4) (\alpha(\k)) \Big], \nonumber \\ \, \nonumber 
\\
I_3(\k') - I_3(\k) &=& 0.
\label{idosb}
\eear
The invariance under the second Reidemeister move of $\alpha_{42}(K)$ in
(\ref{primfourm}) and $\alpha_{43}(K)$ in (\ref{primfourn}) then follows,
after summing up the contributions coming from the crossing numbers in
(\ref{todosrda}) and (\ref{todosrdb}), and the expressions  (\ref{idosa}) 
and
(\ref{idosb}), respectively, and finding out that they cancel.

To prove the invariance under R-III we will study first the behaviour of the
$I_k$ sums (\ref{iuno}$-$\ref{itres}) under R-III. As in the previous case,
we start by writing down the general structure of their variation:
\bear I_k(\k') - I_k(\k) &=& \sum_{i>j \neq \{e,f,g\} \atop i,j \in c(k) }  
\big[
\epsilon_i \epsilon_j(\k) - \epsilon_i  \epsilon_j(\alpha(\k)) \big] \, 
\big(
F^k_{ij} (\k') - F^k_{ij} (\k) \big)  \nonumber \\ &+& \sum_{i \neq  
\{e,f,g\}
\atop i,e \in c(k) } \epsilon_e \big[ \epsilon_i (\k) - \epsilon_i  
(\alpha(\k))
\big] \, \big(  F^k_{ie} (\k') - F^k_{ie} (\k) \big)  \nonumber \\ &+& 
\sum_{i
\neq  \{e,f,g\} \atop i,f \in c(k) } \epsilon_f \big[ \epsilon_i (\k) -
\epsilon_i  (\alpha(\k)) \big] \, \big( F^k_{if} (\k') - F^k_{if} (\k) \big)
\nonumber \\ &+& \sum_{i \neq  \{e,f,g\} \atop i,g \in c(k) } \epsilon_g 
\big[
\epsilon_i (\k) - \epsilon_i  (\alpha(\k)) \big] \, \big( F^k_{ig} (\k') -
F^k_{ig} (\k) \big). \nonumber \\ \, \label{gentres} \eear 
Again, the terms proportional to $\epsilon_e \epsilon_f$, $\epsilon_e 
\epsilon_g$
or $\epsilon_g \epsilon_f$ do not contribute because their signature values 
do
not vary when going from $\k$ to $\alpha(\k)$. The computation of the 
variation
of the different $F^k$ functions appearing in (\ref{gentres}) 
is more complicated than before. 
Instead of changes in the crossings involved in the move, we are now dealing 
with
a change in the configuration of the three crossings affected by the move. This
implies that many of the crossings contributing to the
functions $\chi_2^A$ change.
 For example, the 
value of the subtraction  $\chi_2^A(\alpha(\k_{ij_a}')) - 
\chi_2^A(\alpha(\k_{ij_a}))$ (or any other of the functions evaluated in the  
splitt knot)  depends on the sections of the knot in which the 
crossings $i$ and $j$ lay in between. Let us work out some examples. 
In these examples, we will specify both labels of the crossings: $i_1<i_2$ 
for
one and $j_1<j_2$ for the other. 

If a crossing happens to have all the labels in the 
region $(e^-, e^+)$ we find:
\beq
\chi_2^A(\alpha(\k_{ij_a}')) - \chi_2^A(\alpha(\k_{ij_a})) = 
- \epsilon_e \epsilon_f -  \epsilon_e \epsilon_g - \epsilon_g \epsilon_f = 1 
,
\label{exuno}
\eeq
while if the situation is $i_1 \in (e^-, e^+)$ 
and $i_2, j_1, j_2 \in (f^-, f^+)$:
\beq
\chi_2^A(\alpha(\k_{ij_a}')) - \chi_2^A(\alpha(\k_{ij_a})) = 0.
\label{exdos}
\eeq

In some other cases there are apparently  
non-trivial contributions linear in the signature of the crossings $e$, $f$ 
or
$g$. An example is  $\chi_2^A(\alpha(\k_{ie_a}')) -
\chi_2^A(\alpha(\k_{ie_a}))$  for some crossing $i$ such that $i,e \in  
{\cal
C}_a$ and whose two labels, $i_1<i_2$, lay in the following knot regions: 
$i_1
\in (e^-, e^+)$ and  $i_2 \in (g^+, g^-)$. We then find that: 
\bear
&& \chi_2^A(\alpha(\k_{ie_a}')) - \chi_2^A(\alpha(\k_{ie_a})) = \epsilon_f 
\,
\bigg[ - {\hskip -.5cm  \sum_{j_1 \in (e^-, i_1) \atop j_2 \in (f^-, f^+) }}
\epsilon(j_1, j_2) (\alpha(\k)) + \nonumber \\
 && \; {\hskip -1cm \sum_{j_1 \in (f^-, f^+) \atop j_2 \in (g^+, i_2) }} 
 \epsilon(j_1, j_2) (\alpha(\k)) \bigg] 
- \epsilon_g \, \bigg[ {\hskip -.2cm  \sum_{j_1 \in (e^-, i_1) 
\atop j_2 \in (f^-, f^+) }} \epsilon(j_1, j_2) (\alpha(\k)) - {\hskip -.5cm
\sum_{j_1 \in (f^-, f^+) \atop j_2 \in (g^+, i_2) }}  \epsilon(j_1, j_2)
(\alpha(\k)) \bigg]. \nonumber \\ \, \label{pampa}
\eear
The minus  sign in front of some of the terms inside the brackets is due to 
the fact that in order to close the knot $\k_{ie_a}$, we had to reverse the 
orientation in
some piece of the original knot; this implies a change of sign in the 
signature
functions affected by this reversing. In this example we are reversing the
orientation of the region $(e^-, i_1)$.  The key point is to notice that  to the
contributions inside the brackets sum up to the linking number between some 
specific knots and that this linking number is always zero: 
\bear
\chi_2^A(\alpha(\k_{ie_a}')) - \chi_2^A(\alpha(\k_{ie_a})) &=&  \epsilon_f 
\; 
\cdot {\cal L} \big( \alpha(\k_{ie_a}^{f_+}), \alpha(\k_{ie_a}^{f_-}) \big) 
\nonumber \\ &-&  \epsilon_g \;  \cdot {\cal L} \big( 
\alpha(\k_{ie_a}^{g_+}),
\alpha(\k_{ie_a}^{g_-}) \big) = 0, \label{examp} 
\eear
where ${\cal L}(\k_1,\k_2)$ stands for the linking number between $\k_1$ and
$\k_2$. In the first term the knots are the two pieces into which the knot
$\alpha(\k_{ie_a})$ is divided  when  splitting the $f$-crossing, and in 
the
second those obtained after the splitting of the $g$-crossing. 
As the knot
$\alpha(\k_{ie_a})$ is just an ascending  diagram, these two pieces lay one
on top of
the other, and so their linking number is zero. 
This kind of argument can be
applied to  other contributions of the same type (\ie\ for other choices of
crossings in the sums $I_k$ and other $\chi_2^A$ functions appearing in 
them).
The computation of all the contributions to (\ref{gentres}) of the sums
(\ref{iuno} $-$ \ref{itres}) can  now be done without difficulty, leading to 
the
following formulae for their behaviour under R-III (where again
$i_1<i_2<i_3<i_4$):
 \bear 
 && I_1(\k') - I_1(\k) = \bigg\{ {\hskip -.65cm
\sum_{i_3, i_4 \in (g^+, g^-) \atop i_1 \in (e^-, e^+), i_2 \in (f^-, f^+) 
}} +
{\hskip -.5cm \sum_{ i_1 \in (e^-, e^+), i_4 \in (g^+, g^-) \atop i_2, i_3 
\in
(f^-, f^+)  }} + {\hskip -.5cm \sum_{i_1, i_2 \in (e^-, e^+) \atop i_3 \in 
(f^-,
f^+), i_4 \in (g^+, g^-) }} {\hskip -.5cm \bigg\} } \nonumber \\ 
\, \nonumber \\
&& \times \; \epsilon(i_1, i_3) \epsilon(i_2, i_4) \; + \, 2 \,{\hskip -.5cm
\sum_{i_2 \in (g^+, g^-) \atop i_1 \in (f^-, f^+) }} \epsilon(i_1, i_2),
\nonumber \\ \,\nonumber \\
\,\nonumber \\ 
&& I_2(\k') - I_2(\k) = 3\, \bigg\{ {\hskip -.65cm
\sum_{i_3, i_4 \in (g^+, g^-) \atop i_1 \in (e^-, e^+), i_2 \in (f^-, f^+) 
}} +
{\hskip -.5cm \sum_{ i_1 \in (e^-, e^+), i_4 \in (g^+, g^-) \atop i_2, i_3 
\in
(f^-, f^+)  }} + {\hskip -.5cm \sum_{i_1, i_2 \in (e^-, e^+) \atop i_3 \in 
(f^-,
f^+), i_4 \in (g^+, g^-) }} {\hskip -.5cm \bigg\} } \nonumber \\  \, 
\nonumber 
\\ 
 && \times \; \epsilon(i_1, i_3)  \epsilon(i_2, i_4)  
 + \bigg\{ {\hskip -.65cm
\sum_{i_1 \in (e^-, e^+) \atop i_2, i_3, i_4 \in (f^-, f^+) }} + {\hskip 
-.5cm
\sum_{ i_2, i_3, i_4 \in (g^+, g^-)
 \atop  i_1 \in (f^-, f^+)  }} + {\hskip -.5cm \sum_{i_1 \in (e^-, e^+) 
\atop i_2, i_3, i_4 \in (g^+, g^-) }} {\hskip -.5cm \bigg\}  \epsilon(i_1, 
i_3)
\epsilon(i_2, i_4) } \nonumber \\  
\, \nonumber \\
 && + \; \bigg\{ {\hskip -.65cm
\sum_{i_4 \in (f^-, f^+) \atop i_1, i_2, i_3 \in (e^-, e^+) }} + {\hskip 
-.5cm
\sum_{i_1, i_2, i_3 \in (f^-, f^+)
 \atop  i_4 \in (g^+, g^-)  }} + {\hskip -.5cm \sum_{i_4 \in (g^+, g^-) 
\atop i_1, i_2, i_3 \in (e^-, e^+) }} {\hskip -.5cm \bigg\}  \epsilon(i_1, 
i_3)
\epsilon(i_2, i_4) } 
 + \; 6 \, {\hskip -.5cm
\sum_{i_2 \in (g^+, g^-) \atop i_1 \in (f^-, f^+) }} \epsilon(i_1, i_2),
\nonumber \\  
\,\nonumber \\
\,\nonumber \\
&& I_3(\k') - I_3(\k) = \bigg\{ {\hskip -.65cm \sum_{i_3, i_4 \in (g^+, g^-)
\atop i_1 \in (e^-, e^+), i_2 \in (f^-, f^+) }}   + {\hskip -.5cm 
\sum_{i_1, i_2
\in (e^-, e^+) \atop i_3 \in (f^-, f^+), i_4 \in (g^+, g^-) }} {\hskip -.5cm
\bigg\}  \epsilon(i_1, i_4) \epsilon(i_2, i_3) } \nonumber \\  \, \nonumber 
\\
&& + \; {\hskip -.5cm \sum_{ i_1 \in (e^-, e^+), i_4 \in (g^+, g^-) \atop 
i_2, 
i_3
\in (f^-, f^+)  }} \epsilon(i_1, i_2) \epsilon(i_3, i_4) \label{morena} \\ 
\,
\nonumber \\ 
&& + \; \bigg\{ {\hskip -.65cm \sum_{i_1, i_2 \in (e^-, e^+) \atop 
i_3, i_4 \in (f^-, f^+) }} + {\hskip -.4cm \sum_{  i_3, i_4 \in (g^+, g^-)
 \atop  i_1, i_2 \in (f^-, f^+)  }} + {\hskip -.4cm \sum_{i_1, i_2 
\in (e^-, e^+) \atop  i_3, i_4 \in (g^+, g^-) }} {\hskip -.2cm \bigg\} 
\epsilon(i_1, i_4) \epsilon(i_2, i_3) }  
\nonumber
\eear

Taking into account (\ref{morena}) and the behaviour under R-III
 of the crossing numbers given in (\ref{todosrt}), one can see that all the 
terms
appearing in computing the variation of  (\ref{primfourm}) and 
(\ref{primfourn})
under R-III  cancel, and thus the topological invariance of 
the
combinatorial expressions (\ref{primfourm}) and (\ref{primfourn}) for
$\alpha_{42}(K)$ and $\alpha_{43}(K)$ is established. 

\vfill
\newpage

\section{Conclusions}
\setcounter{equation}{0}

In this paper we have analysed  Chern-Simons gauge theory in the temporal 
gauge.
The main outcome of our work is that we have shown that this gauge is
particularly well suited to obtain combinatorial expressions for Vassiliev
invariants. These are much simpler  than the integral expressions
obtained in covariant gauges or the ones leading to Kontsevich integrals 
which emerge in the light-cone gauge.

One of the crucial ingredients of our work is the observation that in the 
temporal gauge all the
signature-dependent parts of the invariant can be easily extracted. In fact 
we have obtained an explicit general expression for the leading
signature-dependent terms. These terms  are the ones in the expansion
(\ref{nucleos}) and constitute the kernels of the Vassiliev invariants. The
kernels are not Vassiliev invariants. Different kernels may belong to the 
same
knot, but they are well defined on knot projections. As an order-$n$ 
Vassiliev
invariant, an order-$n$ kernel vanish in signed sums of order $n+1$. The
kernel is the only part of the order-$n$ Vassiliev invariant that in 
general 
does 
not
vanishes for signed sums of order less than $n+1$. In other words, an 
order-$n$
kernel differs from an order-$n$ Vassiliev invariant by terms which vanish 
in
signed sums of order $n$.

The kernels contain a large amount of information about the Vassiliev 
invariants.
We have shown how the full invariants can be reconstructed from them. The 
two
main ingredients of the reconstruction procedure are the factorization 
theorem
and the structure of the perturbative series expansion of the vacuum 
expectation
value of the Wilson loop in the temporal gauge. The key observation of the
reconstruction procedure is that combinatorial expressions can be obtained
without actually performing  any of the $D$-integrals. All these integrals 
are
solved in terms of the kernels using the series of relations provided by the
factorization theorem.

In our analysis we have carried along the unknown function $b(\k)$, 
obtaining a
series of consistency relations for it. These relations are necessary 
conditions
to have knot invariants. We have shown that these relations possess a simple
solution, similar to the one that must be introduced in the light-cone 
gauge. 
It
would be very helpful to understand the origin of this function in the 
context
of Chern-Simons gauge theory, and to prove that, indeed, our ansatz is 
correct.
The same type of problem has not been solved in the light-cone gauge.

The reconstruction procedure has been performed up to order four. We have
obtained known combinatorial expressions at orders two and three, and new 
ones
for the two primitive Vassiliev invariants present at order four. The form 
of
these combinatorial expressions suggests some general structure. For example, 
it
seems that at even orders the quantities that enter  the combinatorial
expressions are paired, one of the  terms for the diagram
associated to $\k$, and another for the corresponding ascending diagram with
opposite sign. In addition, the terms involved in the splitting of a knot 
are
evaluated on the ascending diagram. For odd orders, crossing numbers seem 
not to be accompanied by their ascending-diagram counterparts. However, as in 
the even-order cases, the terms involved in the splitting  are evaluated in the
ascending diagram.

We have successfully applied the reconstruction procedure up to order four,
obtaining new combinatorial expressions for Vassiliev invariants. The 
question
to ask now is if the procedure can be generalized to higher orders. We 
conjecture 
that this can be done. Certainly,
the complexity of the combinatorial expressions will increase with the 
order,
but it would be very important to establish if the procedure would work at 
any
order. In other words, it would be very important to possess a 
reconstruction
theorem which would guarantee that from the kernels (\ref{nucleos}) and the
factorization theorem, we can solve for all the $D$-integrals present at 
each
order. Provided we know a basis of primitive group factors, this would 
imply
that there exists a systematic algorithm to obtain combinatorial expressions 
for
Vassiliev invariants at any order.

\vskip2cm
\begin{center} {\bf Acknowledgements}
\end{center}

\vspace{4 mm}

We would like to thank M. Alvarez for helpful discussions.
This work was supported in part by DGICYT under grant PB97-0960, and by
the EU Commission under TMR grant FMAX-CT96-0012.

\vskip 2cm

\newpage

\vskip 1cm                                               
{\Large{\bf APPENDIX}}                                 
\vskip .5cm                                              
\renewcommand{\theequation}{\rm{A}.\arabic{equation}}    
\setcounter{equation}{0}                                 

\vspace{7 mm} In this appendix we present the values of the primitive Vassiliev 
invariants
at orders two, three and four for all prime knots up to nine crossings.
These have  been
computed, with the aid of a Mathematica algorithm, using the formulae
(\ref{primtwoc}), (\ref{primthreeb}), (\ref{primfourm}) and 
(\ref{primfourn}).
In the tables 1 and 2 we present the value of these invariants once their
value for the unknot has been substracted. In other words, the
$\alpha_{ij}(K)$ shown in the tables are the result of the replacement:
\bear
\alpha_{ij}(K) \longrightarrow  \alpha_{ij}(K) - \alpha_{ij}(U).
\nonumber 
\eear
The values for the unknot primitive invariants up to order four are, in the 
normalization and basis that we used:
\bear
\begin{array}{ll}
 \alpha_{21}(U) = - {1 \over 6}, 
&\alpha_{31}(U) = 0,   \\
\alpha_{42}(U) =  {1 \over 360},
&\alpha_{43}(U) = - {1 \over 360}.
\end{array}
\nonumber 
\eear

\begin{table}[hp]
\begin{center}
\begin{tabular}{|c||c|c|c|c|c|c||c|c|c|c|}\cline{1-5} \cline{7-11}
  Knot & $\alpha_{21}$ & $\alpha_{31}$ &
$\alpha_{42}$ & $\alpha_{43}$ & $\;\;\;\;\;\;$
& Knot & $\alpha_{21}$ & $\alpha_{31}$ & $\alpha_{42}$ & $\alpha_{43}$ \\
  \cline{1-5} \cline{7-11}
 $3_1$ & 4 & 8 & 62/3 & 10/3 &  & $8_5$ & $-$4 & $-$24 & $-$62/3 & 86/3 \\
\cline{1-5} \cline{7-11}
 $4_1$ & $-$4 & 0 & 34/3 & 14/3 &  & $8_6$ & $-$8 & $-$24 & 68/3 & 100/3 \\
\cline{1-5} \cline{7-11}
 $5_1$ & 12 & 40 & 174 & 26 &  & $8_7$ & 8 & $-$16 & 124/3 & $-$28/3 \\
\cline{1-5} \cline{7-11}
 $5_2$ & 8 & 24 & 268/3 & 44/3 &  & $8_8$ & 8 & $-$8 & 124/3 & $-$4/3 \\
\cline{1-5} \cline{7-11}
 $6_1$ & $-$8 & $-$8 & 116/3 & 52/3 &  & $8_9$ & $-$8 & 0 & 212/3 & 124/3 \\
\cline{1-5} \cline{7-11}
 $6_2$ & $-$4 & $-$8 & 34/3 & 38/3 &  & $8_{10}$ & 12 & $-$24 & 110 & 10 \\
\cline{1-5} \cline{7-11}
 $6_3$ & 4 & 0 & 14/3 & $-$14/3 &  & $8_{11}$ & $-$4 & $-$16 & $-$14/3 & 62/3 \\
\cline{1-5} \cline{7-11}
 $7_1$ & 24 & 112 & 684 & 100 &  & $8_{12}$ & $-$12 & 0 & 82 & 30 \\
\cline{1-5} \cline{7-11}
 $7_2$ & 12 & 48 & 222 & 34 &  & $8_{13}$ & 4 & $-$8 & 14/3 & $-$38/3 \\
\cline{1-5} \cline{7-11}
 $7_3$ & 20 & 88 & 1510/3 & 242/3 &  & $8_{14}$ & 0 & 0 & 16 & 16 \\
\cline{1-5} \cline{7-11}
 $7_4$ & 16 & 64 & 1016/3 & 184/3 &  & $8_{15}$ & 16 & 56 &776/3 & 112/3 \\
\cline{1-5} \cline{7-11}
 $7_5$ & 16 & 64 & 968/3 & 136/3 &  & $8_{16}$ & 4 & $-$8 & 14/3 & $-$38/3 \\
\cline{1-5} \cline{7-11}
 $7_6$ & 4 & 16 & 158/3 & 34/3 &  & $8_{17}$ & $-$4 & 0 & 82/3 & 62/3 \\
\cline{1-5} \cline{7-11}
 $7_7$ & $-$4 & 8 & $-$14/3 & $-$10/3 &  & $8_{18}$ & 4 & 0 & 62/3 & 34/3 \\
\cline{1-5} \cline{7-11}
 $8_1$ & $-$12 & $-$24 & 66 & 38 &  & $8_{19}$ & 20 & 80 & 1270/3 & 170/3 \\
\cline{1-5} \cline{7-11}
 $8_2$ & 0 & $-$8 & 0 & 24 &  & $8_{20}$ & 8 & 16 & 172/3 & 20/3 \\
\cline{1-5} \cline{7-11}
 $8_3$ & $-$16 & 0 & 520/3 & 200/3 &  & $8_{21}$ & 0 & $-$8 & $-$16 & 8 \\
\cline{1-5} \cline{7-11}
 $8_4$ & $-$12 & 8 & 114 & 54 &  &  &  &  &  &  \\
\cline{1-5} \cline{7-11}
\end{tabular}
\caption{Primitive Vassiliev invariants up to order four
for all prime knots up to eight crossings.}
\end{center}
\label{tablauno}
\end{table}

\begin{table}[hp]
\begin{center}
\begin{tabular}{|c||c|c|c|c|c|c||c|c|c|c|}\cline{1-5} \cline{7-11}
  Knot & $\alpha_{21}$ & $\alpha_{31}$ &
$\alpha_{42}$ & $\alpha_{43}$ & $\;\;\;\;\;\;$
& Knot & $\alpha_{21}$ & $\alpha_{31}$ & $\alpha_{42}$ & $\alpha_{43}$ \\
  \cline{1-5} \cline{7-11}
 $9_1$ & 40 & 240 & 5660/3 & 820/3 &  & $9_{26}$ & 0 & 8& $-$32 & $-$8 \\
\cline{1-5} \cline{7-11}
 $9_2$ & 16 & 80 & 1304/3 & 184/3 &  & $9_{27}$ & 0 & 8 & 16 & 8\\
\cline{1-5} \cline{7-11}
 $9_3$ & 36 & 208 & 1578 & 246 &  & $9_{28}$ & 4 & 0 & $-$34/3 & $-$14/3 \\
\cline{1-5} \cline{7-11}
 $9_4$ & 28 & 152 & 3122/3 & 502/3 &  & $9_{29}$ & 4 & $-$16 & 62/3 & $-$14/3 \\
\cline{1-5} \cline{7-11}
 $9_5$ & 24 & 120 & 780 & 140 &  & $9_{30}$ & $-$4 & $-$8 & 82/3 & 38/3 \\
\cline{1-5} \cline{7-11}
 $9_6$ & 28 & 144 & 2834/3 & 382/3 &  & $9_{31}$ & 8 & 16 & 172/3 & 20/3 \\
\cline{1-5} \cline{7-11}
 $9_7$ & 20 & 96 & 1654/3 & 218/3 &  & $9_{32}$ & $-$4 & 16 & $-$62/3 & 14/3 \\
\cline{1-5} \cline{7-11}
 $9_8$ & 0 & 16 & 48 & 16 &  & $9_{33}$ & 4 & $-$8 & 62/3 & 10/3 \\
\cline{1-5} \cline{7-11}
 $9_9$ & 32 & 176 & 3760/3 & 560/3 &  & $9_{34}$ & $-$4 & 0 & 34/3 & 14/3 \\
\cline{1-5} \cline{7-11}
 $9_{10}$ & 32 & 176 & 3856/3 & 656/3 &  & $9_{35}$ & 28 & 144 & 
   3026/3 & 574/3 \\
\cline{1-5} \cline{7-11}
 $9_{11}$ & 16 & $-$72 & 1160/3 & 208/3 &  & $9_{36}$ & 12 & $-$56 & 270 & 42 \\
\cline{1-5} \cline{7-11}
 $9_{12}$ & 4 & 24 & 302/3 & 58/3 &  & $9_{37}$ & $-$12 & 8 & 82 & 22 \\
\cline{1-5} \cline{7-11}
 $9_{13}$ & 28 & 144 & 2930/3 & 478/3 &  & $9_{38}$ & 24 & 112 & 684 & 100 
\\
\cline{1-5} \cline{7-11}
 $9_{14}$ & $-$4 & 16 & $-$110/3 & $-$34/3 &  & $9_{39}$ & 8 & $-$32 & 460/3 & 
116/3 
\\
\cline{1-5} \cline{7-11}
 $9_{15}$ & 8 & $-$40 & 508/3 & 92/3 &  & $9_{40}$ & $-$4 & $-$8 & 34/3 & 38/3 
\\
\cline{1-5} \cline{7-11}
 $9_{16}$ & 24 & 112 & 668 & 84 &  & $9_{41}$ & 0 & 8 & $-$48 & $-$24 \\
\cline{1-5} \cline{7-11}
 $9_{17}$ & $-$8 & 0 & 116/3 & 28/3 &  & $9_{42}$ & $-$8 & 0 & 164/3 & 76/3 \\
\cline{1-5} \cline{7-11}
 $9_{18}$ & 24 & 120 & 748 & 108 &  & $9_{43}$ & 4 & 16 & 254/3 & 82/3 \\
\cline{1-5} \cline{7-11}
 $9_{19}$ & $-$8 & 8 & 68/3 & 4/3 &  & $9_{44}$ & 0 & 8 & 0 & $-$8 \\
\cline{1-5} \cline{7-11}
 $9_{20}$ & 8 & 32 & 412/3 & 68/3 &  & $9_{45}$ & 8 & $-$32 & 412/3 & 68/3 \\
\cline{1-5} \cline{7-11}
 $9_{21}$ & 12 & $-$48 & 238 & 50 &  & $9_{46}$ & $-$8 & $-$24 & 20/3 & 52/3 \\
\cline{1-5} \cline{7-11}
 $9_{22}$ & $-$4 & 8 & 34/3 & $-$10/3 &  & $9_{47}$ & $-$4 & $-$16 & $-$110/3 & 
$-$34/3 
\\
\cline{1-5} \cline{7-11}
 $9_{23}$ & 20 & 88 & 1462/3 & 194/3 &  & $9_{48}$ & 12 & $-$40 & 190 & 42\\
\cline{1-5} \cline{7-11}
 $9_{24}$ & 4 & 16 & 110/3 & 34/3 &  & $9_{49}$ & 24 & 112 & 700 & 116 \\
\cline{1-5} \cline{7-11}
 $9_{25}$ & 0 & 8 & 64 & 24 &  &  & & & & \\
\cline{1-5} \cline{7-11}
\end{tabular}
\caption{Primitive Vassiliev invariants up to order four 
for all prime  knots with nine crossings.}
\end{center}
\label{jpast}
\end{table}

\end{document}